\documentclass[fleqn]{2023SCGE}
\usepackage{xcolor}

\usepackage{longtable}
\usepackage{graphicx}

\usepackage{graphicx}

\newcommand{\apj}{Astrophys.~J.~}
\newcommand{\apjl}{Astrophys.~J.~Lett.~}
\newcommand{\apjs}{Astrophys.~J.~Suppl.~Ser.~}
\newcommand{\mnras}{Mon.~Not.~R.~Astron.~Soc.~}
\newcommand{\aap}{Astron.~Astrophys.~}
\newcommand{\aj}{Astron.~J.~}

\newcommand{\pasj}{Publ.~Astron.~Soc.~Jpn.~}
\newcommand{\nat}{Nature~}

\newcommand{\farcs}{$.\!\!^{\prime\prime}$}

\begin{document}
\ensubject{subject}

\ArticleType{Article}
\Year{2025}
\Month{December}
\Vol{68}
\No{12}
\DOI{10.1007/s11433-025-2747-4}
\ArtNo{129511}
\ReceiveDate{May 16, 2025}
\AcceptDate{July 11, 2025}
\OnlineDate{September 8, 2025}

\title{Confirming HSC strong lens candidates with DESI Spectroscopy. I. Project overview and first results}{Confirming HSC strong lens candidates with DESI Spectroscopy. I. Project overview and first results}

\author[1]{Yiping Shu}{{yiping.shu@pmo.ac.cn}}%
\author[1, 2]{Shen Li}{}

\AuthorMark{Shu}

\AuthorCitation{Shu Yiping and Li Shen}

\address[1]{Purple Mountain Observatory, Chinese Academy of Sciences, Nanjing, Jiangsu, 210023, People’s Republic of China}
\address[2]{School of Astronomy and Space Sciences, University of Science and Technology of China, Hefei, Anhui, 230026, People’s Republic of China}


\abstract{Accurate redshift determinations of both lenses and sources are critical for confirming strong-lens systems and fully realizing their scientific value. However, the thousands of strong-lens candidates now routinely discovered in wide-field imaging surveys make one-by-one follow-up observations impractical. In this work, we investigate the capability and efficiency of large-scale spectroscopic surveys in confirming strong-lens systems. As a case study, we cross-match strong lens candidates identified from the Hyper Suprime-Cam Subaru Strategic Program with Data Release 1 (DR1) of the Dark Energy Spectroscopic Instrument (DESI). We find that DESI DR1 serendipitously observed putative lenses and/or lensed images in approximately 50\% of these candidates. Analyzing the DESI spectra for $\approx 500$ matched candidates that meet our selection criteria, we determine both lens and source redshifts for 27 systems. Additionally, 76 candidate systems feature lensing galaxies at $z > 0.8$, and one candidate system contains a quasar within its lensing galaxy. Applying this approach to other strong-lens candidates will yield many more confirmations, with a further several-fold increase anticipated from the final DESI data release. Our results highlight the growing importance of large-scale spectroscopic surveys in advancing strong lensing discoveries and science.}



\maketitle


\begin{multicols}{2}
\section{Introduction}\label{section1}

Strong gravitational lensing is the phenomenon of forming multiple images of a background source object by the gravity of a foreground lens object. Owing to its remarkable sensitivity to gravity and cosmology, strong lensing serves as a powerful tool for probing a wide range of astrophysical questions. For instance, by measuring the overall mass distributions of the lensing objects, strong lensing has provided 
\Authorfootnote

\noindent observational evidence for key processes involved in galaxy formation and evolution such as the stellar/AGN feedback, the stellar initial mass function (IMF), merging, and more (e.g., \cite{Koopmans06, Auger10, Bolton12a, Brewer14, Shu2015, Shajib21, Deng2025}). In-depth analyses of strong lens systems have yielded valuable insights into the nature of dark matter (e.g., \cite{Vegetti12, Nierenberg14, Shu16a, Meneghetti2020, Gilman2020, Amruth2023}). Cosmological constraints have also been derived using systems involving variable sources (i.e. quasars or supernovae) or sources at different redshifts (e.g., \cite{Jullo2010, Suyu13, Collett2014, Wong20, Caminha2022, Kelly2023}). 

The small sample size, however, has long been a significant limiting factor for such applications. Despite dedicated efforts over the past few decades, only a few hundred strong lenses have been confirmed to date (i.e. with spectroscopic redshifts for both lenses and sources) (e.g, \cite{Bolton08, Treu11, Auger11, Brownstein12, Courbin12, More12, Stark13, Shu16b, Shu17, Lemon18, Shu18, Shu19}). Thanks to recent wide-field imaging surveys, several thousand promising strong-lens candidates have been discovered (e.g., \cite{Jacobs19b, Petrillo19, Canameras20, holismokesVI, Huang20, Li20, Huang21, Li21, Stein21, Rojas21, Savary21, holismokesVIII, Andika23, holismokesXIII, He25, He25b, holismokesXVI}). These large samples will not only greatly enhance statistical power but also open the door to new and exciting discoveries. Nevertheless, most lensing-related applications critically depend on accurate redshift measurements for both the lens and the source --- a need that has emerged as a new bottleneck. 

Fortunately, large-scale fiber-fed spectroscopic surveys, including the ongoing Dark Energy Spectroscopic Instrument (DESI) Survey (DESI, \cite{DESI}) and the Prime Focus Spectrograph (PFS, \cite{PFS}), as well as planned or proposed missions such as the 4-metre Multi-Object Spectroscopic Telescope (4MOST, \cite{4MOST}), the Extremely Large Spectroscopic Survey Telescope (ESST, \cite{ESST}), the Jiao Tong University Spectroscopic Telescope (JUST, \cite{JUST}), the MegaMapper project \cite{MegaMapper}, The Maunakea Spectroscopic Explorer (MSE, \cite{MSE}), and the MUltiplexed Survey Telescope (MUST, \cite{MUST}), have the capabilities to simultaneously acquire redshifts for thousands to tens of thousands of objects. In fact, there have been proposals of following up strong lens candidates with those spectroscopic surveys (e.g., \cite{Collett23, Huang25}). Naturally, it is also entirely possible that some strong lens candidates may be unintentionally targeted by spectroscopic surveys (i.e. without prior knowledge of their lensing status).

We are therefore interested in investigating the feasibility and efficiency of confirming imaging-selected strong lens candidates using large-scale spectroscopic surveys. As a demonstration, we initiated this project to cross-match strong lens candidates discovered from the Hyper Suprime-Cam Subaru Strategic Program (HSC-SSP, \cite{Aihara2018}) with the recent Data Release 1 (DR1) of the DESI survey \cite{DESI_DR1}. This paper is structured as follows. Section~\ref{sec:data} provides a brief summary of the HSC strong lens candidates and DESI DR1. The cross-matching results are presented in Section~\ref{sec:results} and the \ref{sec:appendix}. Discussions and conclusions are given in Sections~\ref{sec:discussions} and \ref{sec:conclusions}. 

\section{Data}\label{sec:data}
\subsection{Strong lens candidates from the HSC-SSP}

The HSC-SSP survey is a wide-field imaging survey carried out with the Subaru Telescope. It is designed to acquire multi-band ($grizy$) images over more than 1,400 deg$^2$ sky with a typical depth of $i \gtrsim 26$ mag and a typical seeing of 0\farcs7 in the $i$ band. The survey started in March 2014 and its third public data release took place in April 2022 (PDR3, \cite{PDR3}). Thanks to its high image quality and wide sky coverage, several teams have conducted strong-lens searches in the HSC-SSP data, employing a variety of techniques including automated algorithms, machine learning, and crowdsourcing \cite{sugohiI, sugohiII, sugohiIV, sugohiV, sugohiVI, sugohiVII, holismokesVI, holismokesVIII, sugohiVIII, sugohiIX, sugohiX, holismokesXIII, holismokesXVI}. In this work, we used the catalog compiled at \url{https://www-utap.phys.s.u-tokyo.ac.jp/~oguri/sugohi/}, which, at the time of writing, contained 3,961 strong lens candidates discovered from HSC-SSP. For the purpose of this work, we removed 11 systems from this catalog that already have spectroscopic redshifts for both lenses and sources. The rest of the analyses are thus based on the remaining 3,950 HSC strong lens candidates. In the catalog, each candidate is assigned a grade of A, B, or C that is meant to represent definite lens, probable lens, or possible lens based on visual inspections of the HSC imaging data. Breaking it down, there are 319 grade-A, 1262 grade-B, and 2369 grade-C. 

\subsection{DESI DR1}

The DESI survey is a large-scale optical spectroscopic survey being conducted with the Mayall 4-meter Telescope starting from May 2021. Equipped with 5,000 robotic fibers, the DESI survey will measure precise redshifts for $\approx 50$ million galaxies and quasars by the end of its five-year mission. On March 19, 2025, the DESI collaboration made the first release of its Main Survey (DESI DR1), including spectra for almost 19 million unique objects observed during the first year. We downloaded the DESI DR1 redshift catalog \texttt{zall-pix-iron.fits} from \url{https://data.desi.lbl.gov/public/dr1/spectro/redux/iron/zcatalog/v1/}, which contains 28,425,963 entries. In addition to redshifts, the catalog provides the object type (\texttt{spectype}), as determined by best-fit models to the DESI spectra. In the DR1 catalog, 21,696,490 entries have \texttt{spectype=GALAXY}, 1,862,583 have \texttt{spectype=QSO}, and 4,866,890 have \texttt{spectype=STAR}. 

\section{Cross-matching results}\label{sec:results}

We cross-matched the list of HSC strong lens candidates with the DESI DR1 redshift catalog using a matching radius of 6\farcs0. This matching radius was chosen in accordance with the typical separations between lenses and their lensed images. The obtained 3,431 matches corresponded to 2,111 unique HSC strong-lens candidate systems or 2,621 unique DESI spectra. Among these matches, 224 are grade-A, 775 are grade-B, and 1,112 are grade-C. We then downloaded the DESI spectra for the matched cases from the Astro Data Lab \cite{Fitzpatrick14, Nikutta20} for visual inspections. The goals of the visual inspections are twofold: 1) verifying the DESI redshifts provided in the catalog; 2) searching for spectral features (mainly emission or absorption lines) at redshifts different from the reported redshifts, an idea motivated by previous spectroscopic searches for strong lenses (e.g., \cite{Bolton08, Treu11, Auger11, Brownstein12, Shu16}). 

In this work, we focused on three categories chosen based on our specific interests, namely, 
\vspace{-1.0\parsep}
\begin{enumerate}
\itemsep -1.0\parsep
    \item Grade-A candidates with both lens and source redshifts determined by DESI spectra and thus considered as confirmed cases;
    \item Grade-A/B candidates with lensing galaxies spectroscopically confirmed to be at $z > 0.8$;
    \item Candidates containing quasars;      
\end{enumerate}

\subsection{Category 1: Grade-A candidates with secure lens and source redshifts}

For this category, we began with the 210 grade-A candidates that are successfully matched to DESI DR1 and have \texttt{spectype=GALAXY}. By visually inspecting the corresponding 305 unique DESI spectra, we identified eight cases where both the lens and source redshifts are robustly measured by the DESI spectra. In five other cases, the lens redshifts are robustly measured and multiple emission lines from the source galaxies are also clearly detected in the spectra. We measured their source redshifts by simultaneously fitting the detected emission lines with Gaussian profiles. In another four cases, the lens redshifts are robustly measured and single asymmetric emission lines are detected, which we interpreted as Ly$\alpha$ emission from the sources. We measured their source redshifts by fitting the detected emission lines with a skewed Gaussian profile. For one case (114311$-$013935), the source redshift is robustly measured, while the DESI model fit to the lens spectrum was affected by the strong [O\textsc{ii}] emission from the source. By masking regions around the [O\textsc{ii}] doublet and fitting the lens spectrum with a linear combination of galaxy eigenspectra, we successfully measured the lens redshift. Discussions on individual systems are presented in Section~\ref{sec:C1_Appendix}, and a summary table is given in Table~\ref{tab1}. Image cutouts and DESI spectra for this category are presented in Figure~\ref{fig:C1}.

\subsection{Category 2: Grade-A/B candidates with $z > 0.8$ lensing galaxies}

For this category, we began with the 241 grade-A/B candidates that are successfully matched to DESI DR1 and have \texttt{spectype=GALAXY} and reported DESI redshifts above 0.8. By visually inspecting the corresponding 321 unique DESI spectra, we identified six cases where the lens redshifts are robustly measured and single asymmetric emission lines are detected, which we interpreted as Ly$\alpha$ emission from the sources. We measured their source redshifts by fitting the detected emission lines with a skewed Gaussian profile. In one case, the lens redshift is robustly measured and two emission lines (resembling the [O\textsc{ii}] doublet) from the source galaxy are detected in the spectra. We measured its source redshift by simultaneously fitting the detected emission feature with two Gaussian profiles. In another 76 cases, the redshifts of the putative lensing galaxies are robustly measured. Discussions on selected systems are presented in Section~\ref{sec:C2_Appendix}, and a summary table is given in Table~\ref{tab2}. Image cutouts and DESI spectra for this category are presented in Figure~\ref{fig:C2}.

\subsection{Category 3: Candidates with QSOs as sources/lenses}\label{sec:C3}

For this category, we began with the 69 strong-lens candidates that are successfully matched to the DESI DR1 and have \texttt{spectype=QSO}. By visually inspecting the corresponding 77 unique DESI spectra, we identified two cases (121533$-$005842 and 122018$+$011253) where both the lens and source redshifts are robustly measured by the DESI spectra and are considered as confirmed lensed QSO systems. In six other cases, one of the two or four point sources is spectroscopically confirmed as a broad-line quasar, and a putative lensing galaxy is visible after subtracting light from all point sources. In one case (130733$+$001122), although no lensing galaxy is detected after subtracting the two point sources, DESI spectra are available for both point sources, confirming them being broad-line quasars. In addition, these DESI spectra share strong similarities in the continua, broad emission lines, and absorption features. In another case (092121$+$031744), the putative lensing galaxy is spectroscopically confirmed to contain a quasar at $z = 0.346027$. Discussions on individual systems are presented in Section~\ref{sec:C3_Appendix}, and a summary table is given in Table~\ref{tab3}. Image cutouts and DESI spectra for this category are presented in Figure~\ref{fig:C3}.

\section{Discussions}\label{sec:discussions}

Combining the three categories, we spectroscopically confirmed 27 HSC strong-lens systems using DESI DR1. Among them, eight systems contain high-redshift lensing galaxies (lens redshifts ranging from 0.81443 to 0.9983), and two systems are strongly-lensed quasars. In addition, 76 HSC strong-lens candidates were spectroscopically confirmed to contain $z > 0.8$ lensing galaxies and seven were identified as lensed quasar candidates. We also discovered a rare configuration where the putative lensing galaxy contains a quasar. It should be noted that this is not an exhaustive list, as our selective visual inspections were limited to a subset ($\approx 25\%$) of the 2,111 unique HSC strong-lens candidates that were matched to DESI DR1. Moreover, the full DESI mission will deliver spectra for $\approx$50 million unique galaxies and quasars, representing a 3.4-fold increase over DESI DR1. Based on simple scaling estimates, the final DESI data release (expected within the next few years) is projected to confirm $\approx$100 HSC strong-lens systems and provide lens or source redshift measurements for nearly all existing HSC strong-lens candidates. 

\begin{figure}[H]
    \centering
    \includegraphics[width=0.45\textwidth]{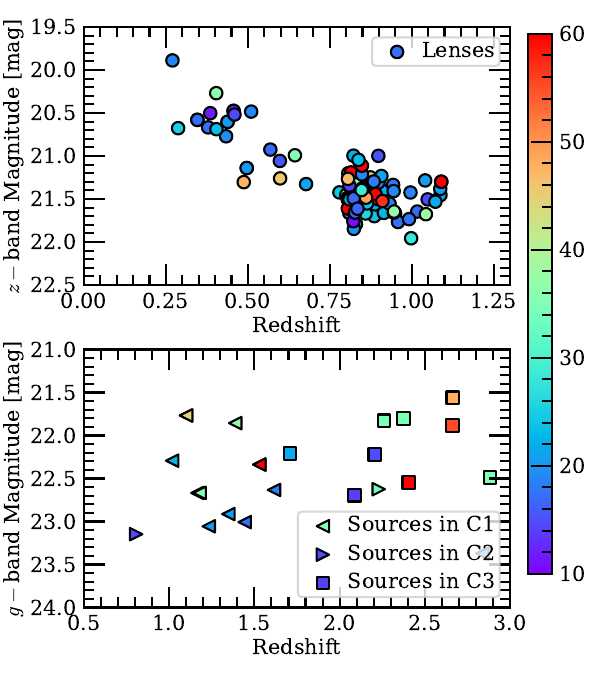}
    \vspace{-0.2cm}
    \caption{Redshift and magnitude distributions of the lenses (top) and lensed images (bottom) in this work. Sources from categories 1, 2, and 3 are represented by left-pointing triangles, right-pointing triangles, and squares, respectively. All symbols are color-coded according to the total exposure time of the spectroscopic observations (from 10 minutes to 60 minutes). 
    }
    \label{fig:z_magnitude_relation}
\end{figure}

The redshift and magnitude distributions of the lenses and lensed images presented in this work are shown in Figure~\ref{fig:z_magnitude_relation}. The $z-$band magnitudes of the lenses, dominated by elliptical galaxies, range from $\approx$22.0 to $\approx$19.9 mag. The typical exposure time of the associated DESI observations is $\sim 20$ minutes. For about ten high-redshift lenses, the exposure times were increased to one hour or longer. The $g-$band magnitudes of the sources, dominated by emission-line galaxies or quasars, range from $\approx$23.2 to $\approx$21.5 mag. The exposure time of the associated DESI observations varied between roughly 15 and 60 minutes, with a median of $\approx 28$ minutes. In addition, we note that the DESI observations are subject to specific target selection functions, the details of which are documented in a series of papers (e.g., \cite{Chaussidon2023, Myers2023, Zhou2023, Raichoor2023, Hahn2023}).

The high-redshift lens systems confirmed in this work will be particularly useful for probing galaxy evolution at high redshifts. Numerous studies have leveraged strong lensing galaxies to quantify the redshift evolution of galaxy mass distributions (e.g., \cite{Koopmans06, Bolton12a, Sonnenfeld13, Dye14, Li18, Filipp2023, Sahu2024}). These observational constraints, when combined with cosmological hydrodynamical simulations, can provide powerful tests of fundamental astrophysical processes such as dark matter properties, feedback mechanisms, and merger histories (e.g., \cite{Naab09, Duffy10, Remus13, Velliscig14, Remus17, Wang19}). 
However, prior to this study, only about ten $z > 0.8$ strong lens systems\footnote{Excluding lensed quasars which are usually not used for constraining the lens mass distribution evolution.} had been confirmed (i.e. with spectroscopically measured lens and source redshifts) (e.g., \cite{Ratnatunga95, Crampton02, Cabanac05, Faure11, Dye14, sugohiIV, Jaelani20, Tran22, Li25}). As a result, our understanding of the evolution of galaxy mass distribution has largely been confined to $z < 0.8$, which only covers about half of the Universe's history. The inclusion of the eight $z > 0.8$ strong-lens systems from this work thus represents a significant increase in the number of high-redshift tracers, with 76 more $z > 0.8$ strong-lens candidates awaiting confirmation. 

\vspace{-0.05cm}
\vspace{-0.05cm}

The quasar lens candidate identified in this work, 092121$+$031744, provides a rare opportunity of precisely measuring the total mass (including dark matter) of a quasar host galaxy, which opens up a new avenue for probing the connections between supermassive black holes (SMBHs) and their host galaxies. Studies of quasar host galaxies have revealed tight correlations between the masses of central SMBHs and bulge properties, such as the stellar mass and stellar velocity dispersion (e.g., \cite{Magorrian98, Ferrarese00, Gebhardt00}). Nonetheless, commonly-used galaxy properties in those studies are inferred from the bulge light that is often heavily contaminated by the quasar emission, especially at high redshifts. In addition, stellar mass estimations are subject to significant uncertainties arising from assumptions about the stellar IMF, star formation history, age-dust-metallicity degeneracy, and more. On the other hand, when a quasar host galaxy acts as a strong lens, its total mass (in the central region) can be determined with high precision and accuracy through lensing, typically to within $\lesssim 5\%$. It is therefore of great interest to explore potential correlations between the central SMBHs and the total masses in the central regions of galaxies. Quasar lenses are, however, quite rare, with only a handful known to date (e.g., \cite{Courbin12}). It was not until \cite{Millon23} that the first such attempt was carried out, based on the analysis of a single quasar lens. A prediction by \cite{Taak20} suggested that $\sim 100$ quasar lenses would be detectable in the HSC survey, and the discovery of 092121$+$031744 thus marks the beginning of this effort.

\section{Conclusions}\label{sec:conclusions}

In this work, we leveraged the recently released DESI DR1 to determine lens and source redshifts for a sample of high-quality strong-lens candidates discovered in the HSC survey. Of the 3,950 HSC strong-lens candidates selected, 2,111 have at least one DESI fiber allocated within 6$^{\prime \prime}$. Upon visually inspecting DESI DR1 spectra and HSC image cutouts for $\approx 500$ strong-lens candidates that meet our specific criteria, we measured both lens and source redshifts for 27 strong-lens systems, thereby confirming their lensing nature. Among these, eight systems feature high-redshift lensing galaxies (lens redshifts from 0.81443 to 0.9983) and two are identified as strongly-lensed quasars. Additional 76 strong-lens candidates were verified to host $z > 0.8$ lensing galaxies. We further confirmed seven candidates with quasar sources and one unusual candidate involving a quasar lens. This dataset is especially valuable for studying galaxy evolution at high redshifts. 

A basic scaling estimate suggests that, with the final DESI data release, $\approx 100$ HSC strong-lens systems could be confirmed, and nearly all existing HSC strong-lens candidates will have their lens or source redshifts determined. It is noteworthy that the HSC strong-lens candidates were selected from $\approx 1,500$ square degrees of sky, while the DESI spectroscopic survey will cover $\approx 14,000$ square degrees. Many more strong lenses---including rare configurations such as quasar lenses---are expected to be confirmed when similar analyses are applied to thousands of additional strong-lens candidates within the DESI footprint (e.g., \cite{Canameras20, Huang20, Huang21, Li20, Li21, Dawes23, Storfer24, He25b}). 

In fact, the number of strong-lens candidates will soon reach $\sim 10^5$ with the advent of the Euclid mission \cite{Euclid}, the \textit{Vera Rubin} Observatory Legacy Survey of Space and Time \cite{Ivezi2019}, the China Space Station survey Telescope \cite{Zhan2021}, and others. The traditional approach of validating candidates through individual, case-by-case follow-up observations is becoming impractical. Our results has demonstrated that large-scale spectroscopic surveys, especially next-generation missions equipped with $\sim 20,000$ fibers, will play a crucial role in the spectroscopic confirmation of large samples of strong lenses. 

\Acknowledgements{We thank Dr. Hu Zou and Dr. Weijian Guo for helpful discussions on DESI data and observations. This work is supported by the China Manned Space Program with grant no. CMS-CSST-2025-A20. This research uses services or data provided by the SPectra Analysis and Retrievable Catalog Lab (SPARCL) and the Astro Data Lab, which are both part of the Community Science and Data Center (CSDC) program at NSF National Optical-Infrared Astronomy Research Laboratory. NOIRLab is operated by the Association of Universities for Research in Astronomy (AURA), Inc. under a cooperative agreement with the National Science Foundation.}

\InterestConflict{The authors declare that they have no conflict of interest.}




\begin{appendix}




\renewcommand{\thesection}{Appendix}

\section{}
\label{sec:appendix}

\subsection{\label{sec:C1_Appendix} Notes on individual systems in Category 1}

\subsubsection{HSC\,J0015$+$0137}
This system was first discovered by \cite{holismokesXVI}. Its HSC color-composite image shows an orange elliptical galaxy with four purple lensed images in a cross configuration. Two DESI fibers were allocated to this system, one centered on the lens and the other on one of the lensed images. Best-fit models to the two DESI spectra appear robust, and the lens and source redshifts are thus determined to be $0.37884 \pm 0.00005$ and $1.165247 \pm 0.000013$. 

\subsubsection{012018$+$001125}
This system was first discovered by \cite{holismokesVIII}. Its HSC color-composite image shows a red elliptical galaxy surrounded by a green arclet and a counter image. One DESI fiber was allocated to this system, and was centered on the lens. The best-fit model securely determines the lens redshift to be $0.59852 \pm 0.00011$. In addition, we identified one strong emission line at $\approx$4862\AA, which we interpreted as Ly$\alpha$ emission from the source. Modeling this emission line with a skewed Gaussian profile using \texttt{pyplatefit}\footnote{We used \texttt{pyplatefit} for line and spectrum fittings in the rest of the paper}, we obtained a good fit with a $\chi_r^2 = 1.104$. The signal-to-noise ratio (SNR) of this line is 7.46. We therefore adopted a source redshift of $2.9994 \pm 0.0009$.

\subsubsection{015310$-$042315}
This system was first discovered by \cite{holismokesVIII}. Its HSC color-composite image shows an orange elliptical galaxy with two elongated arclets. One DESI fiber was allocated to this system, and was centered on the lens. The best-fit model securely determines the lens redshift to be $0.28725 \pm 0.00007$. In addition, we identified five strong emission lines at approximately 6143\AA, 6147\AA, 8014\AA, 8176\AA, and 8255\AA, which resemble the [O\textsc{ii}] doublet, H$\beta$, [O\textsc{iii}] doublet at $z \approx 0.6482$. Fitting these five lines simultaneously with Gaussian profiles yielded a redshift of $0.64816 \pm 0.00003$, which we adopted as the source redshift. 

\subsubsection{083933$-$014044}
This system was first discovered by \cite{sugohiV}. Its HSC color-composite image shows an orange elliptical galaxy surrounded by an orange, elongated arc and its counter image. Three DESI fibers were allocated to this system, one centered on the lens and two on the tip of the arc. Best-fit models to the three DESI spectra appear robust, with redshifts of $0.26991 \pm 0.00006$, $0.70907 \pm 0.00006$ and $0.70925 \pm 0.00007$, respectively. We averaged the latter two measurements, resulting in lens and source redshifts of $0.26991 \pm 0.00006$ and $0.70916 \pm 0.00005$.

\subsubsection{085046$+$003905}
This system was first discovered by \cite{sugohiVI}. Its HSC color-composite image shows a red elliptical galaxy surrounded by four blue lensed images in a cross configuration. Two DESI fibers were allocated to this system, one centered on the lens and the other on one of the lensed images. The best-fit model securely determines the lens redshift to be $0.8844 \pm 0.0003$. The best-fit model of the spectrum of the lensed image misinterpreted a strong emission feature at $\approx$4665\AA\, as a broad Ly$\alpha$ line and suggested a \texttt{spectype} of \texttt{QSO}. However, that emission feature actually contains two narrow emission lines, which we interpreted as double-peaked Ly$\alpha$ emission. Modeling this feature with two skewed Gaussian profiles, we obtained a good fit with $\chi_r^2 = 1.178$. The SNRs of the blue and red peaks are 7.05 and 17.35 respectively. We inferred a source redshift of $2.83777 \pm 0.00011$ from the center of the two peaks.

\subsubsection{090430$+$042648}
This system was first discovered by \cite{sugohiV}. Its HSC color-composite image shows an orange elliptical galaxy with two purple, elongated arclets to its west. Three DESI fibers were allocated to this system, two centered on the lens and the other on one arclet. Best-fit models to the three DESI spectra appear robust, with two lens spectra having redshifts of $0.45648 \pm 0.00005$ and $0.45643 \pm 0.00005$, respectively. We adopt their average value, $0.45646 \pm 0.00004$, as the redshift of the lens. The source spectrum has a redshift of $1.01730 \pm 0.00003$. Therefore, the lens and source redshifts are determined to be $0.45646 \pm 0.00004$ and $1.01730 \pm 0.00003$, respectively.

\subsubsection{090938$+$002842}
This system was first discovered by \cite{sugohiVI}. Its HSC color-composite image shows a red elliptical galaxy surrounded by a blue arclet and a counter image. Two DESI fibers were allocated to this system, one centered on the lens and the other on the counter image. The best-fit model securely determines the lens redshift to be $0.77951 \pm 0.00009$. The fit to the spectrum of the counter image was affected by contamination from the lens light and gave an incorrect redshift. We noticed one emission line at $\approx$4311\AA{} in both DESI spectra, which we interpreted as Ly$\alpha$ emission from the source. Independently modeling this emission line in the two spectra with a skewed Gaussian profile, we obtained good fits with $\chi_r^2 = 1.035$ and $\chi_r^2 = 1.244$. The SNRs of this line are 7.75 and 5.61. The best-fit redshifts are also consistent. We took the mean redshift of $2.5459 \pm 0.0003$ as the source redshift. 

\subsubsection{094348$+$005926}
This system was first discovered by \cite{sugohiV}. Its HSC color-composite image shows an orange elliptical galaxy surrounded by two extended, purple arclets. Four DESI fibers were allocated to this system, two centered on the lens and the other two on the brighter arclet. The best-fit models to the four DESI spectra appear robust, with redshifts of $0.43805 \pm 0.00011$, $0.43829 \pm 0.00008$, $1.098890 \pm 0.000005$, and $1.098928 \pm 0.000005$, respectively. We averaged the values, resulting in lens and source redshifts of $0.43817 \pm 0.00007$ and $1.098909 \pm 0.000004$, respectively. 

\subsubsection{HSC\,J1004$-$0031}
This system was first discovered by \cite{holismokesVI}. Its HSC color-composite image shows an orange elliptical galaxy surrounded by four lensed images in a cross configuration. Two DESI fibers were allocated to this system, one centered on the lens and the other on one lensed image. Best-fit models to the two DESI spectra appear robust, and the lens and source redshifts are thus determined to be $0.56877 \pm 0.00012$ and $1.44286 \pm 0.00004$.

\subsubsection{HSC\,J1014$+$0332}
This system was first discovered by \cite{holismokesXVI}. Its HSC color-composite image shows a red elliptical galaxy surrounded by four blue lensed images in a cross configuration. One DESI fiber was allocated to this system, and was centered on the lens. The best-fit model securely determines the lens redshift to be $0.67770 \pm 0.00008$. In addition, we noticed one emission line at $\approx$3848\AA, which we interpreted as Ly$\alpha$ emission from the source. Modeling this emission line with a skewed Gaussian profile, we obtained a good fit with a $\chi_r^2 = 0.985$. The SNR of this line is 3.36. We thus adopted a source redshift of $2.165 \pm 0.002$.

\subsubsection{114311$-$013935}
This system was first discovered by \cite{sugohiII}. Its HSC color-composite image shows an orange elliptical galaxy surrounded by four purple arclets. Two DESI fibers were allocated to this system, one centered on the lens and the other on one of the arclets. The spectrum of the arclet shows very strong [O\textsc{ii}] emission, which firmly establishes its redshift at $1.38763 \pm 0.00002$. The fit to the spectrum of the lens mis-identified the arclet [O\textsc{ii}] emission as H$\alpha$, and gave an incorrect lens redshift. To infer the lens redshift, we masked regions around the [O\textsc{ii}] doublet and fitted the DESI spectrum with a linear combination of galaxy eigenspectra following \cite{Bolton12b}. We measured the lens redshift to be $0.6436 \pm 0.0002$, which matches well with the characteristic absorption features in the DESI spectrum, including Ca K and H lines, H$\delta$, G-band, and Mg I b. 

\subsubsection{122450$-$004215}
This system was first discovered by \cite{Petrillo19}. Its HSC color-composite image shows an orange elliptical galaxy surrounded by two pink, elongated arclets. Two DESI fibers were allocated to this system, one centered on the lens and the other on the brighter arclet. The best-fit model securely determines the lens redshift to be $0.40355 \pm 0.00005$. The spectrum of the arclet clearly shows two emission lines at approximately 8119\AA{} and 8125\AA, which resemble the [O\textsc{ii}] doublet at $z \approx 1.1784$. The fit to the spectrum of the arclet, however, mis-identified the [O\textsc{ii}] doublet as a lower-redshift H$\alpha$. Fitting these two lines simultaneously with Gaussian profiles yielded a redshift of $1.17846 \pm 0.00004$, which we took as the source redshift. 

\subsubsection{123636$-$003539}
This system was first discovered by \cite{holismokesVIII}. Its HSC color-composite image shows an orange elliptical galaxy with three pink lensed images in a cusp configuration. One DESI fiber was allocated to this system, and was centered on the lens. The best-fit model securely determines the lens redshift to be $0.51016 \pm 0.00009$. In addition, we noticed four strong emission lines at approximately 6572\AA, 6577\AA, 8747\AA, and 8830\AA, which nicely resemble the [O\textsc{ii}] doublet and [O\textsc{iii}] doublet at $z \approx 0.7632$. Fitting these four lines simultaneously with Gaussian profiles yielded a redshift of $0.763224 \pm 0.000013$, which we took as the source redshift. 

\subsubsection{141649$+$013822}
This system was first discovered by \cite{sugohiVI}. Its HSC color-composite image shows an orange elliptical galaxy surrounded by two white-ish arclets. Two DESI fibers were allocated to this system, one centered on the lens and the other on the brighter arclet. The best-fit model securely determines the lens redshift to be $0.43371 \pm 0.00006$. The fit to the spectrum of the arclet was affected by contamination from the lens light and gave an incorrect redshift. Nevertheless, we noticed two emission lines at approximately 8621\AA{} and 8627\AA{} in both DESI spectra, which resemble the [O\textsc{ii}] doublet at $z \approx 1.313$. Fitting these two lines simultaneously with Gaussian profiles yielded a redshift of $1.31300 \pm 0.00002$, which we took as the source redshift. 

\subsubsection{155957$+$441543}
This system was first discovered by \cite{sugohiI}. Its HSC color-composite image shows an orange elliptical galaxy surrounded by four compact, orange-red blobs in a cusp configuration. Three DESI fibers were allocated to this system, one centered on the lens and the other two on one of the lensed images. Best-fit models to the three DESI spectra appear robust, with redshifts of $0.59761 \pm 0.00009$, $1.52910 \pm 0.00012$, and $1.52896 \pm 0.00008$, respectively. We averaged the latter two measurements, resulting in lens and source redshifts of $0.59761 \pm 0.00009$ and $1.52903 \pm 0.00007$. 

\subsubsection{222638$-$003449}
This system was first discovered by \cite{sugohiV}. Its HSC color-composite image shows an orange elliptical galaxy surrounded by four purple arclets close to a fold configuration. Two DESI fibers were allocated to this system, one centered on the lens and the other on one arclet. The best-fit model securely determines the lens redshift to be $0.40425 \pm 0.00010$. The spectrum of the arclet has low SNRs in general, but shows two emission lines at approximately 8747\AA{} and 8754\AA, which resemble the [O\textsc{ii}] doublet at $z \approx 1.347$. Fitting these two lines together with Gaussian profiles yielded a redshift of $1.34701 \pm 0.00007$, which we took as the source redshift. 

\subsubsection{224154$+$000331}
This system was first discovered by \cite{sugohiVI}. Its HSC color-composite image shows an orange elliptical galaxy surrounded by two compact, blue arclets. Three DESI fibers were allocated to this system, two centered on the lens and the third on the brighter arclet. Best-fit models to the three DESI spectra appear robust, with redshifts of $0.49676 \pm 0.00008$, $0.49643 \pm 0.00007$, and $1.61508 \pm 0.00005$, respectively. We averaged the former two measurements, resulting in lens and source redshifts of $0.49660 \pm 0.00005$ and $1.61508 \pm 0.00005$. 

\subsubsection{224221$+$001144}
This system was first discovered by \cite{sugohiI}. Its HSC color-composite image shows an orange elliptical galaxy with one elongated arclet and one counter image. Three DESI fibers were allocated to this system, one centered on the lens, one on the arclet, and one on the counter image. The best-fit model to the spectrum of the arclet is questionable, but the best-fit models to the other two spectra appear robust. The lens and source redshifts are thus determined to be $0.38531 \pm 0.00007$ and $1.23193 \pm 0.00003$.

\subsection{\label{sec:C2_Appendix} Notes on individual systems in Category 2}

\subsubsection{083651$+$003038}
This system was first discovered by \cite{holismokesVIII}. Its HSC color-composite image shows a red elliptical galaxy surrounded by three blue lensed images and a faint counter image in a fold configuration. One DESI fiber was allocated to this system, and was centered on the lens. The best-fit model securely determines the lens redshift to be $0.83603 \pm 0.00010$. In addition, we identified one strong emission line at $\approx$4940\AA, which we interpreted as Ly$\alpha$ emission from the source. Modeling this line with a skewed Gaussian profile, we obtained a good fit with a $\chi_r^2 = 1.045$. The SNR of this line is 7.76. We thus adopted a source redshift of $3.0629 \pm 0.0003$.

\subsubsection{090402$+$031403}
This system was first discovered by \cite{sugohiX}. Its HSC color-composite image shows a red elliptical galaxy surrounded by two purple arclets. One DESI fiber was allocated to this system, and was centered on the lens. The best-fit model securely determines the lens redshift to be $0.84681 \pm 0.00010$. In addition, we identified one strong emission line at $\approx$5293\AA, which we interpreted as Ly$\alpha$ emission from the source. Modeling this emission line with a skewed Gaussian profile, we obtained a good fit with a $\chi_r^2 = 0.939$. The SNR of this line is 8.88. We thus adopted a source redshift of $3.3528 \pm 0.0002$.

\subsubsection{090548$+$004743}
This system was first discovered by \cite{sugohiVI}. Its HSC color-composite image shows a red elliptical galaxy surrounded by two blue lensed images. Three DESI fibers were allocated to this system, two centered on the lens and the other on the lensed image to the west. The best-fit models of the two lens spectra robustly determined the lens redshift to be $0.85353 \pm 0.00012$ and $0.85351 \pm 0.00009$, respectively. We adopted the average value of $0.85352 \pm 0.00008$ as the redshift of the lens. The fit to the spectrum of the lensed image was not reliable. Nevertheless, we identified one strong emission line at $\approx$4756\AA{} in all three spectra, which we interpreted as Ly$\alpha$ emission from the source. Independently modeling this emission line in the three spectra with a skewed Gaussian profile, we obtained good fits with $\chi_r^2 = 1.123$, $\chi_r^2 = 0.796$ and $\chi_r^2 = 0.797$. The SNRs of this line are 13.19, 14.30 and 8.27. The best-fit redshifts are almost identical. We adopted the mean redshift of $2.91165 \pm 0.00010$ as the source redshift. 

\subsubsection{HSC\,J1104$-$0052}
This system was first discovered by \cite{holismokesXVI}. Its HSC color-composite image shows a red elliptical galaxy surrounded by one blue, extended arclet and its counter image. One DESI fiber was allocated to this system, and was centered on the lens. The best-fit model securely determines the lens redshift to be $0.87299 \pm 0.00017$. In addition, we identified one strong emission line at $\approx$4201\AA, which we interpreted as Ly$\alpha$ emission from the source. Modeling this emission line with a skewed Gaussian profile, we obtained a good fit with a $\chi_r^2 = 1.053$. The SNR of this line is 9.02. We thus adopted a source redshift of $2.45593 \pm 0.00019$.

\subsubsection{HSC\,J1107$-$0037}
This system was first discovered by \cite{holismokesXVI}. Its HSC color-composite image shows a red elliptical galaxy surrounded by one purple, extended arclet and its counter image. One DESI fiber was allocated to this system, and was centered on the lens. The best-fit model securely determines the lens redshift to be $0.81443 \pm 0.00010$. In addition, we identified an emission feature at $\approx$8989\AA{} that resembles the [O\textsc{ii}] doublet at $z \approx 1.411$. Fitting this feature with two Gaussian profiles yielded a redshift of $1.4106 \pm 0.0003$, which we took as the source redshift.

\subsubsection{115630$-$020027}
This system was first discovered by \cite{sugohiVI}. Its HSC color-composite image shows a red elliptical galaxy surrounded by two blue arclets. Two DESI fibers were allocated to this system, one centered on the lens and the other on the lensed image to the north. The best-fit model securely determines the lens redshift to be $0.9983 \pm 0.0002$. The fit to the spectrum of the lensed image was affected by contamination of the lens light. Nevertheless, we identified one strong line at $\approx$3926\AA{} in both spectra, which we interpreted as Ly$\alpha$ emission from the source. Independently modeling this emission line in the two spectra with a skewed Gaussian profile, we obtained good fits with $\chi_r^2 = 0.847$ and $\chi_r^2 = 1.032$. The SNRs of this line are 7.95 and 6.24. The best-fit redshifts are almost identical. We adopted the mean redshift of $2.2298 \pm 0.0005$ as the source redshift.

\subsubsection{120657$-$010241}
This system was first discovered by \cite{sugohiX}. Its HSC color-composite image shows a red elliptical galaxy surrounded by two blue arclets. One DESI fiber was allocated to this system, and was centered on the lens. The best-fit model securely determines the lens redshift to be $0.83004 \pm 0.00015$. In addition, we identified an emission feature at $\approx$4588\AA, which we interpreted as Ly$\alpha$ emission from the source. Modeling this emission line with a skewed Gaussian profile, we obtained a good fit with a $\chi_r^2 = 1.077$. The SNR of this line is 6.27. We thus adopted a source redshift of $2.7732 \pm 0.0003$.

\subsection{\label{sec:C3_Appendix} Notes on individual systems in Category 3}

\subsubsection{004738$+$032232}
This system was first discovered by \cite{sugohiIX}. Its DESI Legacy Surveys color-composite image shows an orange, disk-like galaxy surrounded by two point sources\footnote{For this object, only $g$-band imaging data is available in HSC PDR3.}. One DESI fiber was allocated to this system, and was centered on one of two point sources. The spectrum is classified as \texttt{spectype=QSO}, and our visual inspection confirmed that it is indeed a broad-line quasar. The best-fit model securely determines the source redshift to be $2.0864 \pm 0.0009$. 

\subsubsection{013827$+$031518}
This system was first discovered by \cite{sugohiIX}. Its DESI Legacy Surveys color-composite image shows two point sources\footnote{For this object, only $g$-band imaging data is available in HSC PDR3.}. In the study by \cite{sugohiIX}, an extended galaxy between the two point sources became apparent upon subtracting the light of the point sources, which we assumed to be the lensing galaxy. One DESI fiber was allocated to this system, and was centered on one of the two point sources. The spectrum is classified as \texttt{spectype=QSO}, and our visual inspection confirmed that it is indeed a broad-line quasar. The best-fit model to the DESI spectra appears robust, and thus the redshift of the source is determined to be $2.4057 \pm 0.0007$. 

\subsubsection{014152$+$003956}
This system was first discovered by \cite{sugohiIX}. Its HSC color-composite image shows a red elliptical galaxy surrounded by two purple point sources. 
One DESI fiber was allocated to this system, and was centered on the galaxy. The spectrum is classified as \texttt{spectype=QSO}, and our visual inspection confirmed the presence of broad emission lines, interpreted as originating from the background quasar. The best-fit model securely determines the source redshift to be $2.2062 \pm 0.0003$.

\subsubsection{091151$-$004406}
This system was first discovered by \cite{sugohiIX}. Its HSC color-composite image shows two pink point sources. In the study by \cite{sugohiIX}, an extended galaxy between the two point sources became apparent upon subtracting the light of the point sources, which we assumed to be the lensing galaxy. One DESI fiber was allocated to this system, and was centered between the two point sources. The spectrum is classified as \texttt{spectype=QSO}, and our visual inspection confirmed that it is indeed a broad-line quasar. The best-fit model securely determines the source redshift to be $2.37547 \pm 0.00015$.

\subsubsection{092121$+$031744}
This system was first discovered by \cite{sugohiV}. Its HSC color-composite image shows an orange elliptical galaxy surrounded by a purple, elongated arc and a possible counter image. Two DESI fibers were allocated to this system, both centered on the elliptical galaxy. Interestingly, the spectra are classified as \texttt{spectype=QSO}, and our visual inspections confirmed that it is indeed a quasar. The best-fit models to the two DESI spectra appear robust and suggest lens redshifts of $0.346027 \pm 0.000018$ and $0.346027 \pm 0.000012$, respectively. We therefore adopted the average value of $0.346027 \pm 0.000011$ as the lens redshift. This represents a rare case in which a quasar acts as the lens.

\subsubsection{095406$-$002225}
This system was first discovered by \cite{sugohiIX}. Its HSC color-composite image shows a red galaxy surrounded by two purple point sources. One DESI fiber was allocated to this system, and was centered on the brighter point source. The spectrum is classified as \texttt{spectype=QSO}, and our visual inspection confirmed that it is indeed a broad-line quasar. The best-fit model securely determines the source redshift to be $2.25975 \pm 0.00019$.

\subsubsection{121533$-$005842}
This system was first discovered by \cite{sugohiX}. Its HSC color-composite image shows an orange elliptical galaxy surrounded by two pink point sources. Two DESI fibers were allocated to this system, one centered on the galaxy and the other on the brighter point source. The spectrum of the point source is classified as \texttt{spectype=QSO}, and our visual inspection confirmed that it is indeed a broad-line quasar. Best-fit models to the two DESI spectra appear robust, and the lens and source redshifts are thus determined to be $z = 0.45912 \pm 0.00008$ and $2.8817 \pm 0.0003$. 

\subsubsection{122018$+$011253}
This system was first discovered by \cite{holismokesVIII}. Its HSC color-composite image shows an orange elliptical galaxy surrounded by two blue point sources. Two DESI fibers were allocated to this system, one centered on the galaxy and the other on the brighter point source. The spectrum of the point source is classified as \texttt{spectype=QSO}, and our visual inspection confirmed that it is indeed a broad-line quasar. Best-fit models to the two DESI spectra appear robust, and the lens and source redshifts are thus determined to be $0.48755 \pm 0.00008$ and $1.7081 \pm 0.0009$.

\subsubsection{130733$+$001122}
This system was first discovered by \cite{sugohiIX}. Its HSC color-composite image shows two blue point sources. One DESI fiber was allocated to each of the two point sources. The two spectra are classified as \texttt{spectype=QSO}, and our visual inspection confirmed the presence of broad emission lines. Strong similarities are also observed between the two spectra in their continua, broad emission lines, and absorption features. The best-fit models suggest the redshifts to be $2.6641 \pm 0.0002$ and $2.6636 \pm 0.0002$, respectively. These observations are all consistent with a lensed quasar scenario. One caveat is that, due to the small separation between the two point sources (measured at $\approx$0\farcs84), substantial fiber crosstalk is expected. However, the two point sources are similarly bright (Gaia G-band magnitudes of 19.96 and 20.80), and additional spectral features would likely have been detected if one were not a quasar or if their redshifts differed significantly. We therefore considered this system as a lensed quasar candidate, and adopted the average value of $2.66385 \pm 0.00014$ as the source redshift. Adopting an Einstein radius of 0\farcs4, the lensing mass would be in the range of $10^{10.5}$-$10^{11.1} M_\odot$ for a lens redshift between 0.5 and 1.5. Assuming the total stellar mass is comparable to the lensing mass, we estimated the $z-$band magnitude of the lensing galaxy to be 22.3-24.9 mag over the same redshift range\footnote{For the magnitude estimations, we assumed a single stellar population model from \cite{BC03} with solar metallicity and a Salpeter IMF \cite{Salpeter1955}.}. Considering the $z-$band depth of $\approx$25.2 mag in HSC PDR3 (5$\sigma$, point sources, \cite{PDR3}), the lensing galaxy would need to lie toward the higher end of the assumed redshift range to be consistent with the non-detection.

\subsubsection{150112$+$422113}
This system was first discovered by \cite{sugohiV}. Its HSC color-composite image shows an orange elliptical galaxy with three blue point sources on one side and another faint point source on the opposite side, resembling a cusp configuration. One DESI fiber was allocated to this system, and was centered on the point source north of the galaxy. The spectrum is classified as \texttt{spectype=QSO}, and our visual inspection confirmed that it is indeed a broad-line quasar. The best-fit model securely determines the source redshift to be $2.6621 \pm 0.0004$. 


\subsection{\label{sec:figures} Cutouts and spectra for the three categories}
\setcounter{figure}{0}
\begin{figure*}[htbp]
\centering
\includegraphics[width=0.49\textwidth]{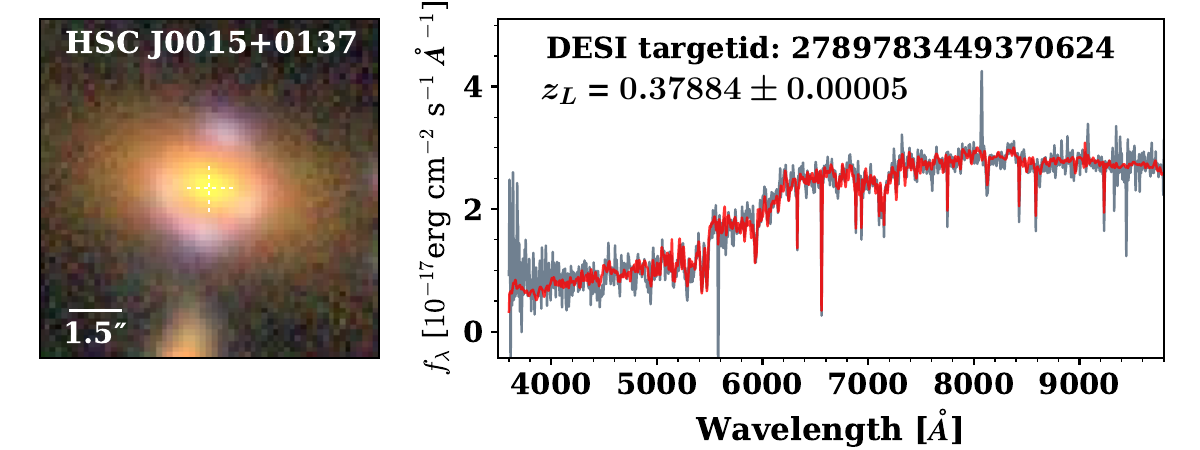}
\includegraphics[width=0.49\textwidth]{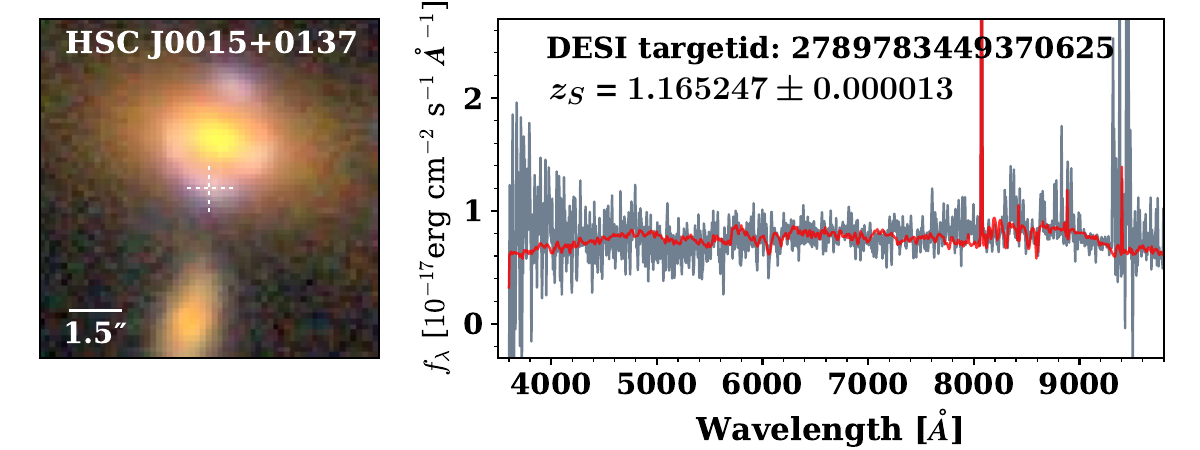}
\includegraphics[width=0.49\textwidth]{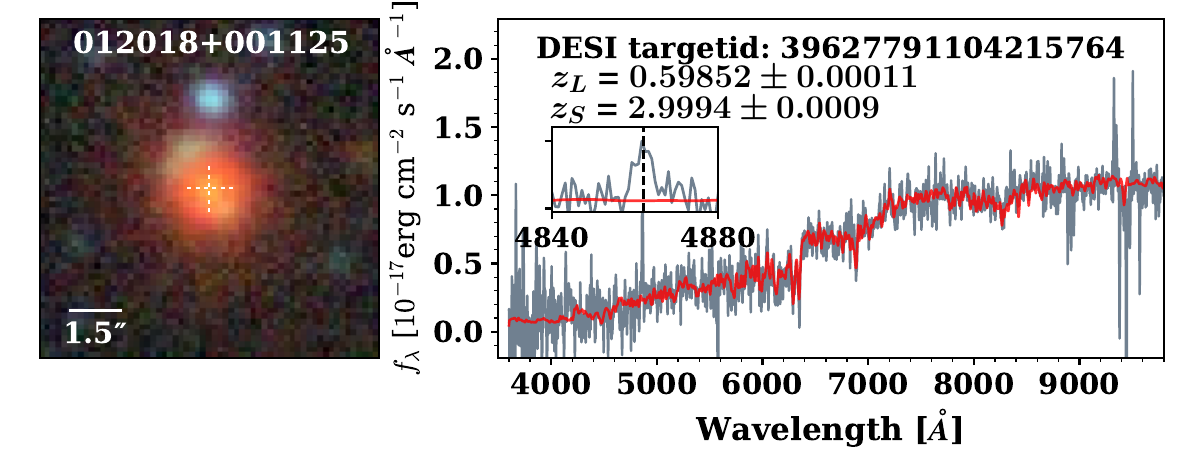}
\includegraphics[width=0.49\textwidth]{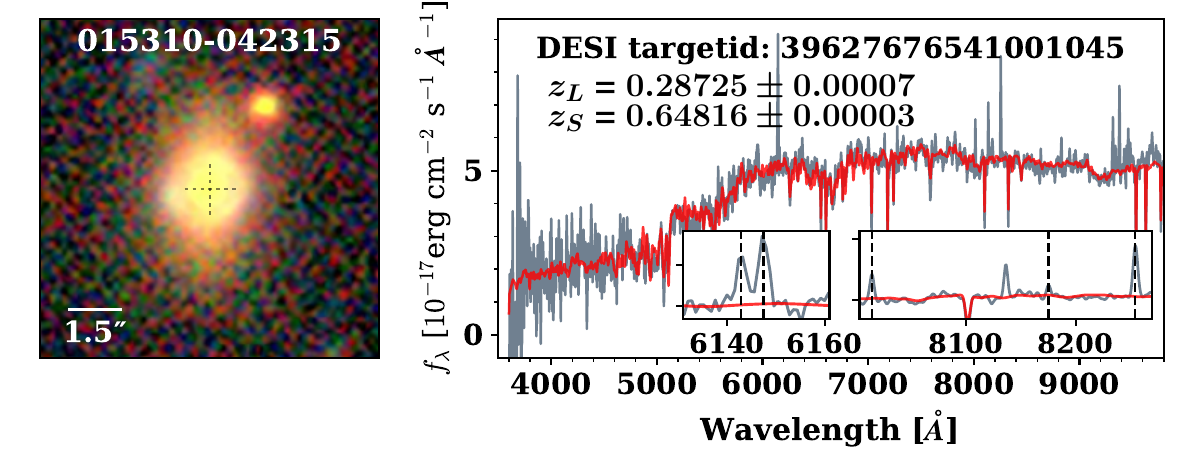}
\includegraphics[width=0.49\textwidth]{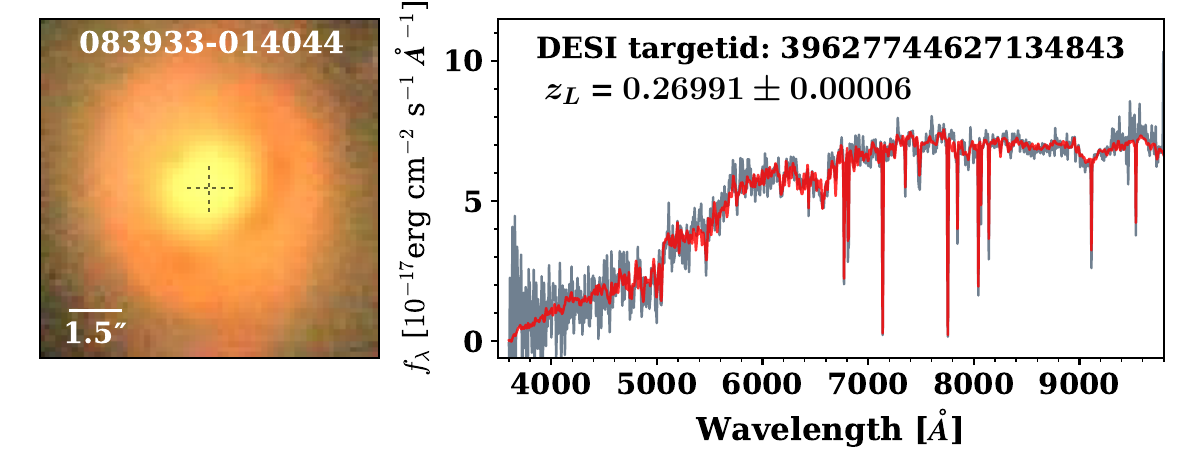}
\includegraphics[width=0.49\textwidth]{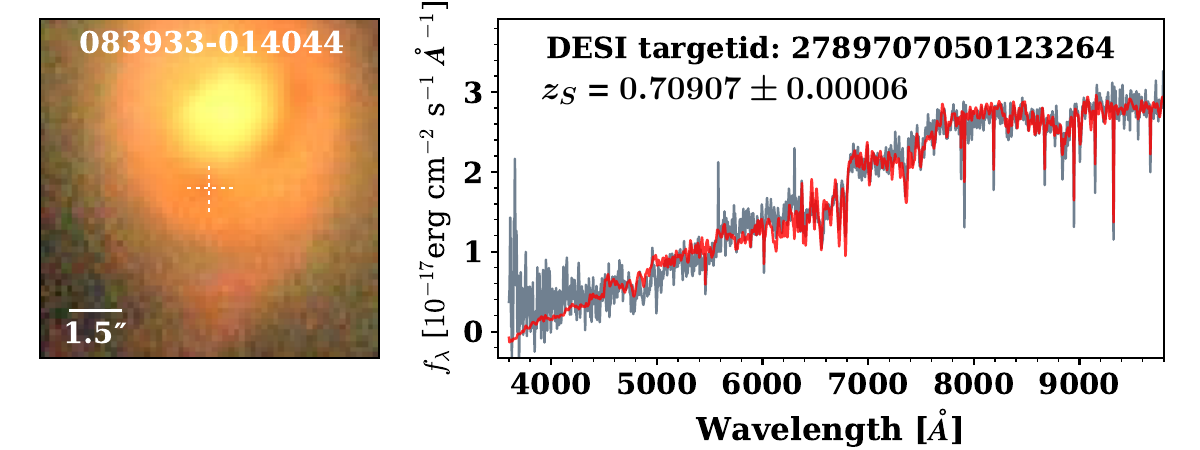}
\includegraphics[width=0.49\textwidth]{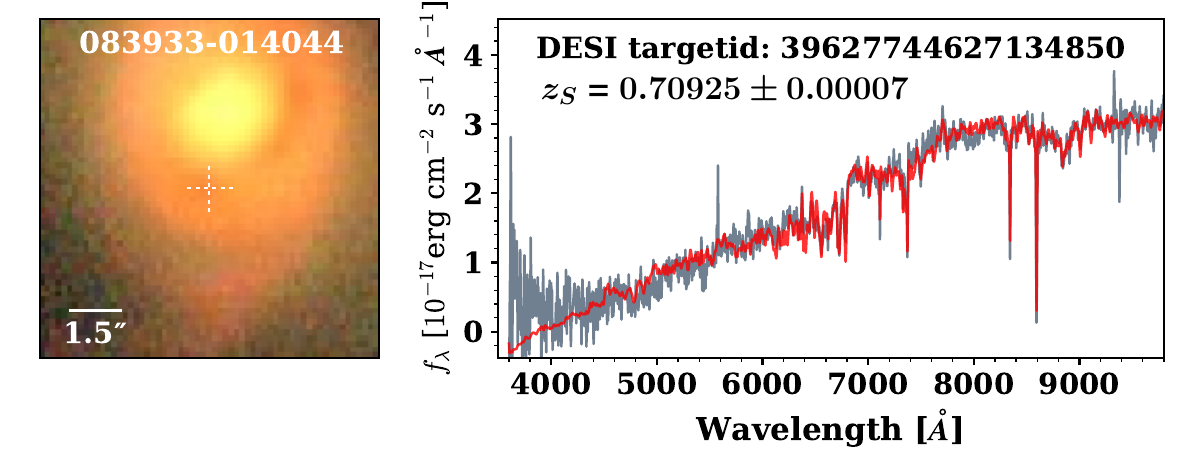}
\includegraphics[width=0.49\textwidth]{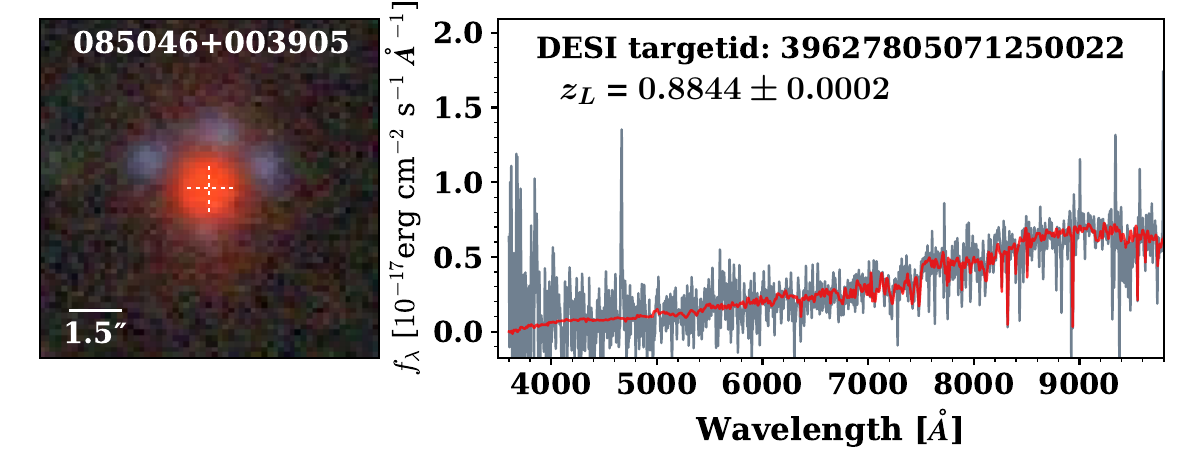}
\includegraphics[width=0.49\textwidth]{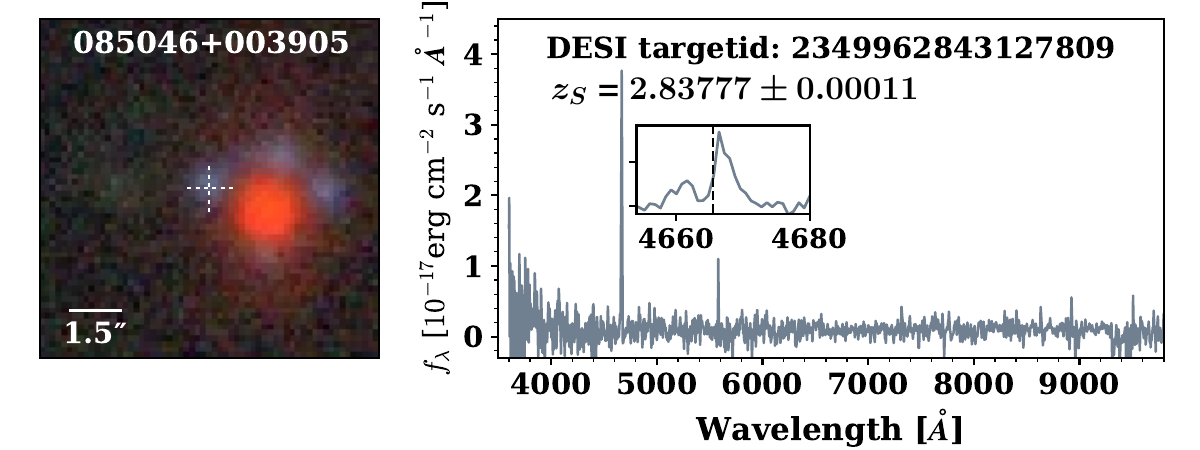}
\includegraphics[width=0.49\textwidth]{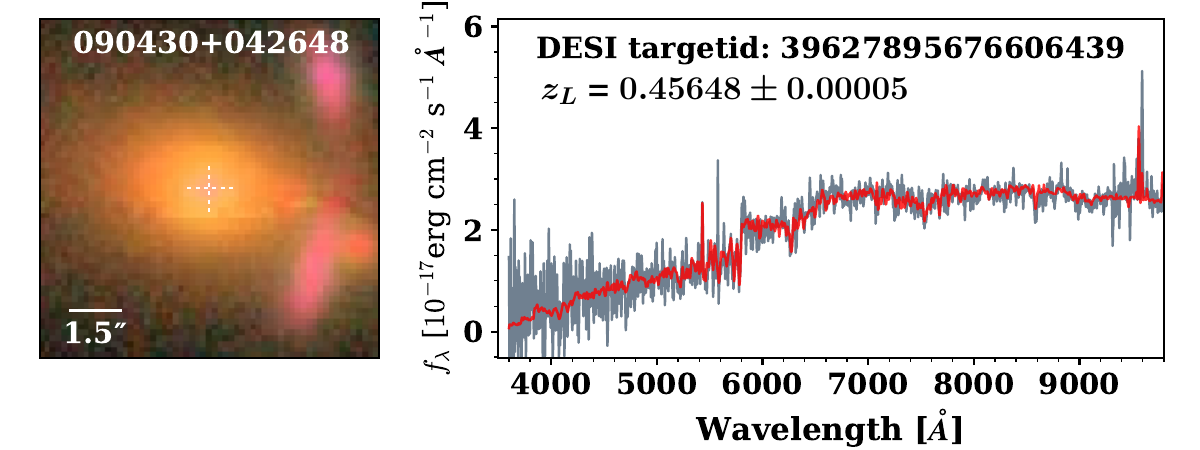}
\includegraphics[width=0.49\textwidth]{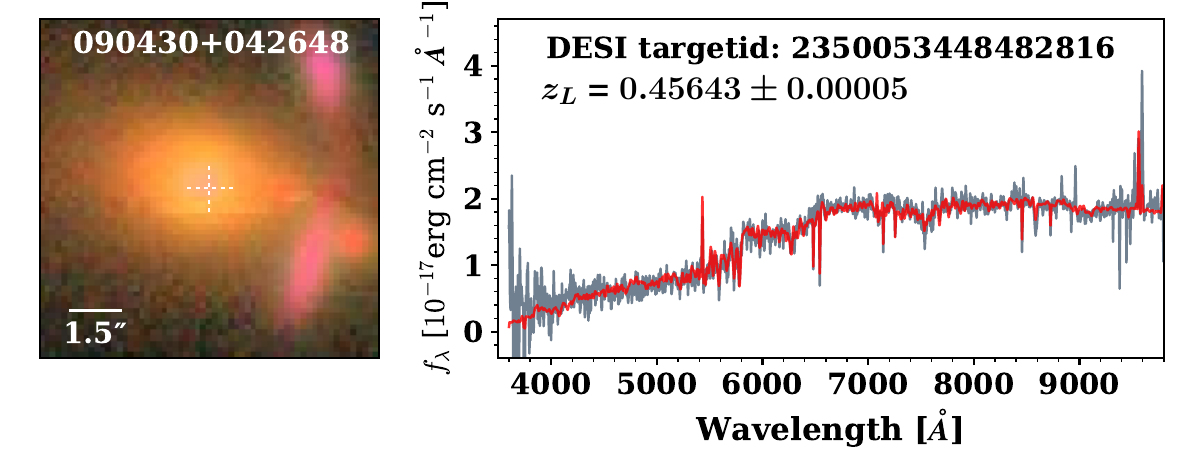}
\includegraphics[width=0.49\textwidth]{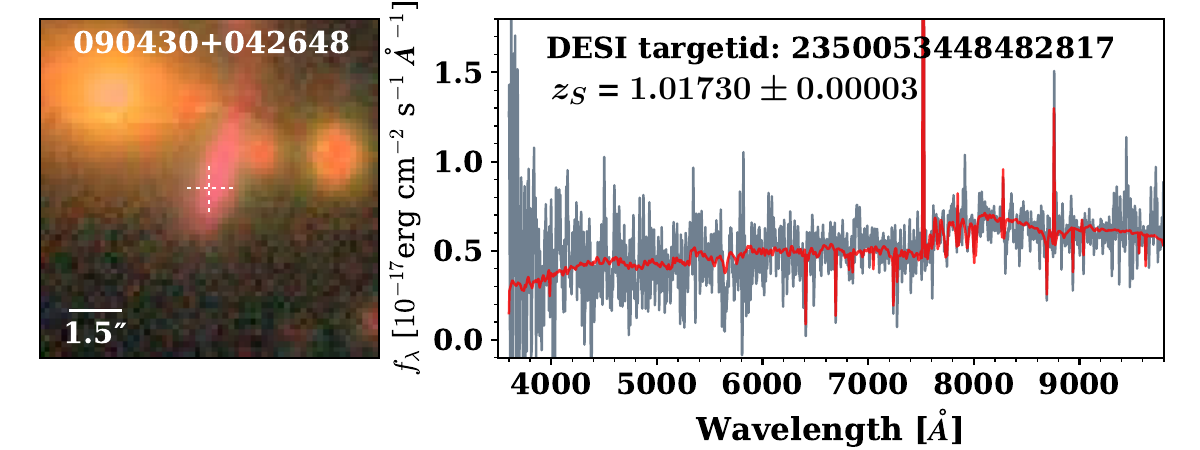}
\includegraphics[width=0.49\textwidth]{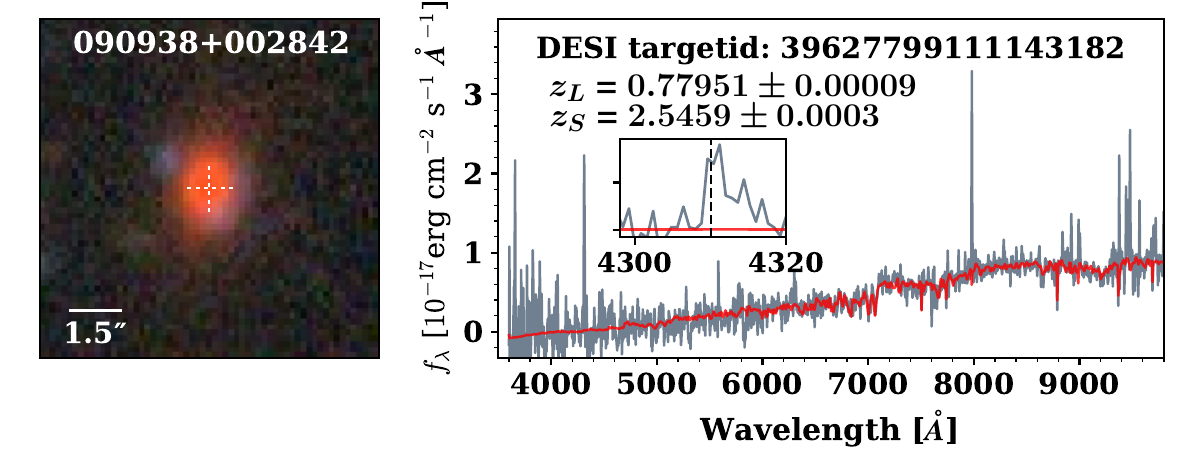}
\includegraphics[width=0.49\textwidth]{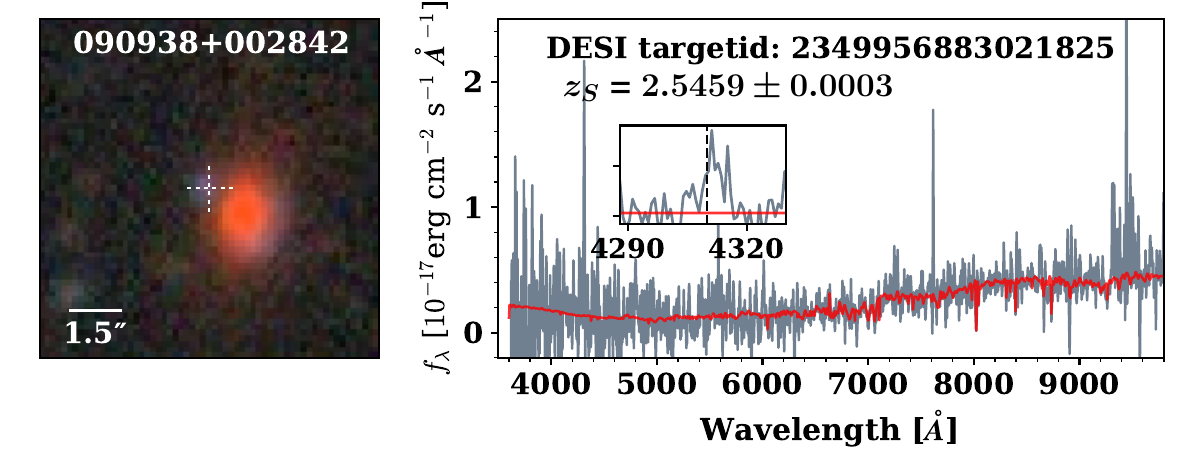}
\caption{Color-composite cutouts and DESI spectra for the 18 systems in Category 1. Each cutout is $10^{\prime \prime} \times 10^{\prime \prime}$, with north up and east to the left. A 1\farcs5 scale bar is shown to indicate the size of the DESI fiber, and the white/black dotted cross corresponds to the center of the fiber. For the spectrum panel, the gray line represents the observed DESI spectrum (smoothed by a Gaussian kernel of 3 pixels) and the red line represents the best-fit model. For some cases, zoom-in insets are included to show emission lines used to determine the source redshifts. }
\label{fig:C1}
\end{figure*}

\begin{figure*}[htbp]
\ContinuedFloat
\centering
\includegraphics[width=0.49\textwidth]
{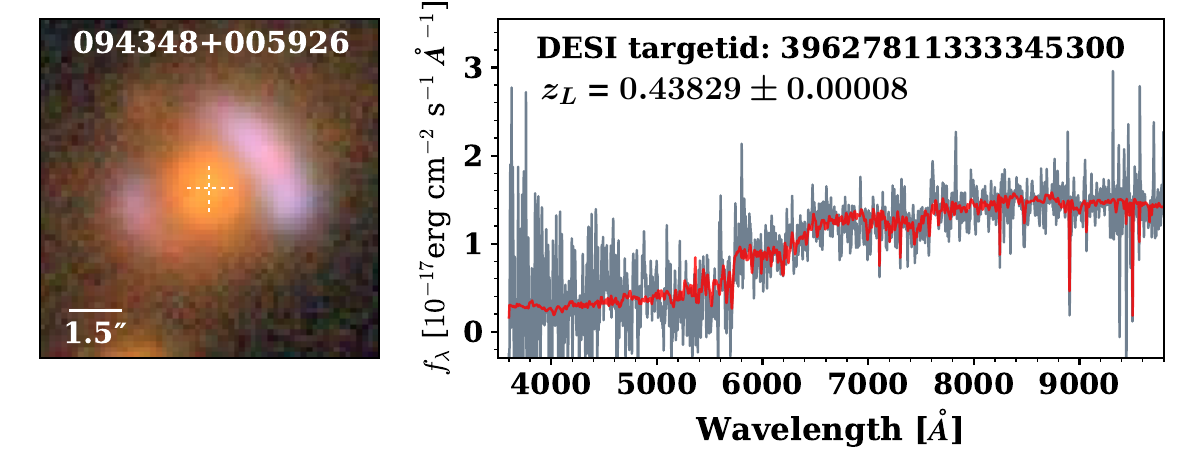}
\includegraphics[width=0.49\textwidth]{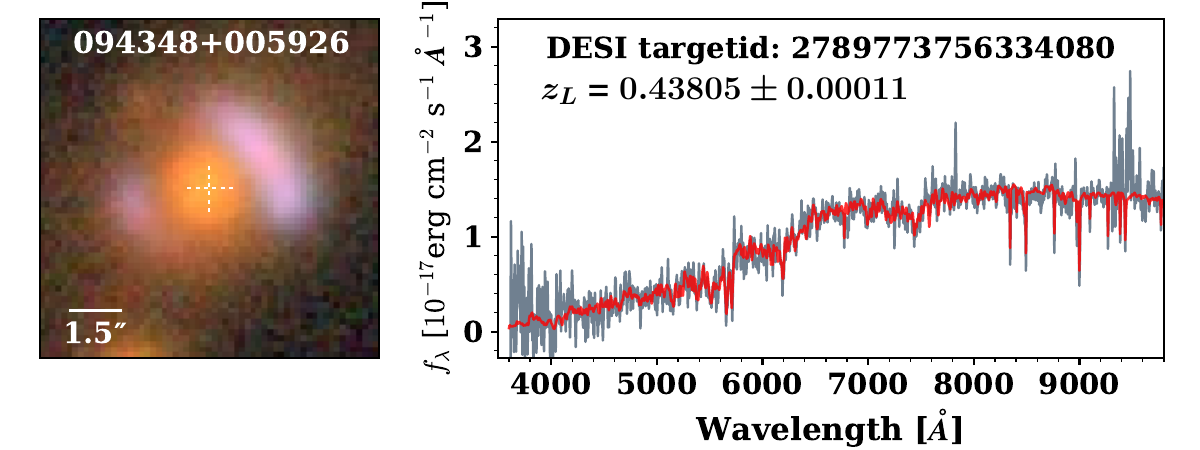}
\includegraphics[width=0.49\textwidth]{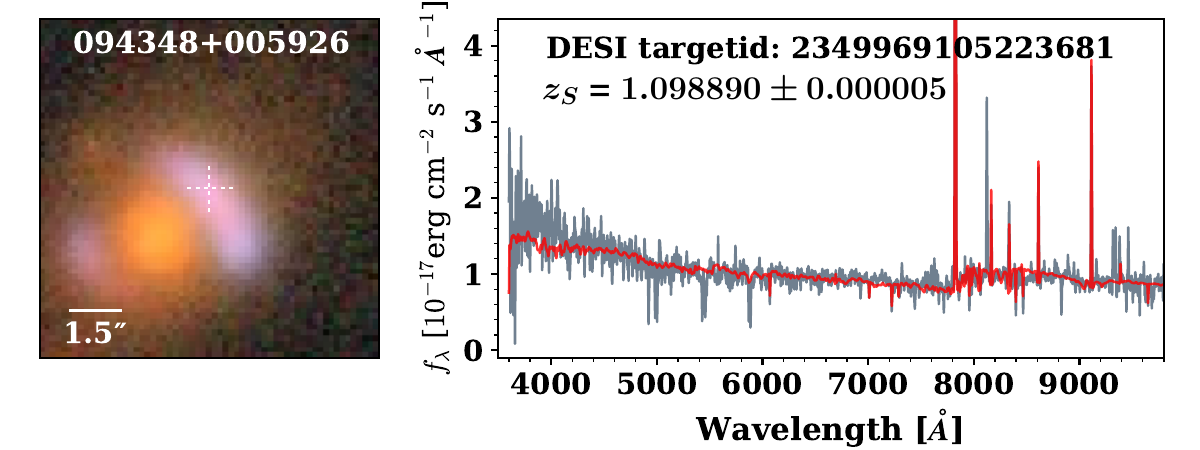}
\includegraphics[width=0.49\textwidth]{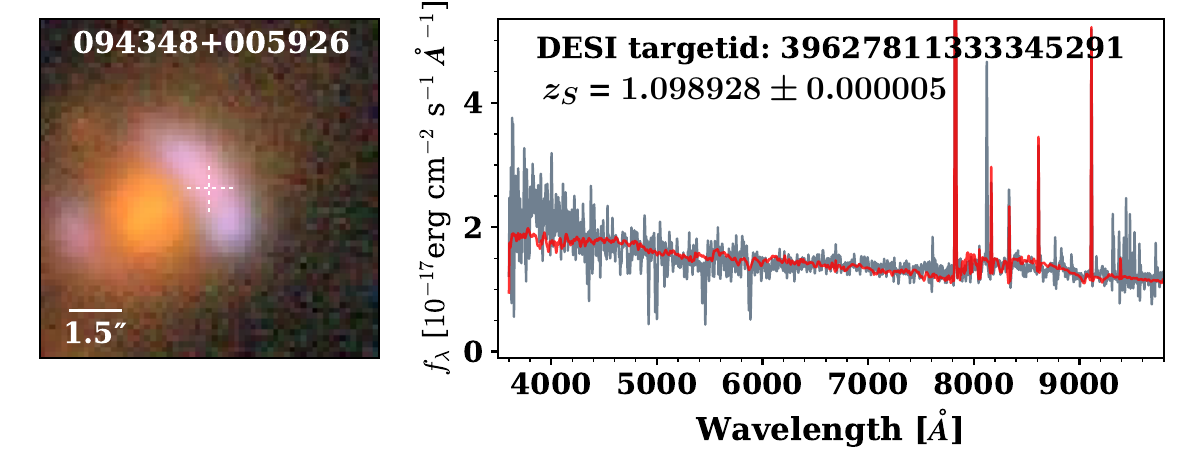}
\includegraphics[width=0.49\textwidth]{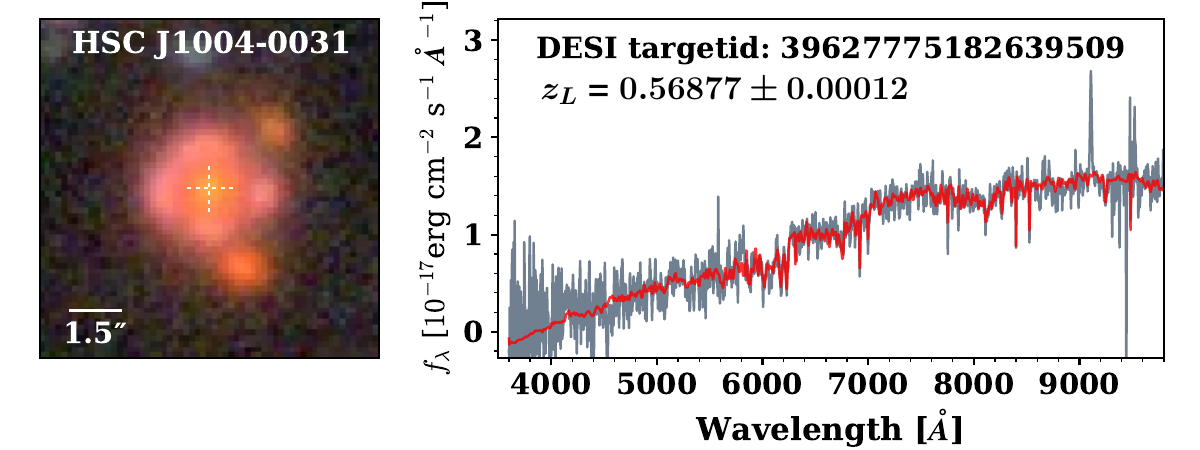}
\includegraphics[width=0.49\textwidth]{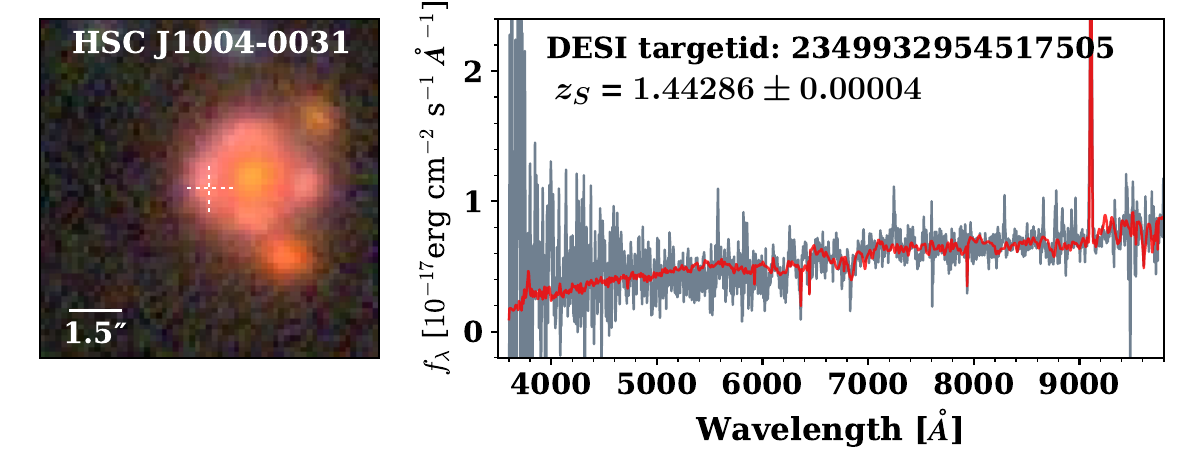}
\includegraphics[width=0.49\textwidth]{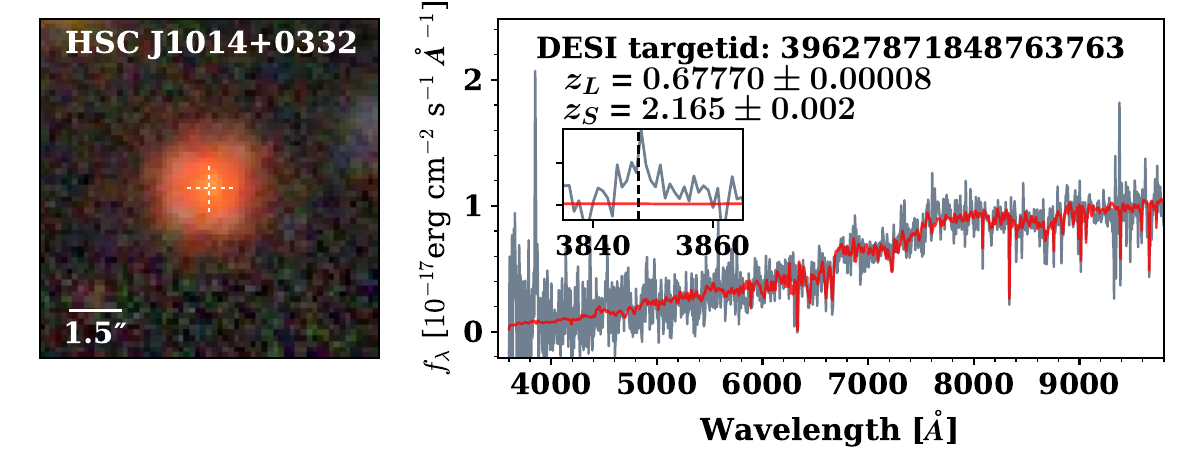}
\includegraphics[width=0.49\textwidth]{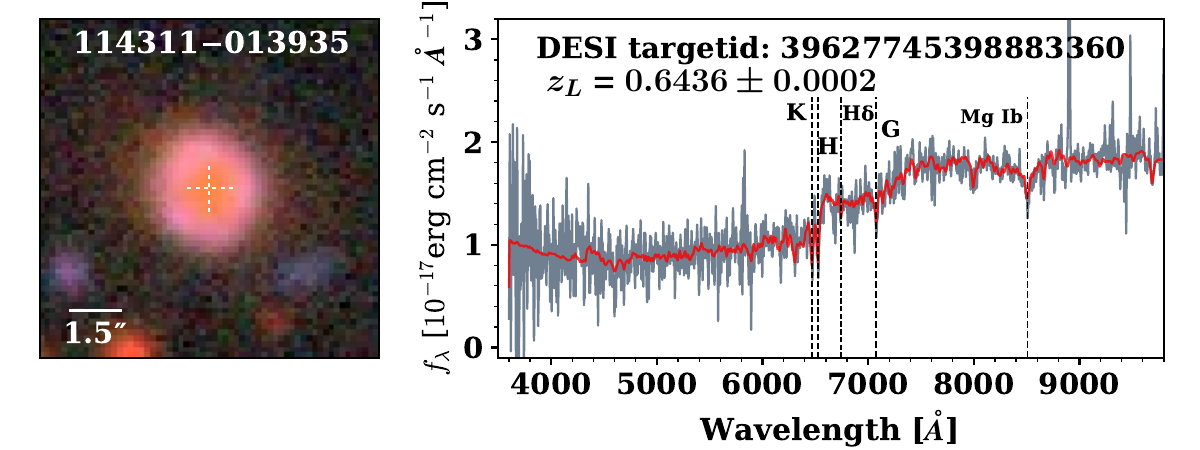}
\includegraphics[width=0.49\textwidth]{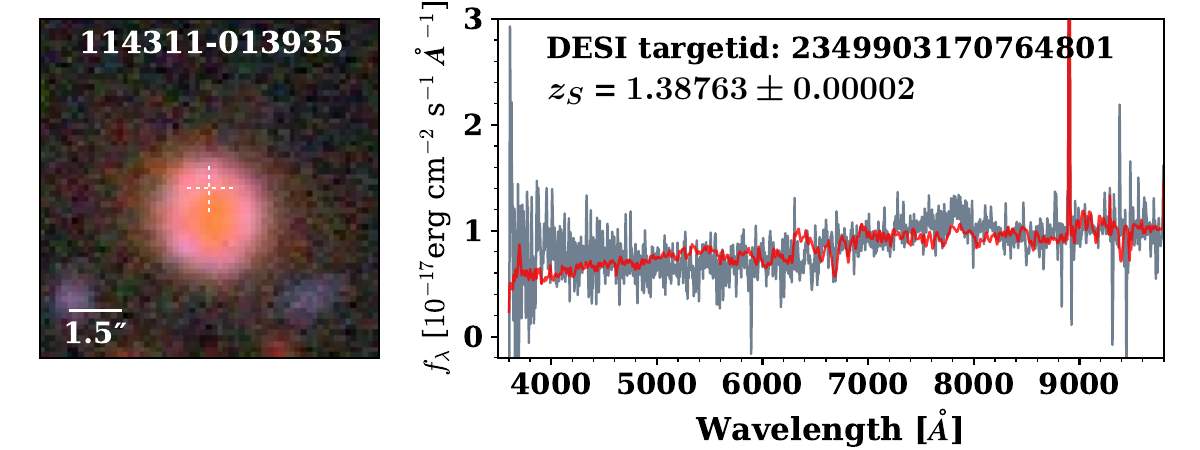}
\includegraphics[width=0.49\textwidth]{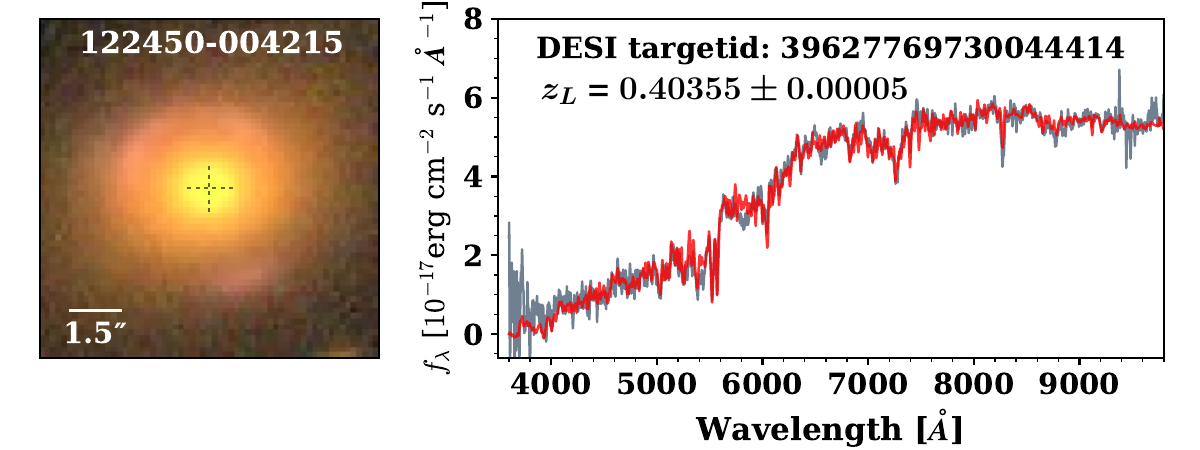}
\includegraphics[width=0.49\textwidth]
{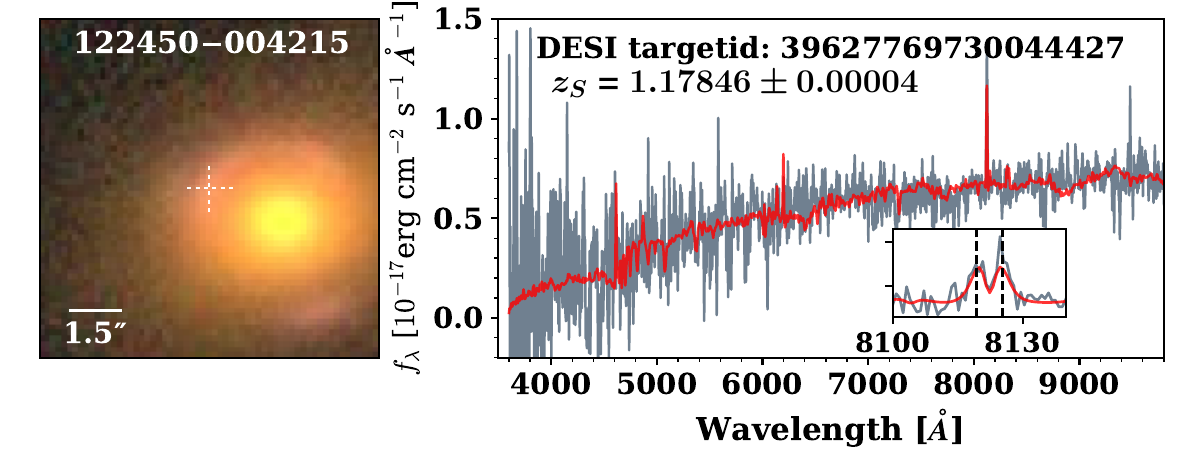}
\includegraphics[width=0.49\textwidth]{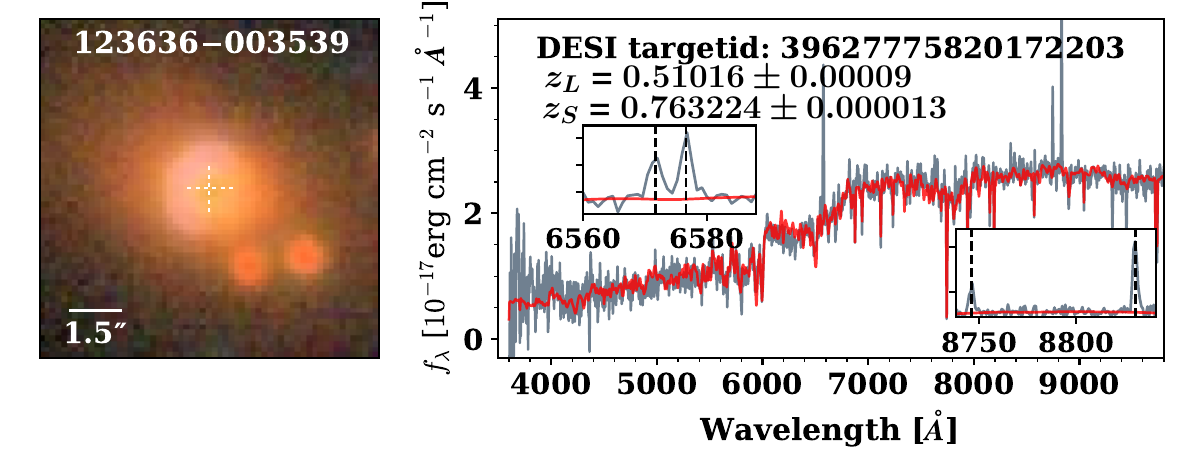}
\includegraphics[width=0.49\textwidth]{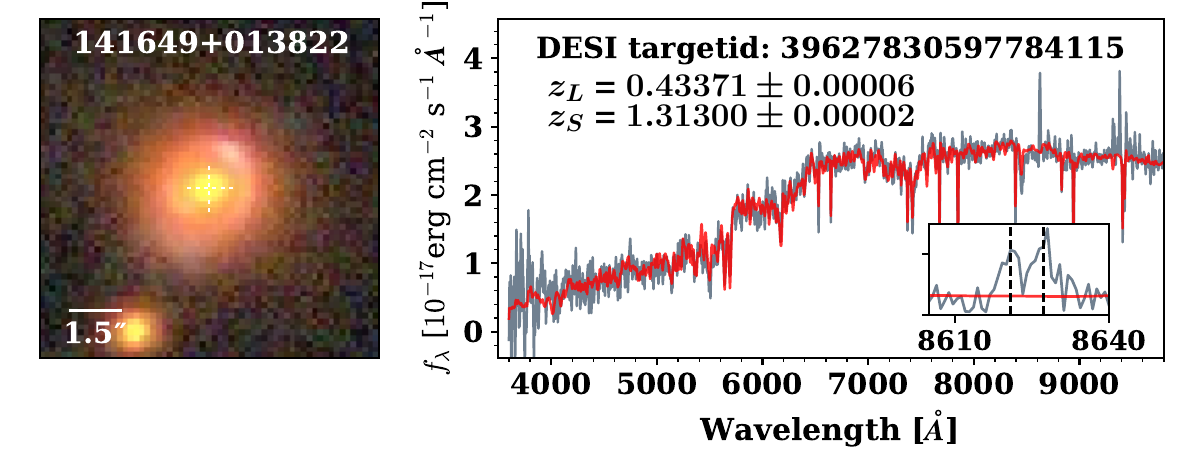}
\includegraphics[width=0.49\textwidth]{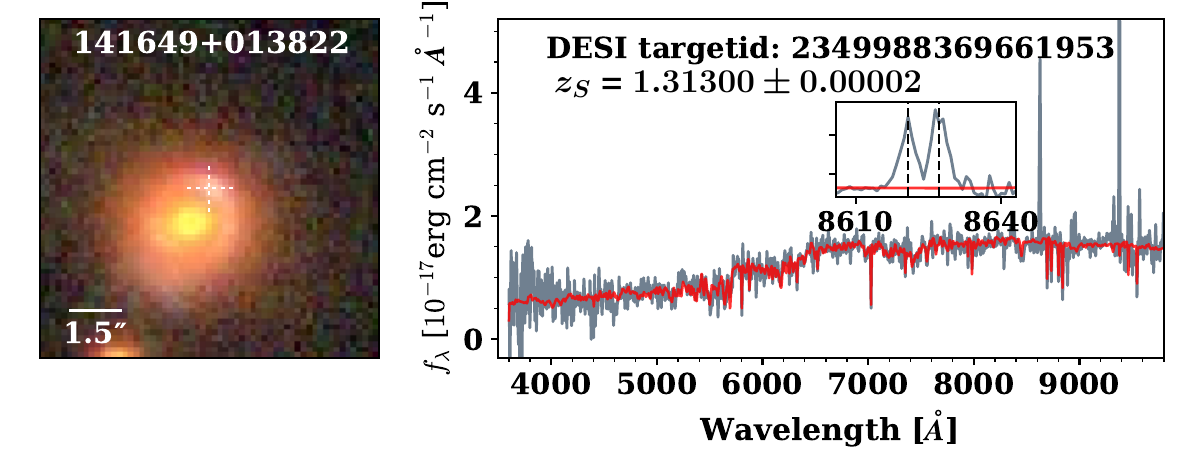}
\caption{\textit{Continued}. For 114311$-$013935, our model to the lens galaxy spectrum is shown.}
\end{figure*}

\begin{figure*}[htbp]
\ContinuedFloat
\centering
\includegraphics[width=0.49\textwidth]
{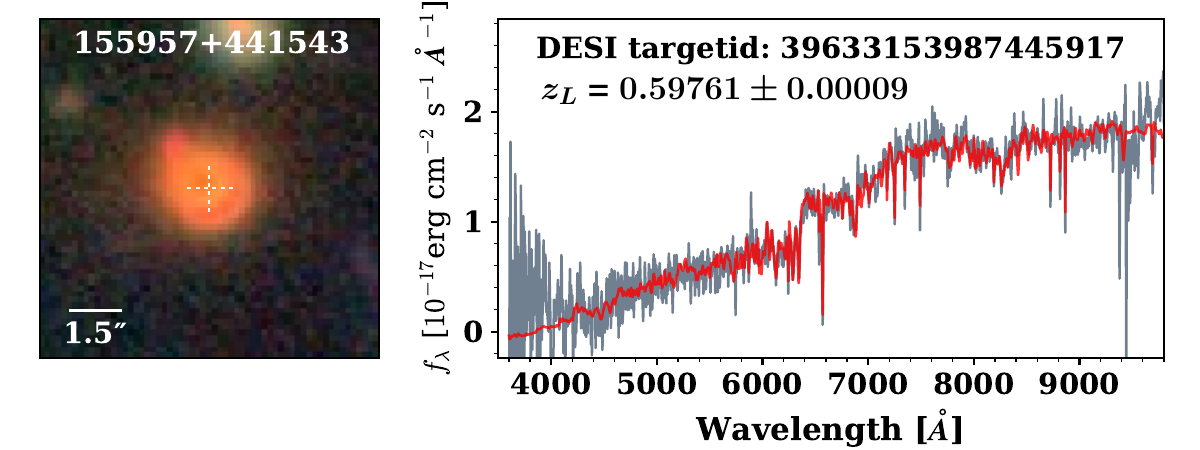}
\includegraphics[width=0.49\textwidth]
{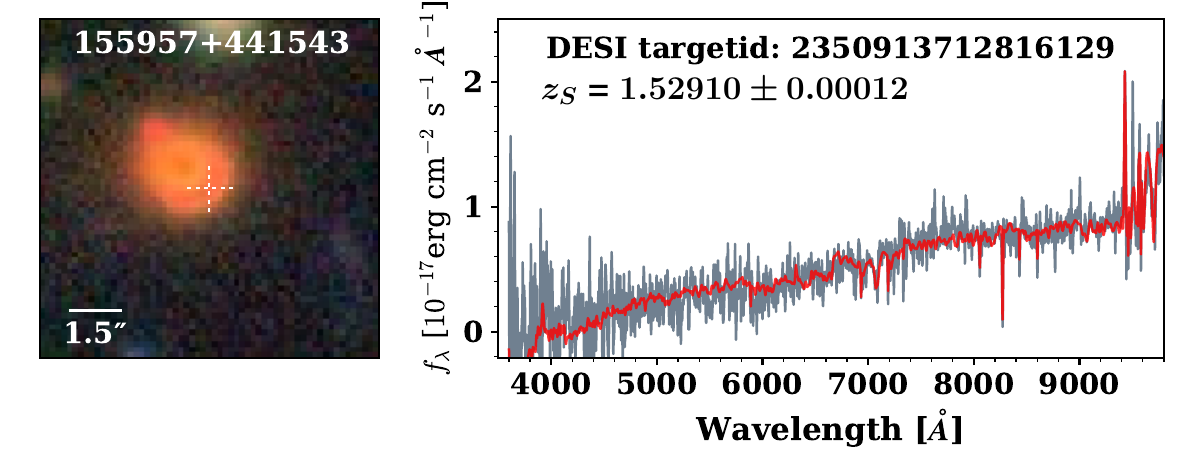}
\includegraphics[width=0.49\textwidth]
{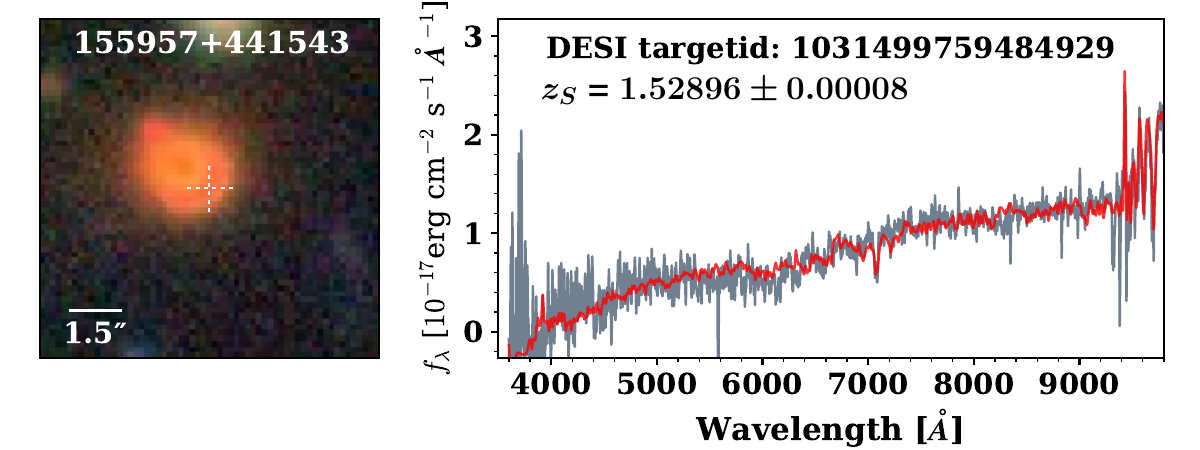}
\includegraphics[width=0.49\textwidth]
{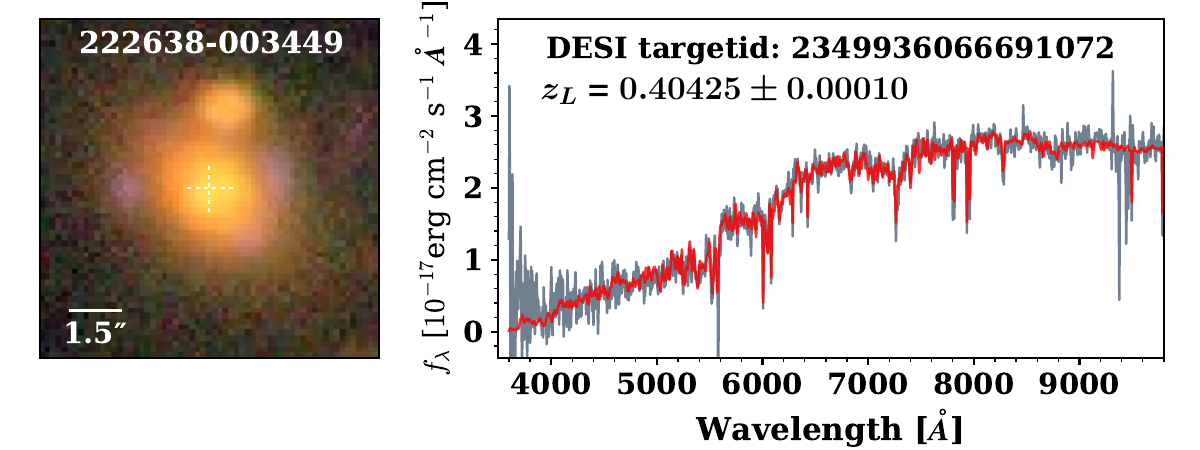}
\includegraphics[width=0.49\textwidth]
{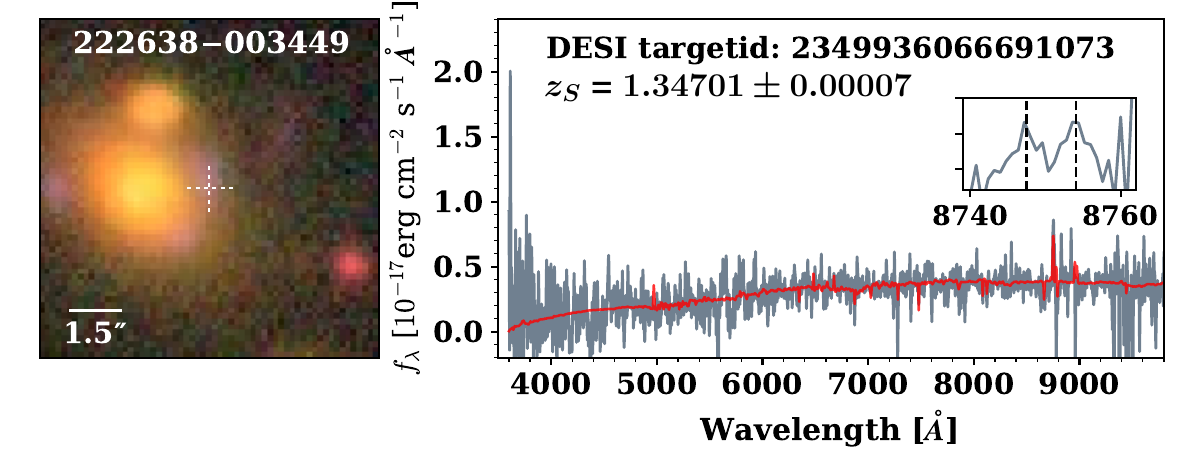}
\includegraphics[width=0.49\textwidth]
{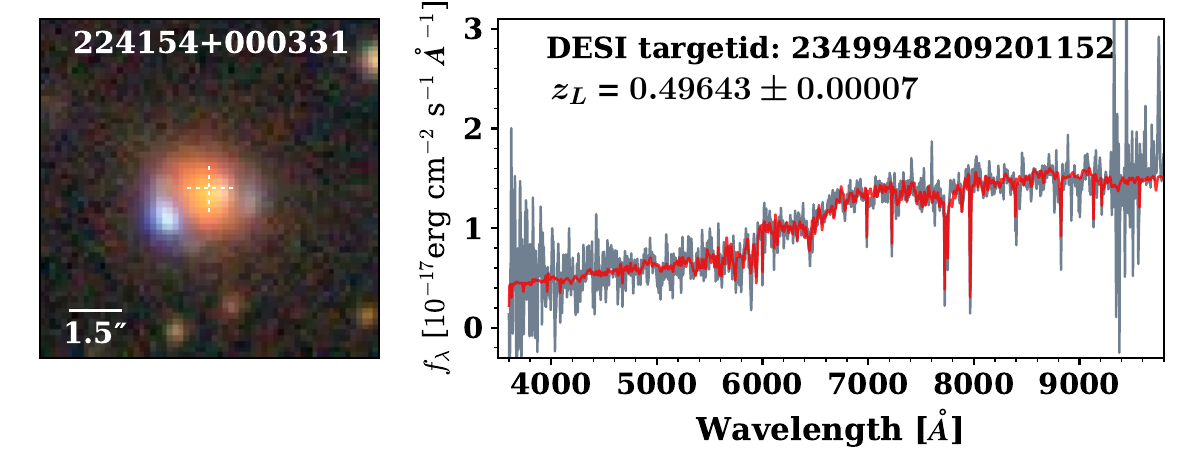}
\includegraphics[width=0.49\textwidth]
{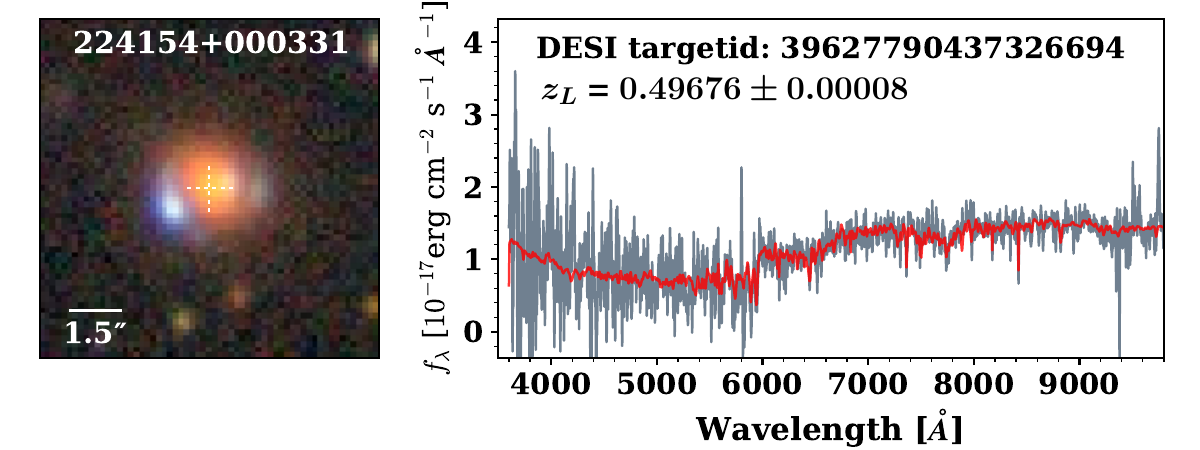}
\includegraphics[width=0.49\textwidth]
{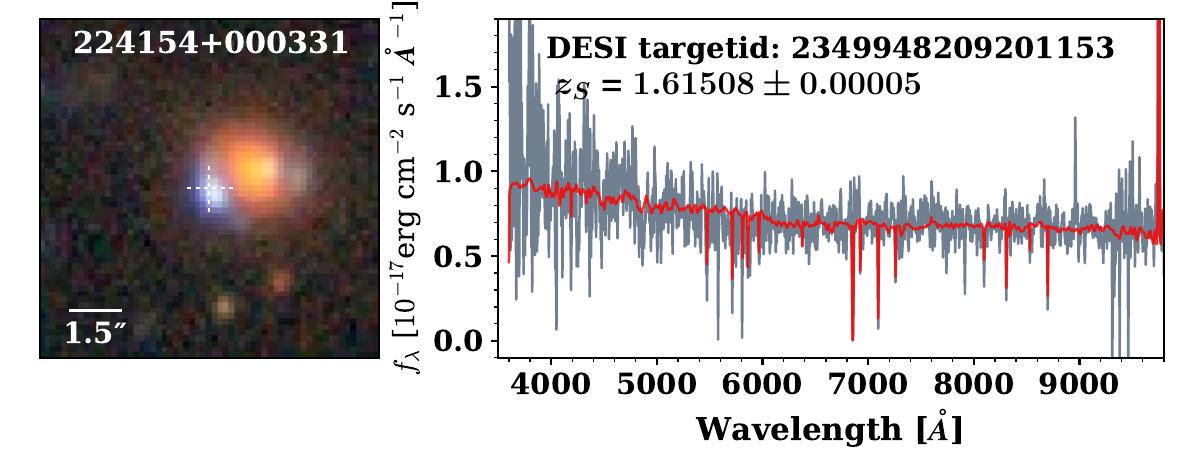}
\includegraphics[width=0.49\textwidth]
{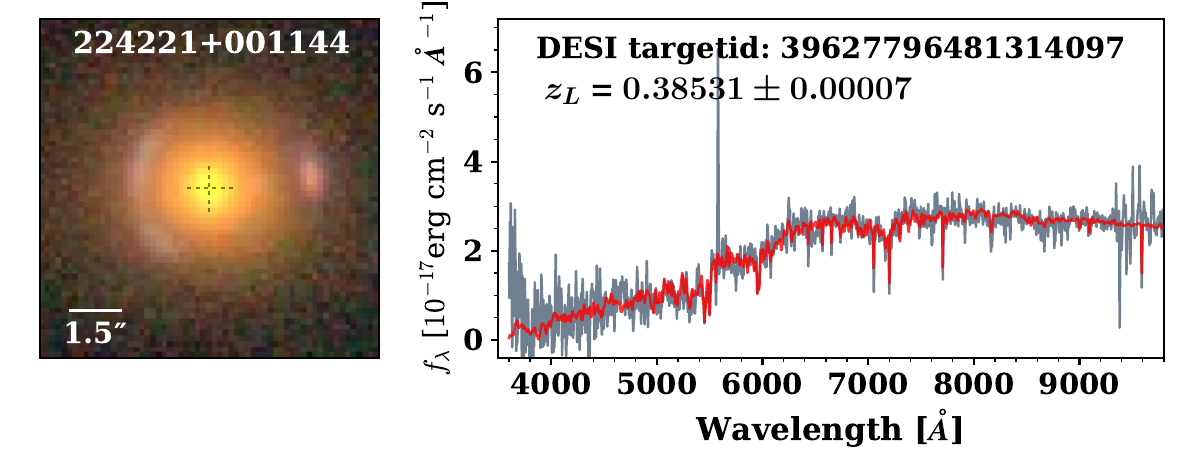}
\includegraphics[width=0.49\textwidth]
{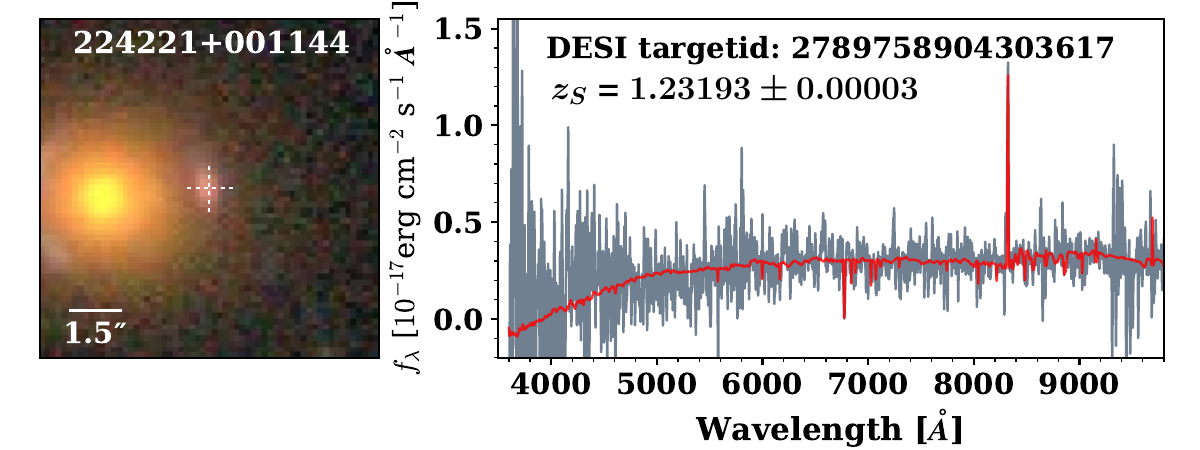}
\includegraphics[width=0.49\textwidth]
{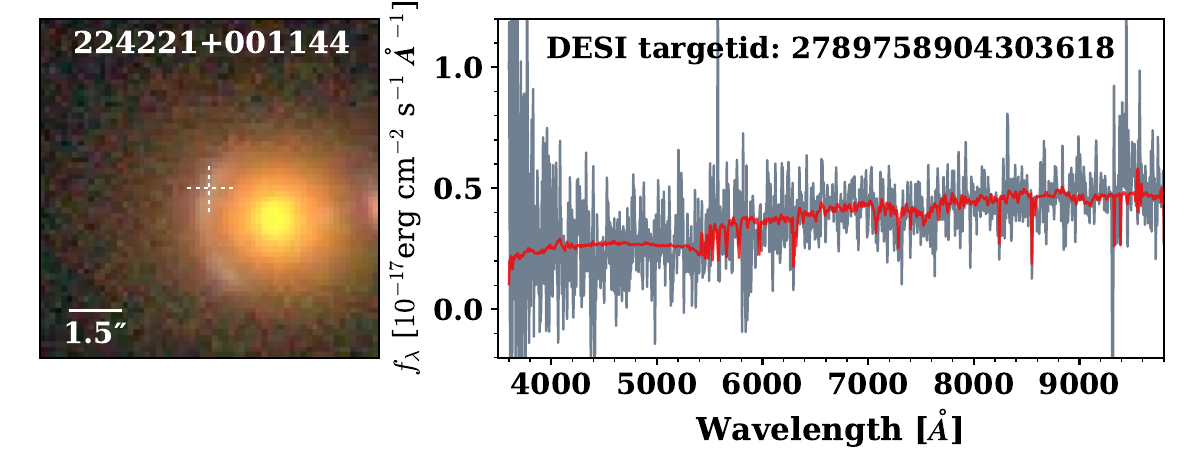}
\hspace*{0.49\textwidth}
\caption{\textit{Continued.}}
\end{figure*}

\begin{figure*}[htbp]
\centering
\includegraphics[width=0.49\textwidth]{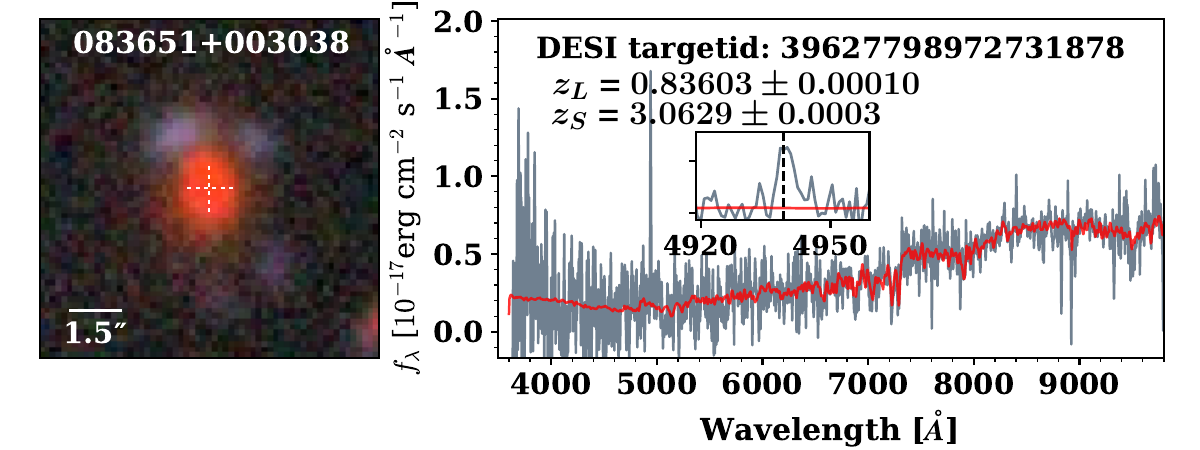}
\includegraphics[width=0.49\textwidth]{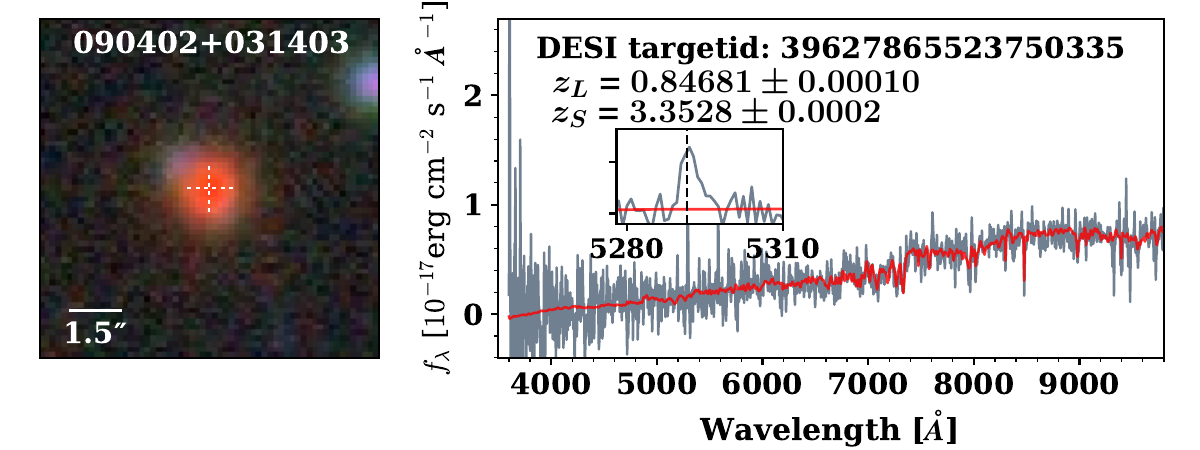}
\includegraphics[width=0.49\textwidth]{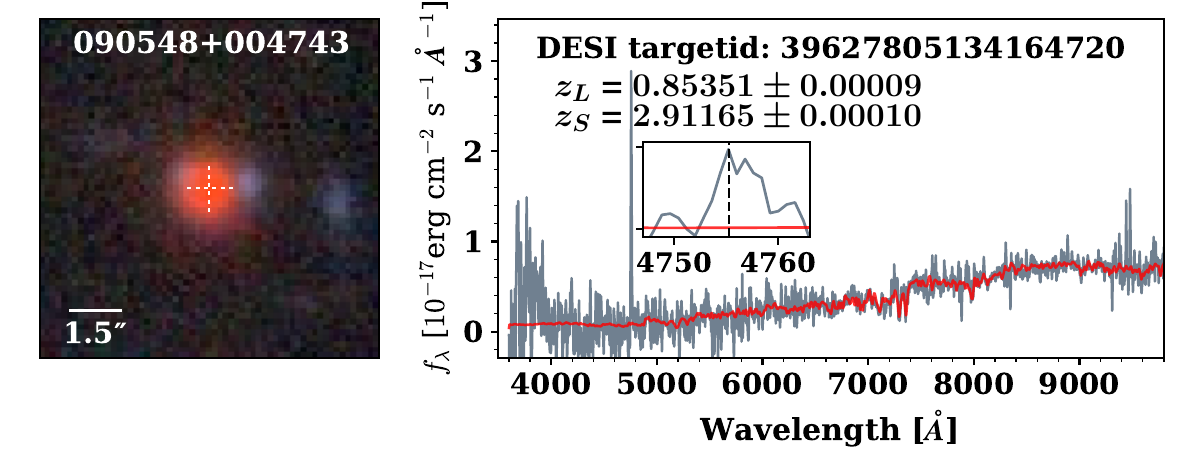}
\includegraphics[width=0.49\textwidth]{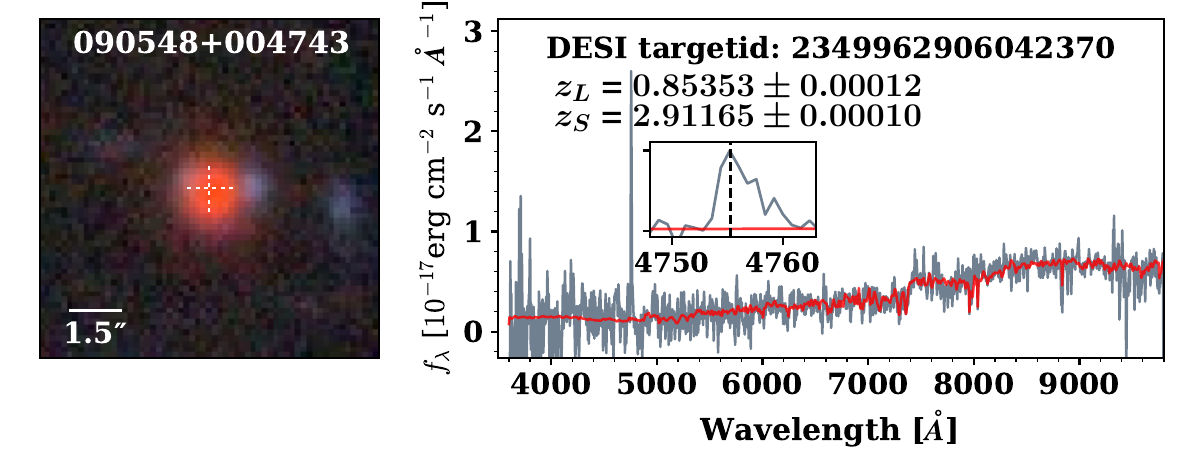}
\includegraphics[width=0.49\textwidth]{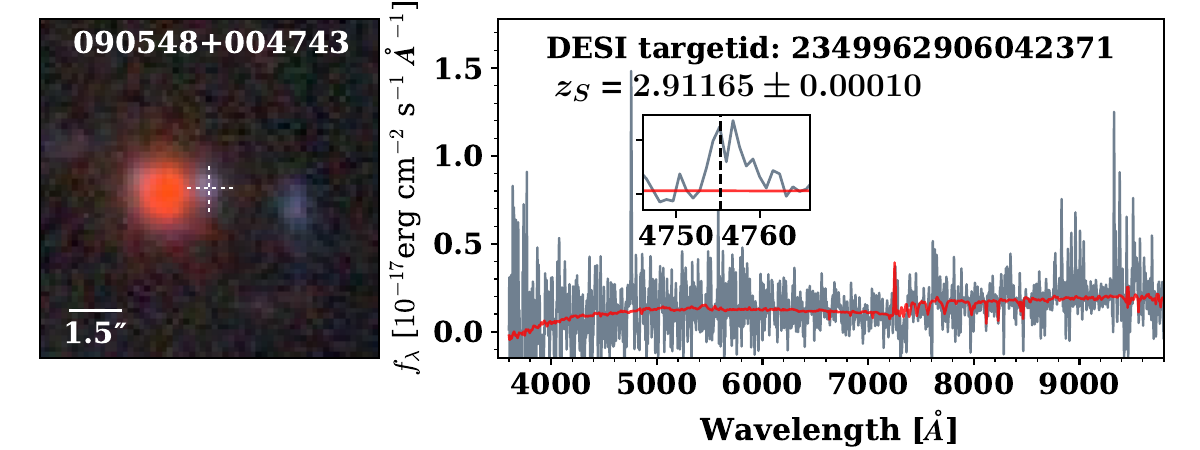}
\includegraphics[width=0.49\textwidth]{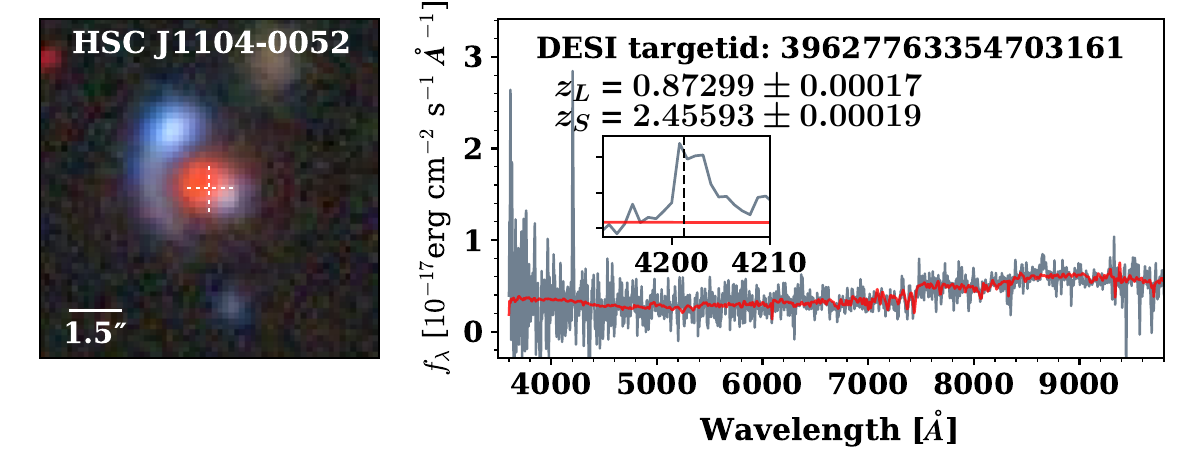}
\includegraphics[width=0.49\textwidth]{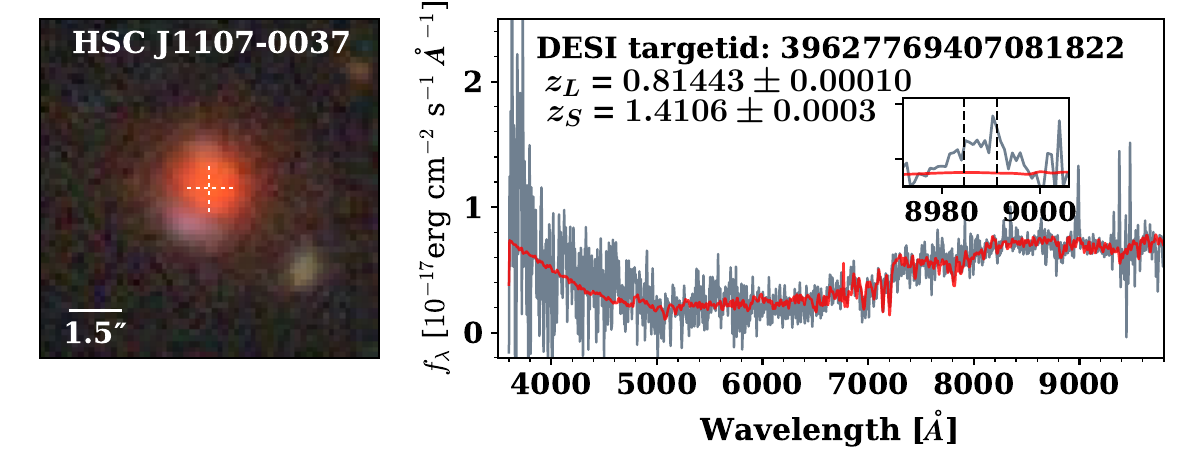}
\includegraphics[width=0.49\textwidth]{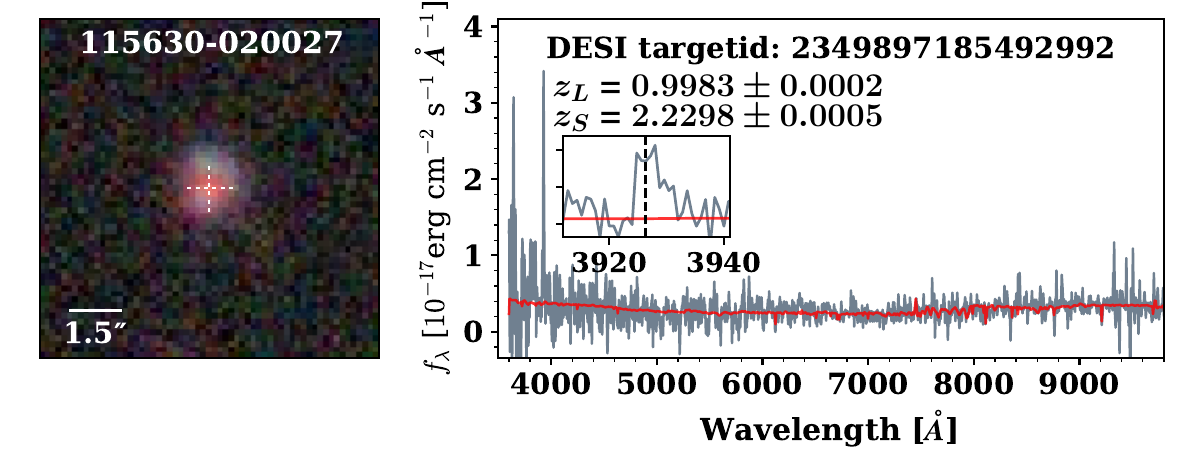}
\includegraphics[width=0.49\textwidth]{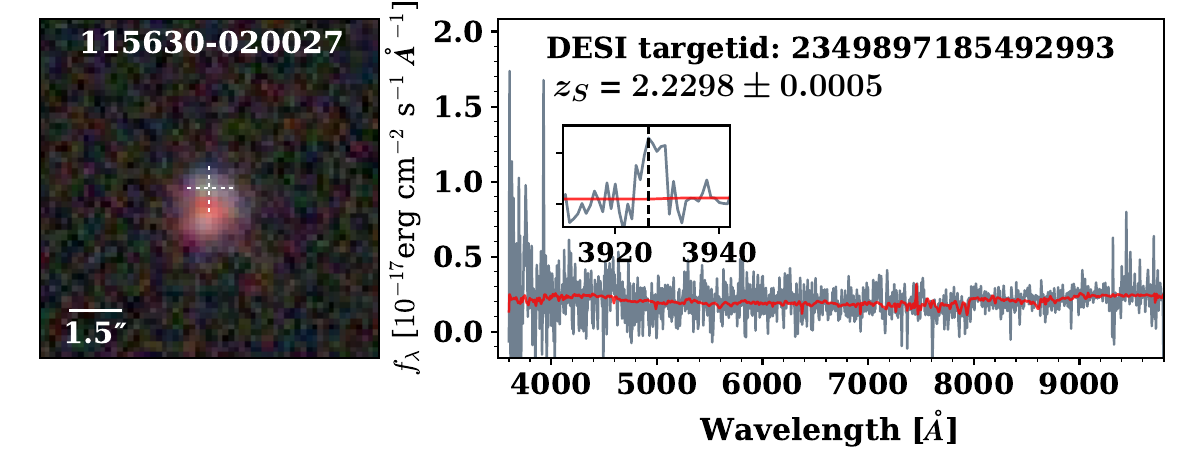}
\includegraphics[width=0.49\textwidth]{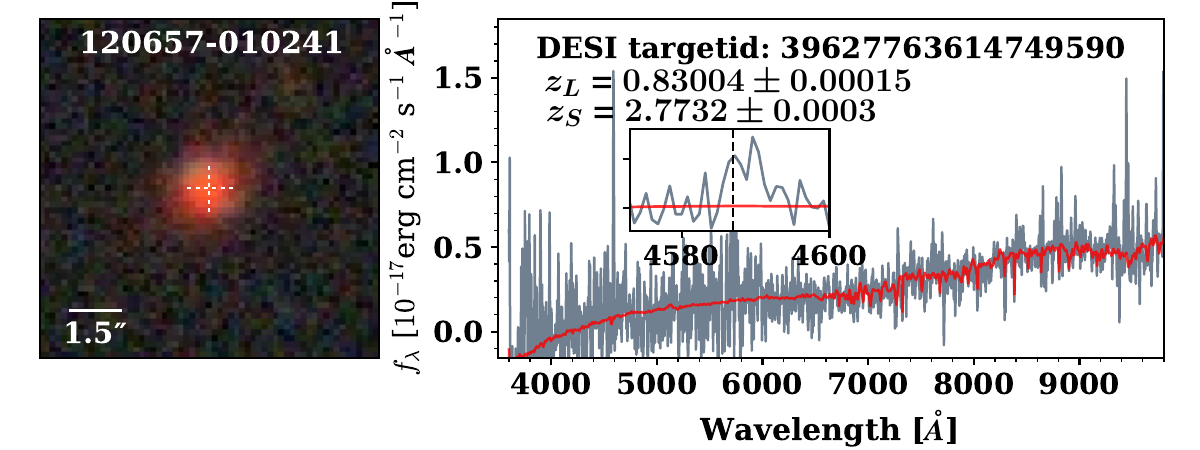}
\includegraphics[width=0.49\textwidth]
{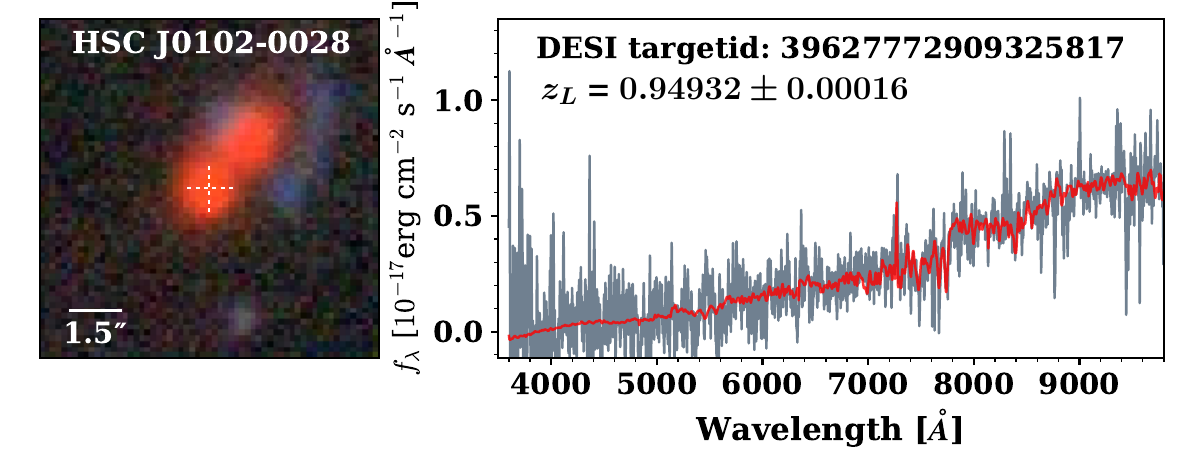}
\includegraphics[width=0.49\textwidth]
{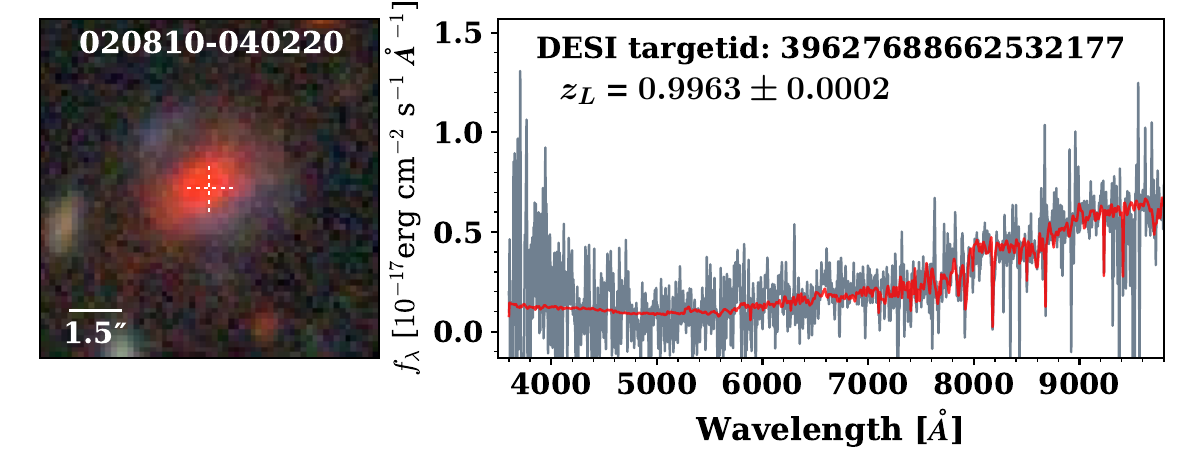}
\includegraphics[width=0.49\textwidth]
{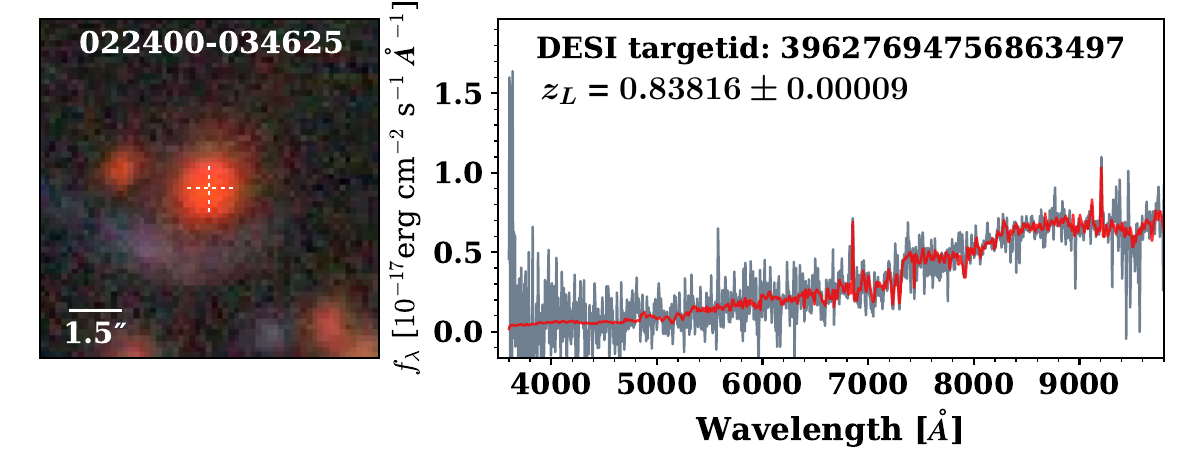}
\includegraphics[width=0.49\textwidth]
{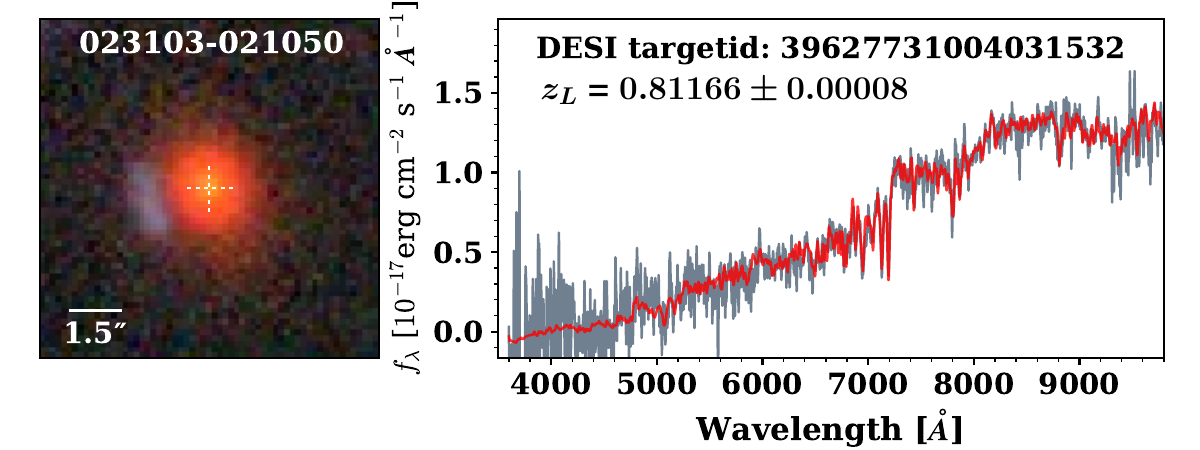}
\caption{Color-composite cutouts and DESI spectra for the 83 systems in Category 2. The firs ten figure sets correspond to the seven systems with both lens and source redshifts measured. Symbols and lines follow the same conventions as in Figure~\ref{fig:C1}.}
\label{fig:C2}
\end{figure*}

\begin{figure*}[htbp]
\ContinuedFloat
\centering
\includegraphics[width=0.49\textwidth]
{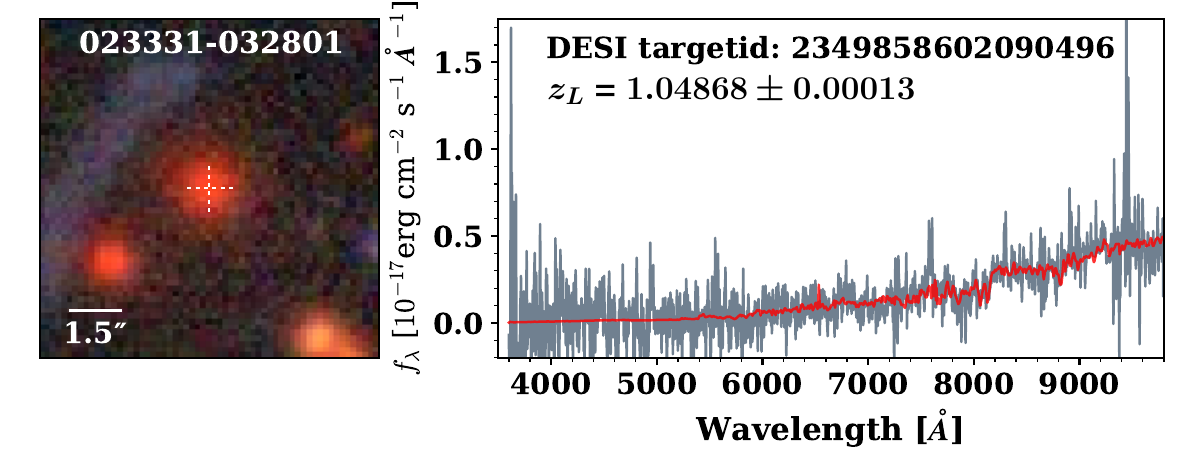}
\includegraphics[width=0.49\textwidth]
{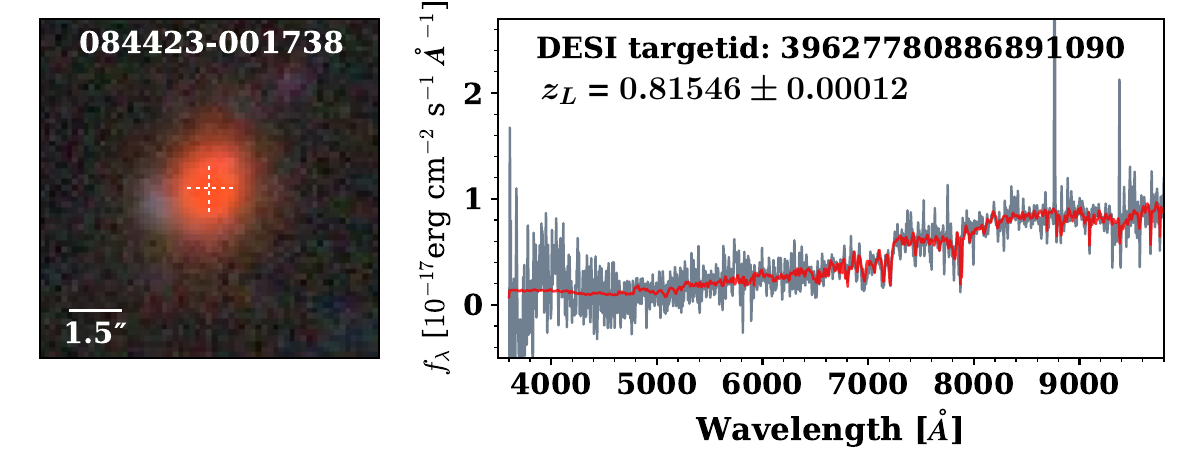}
\includegraphics[width=0.49\textwidth]
{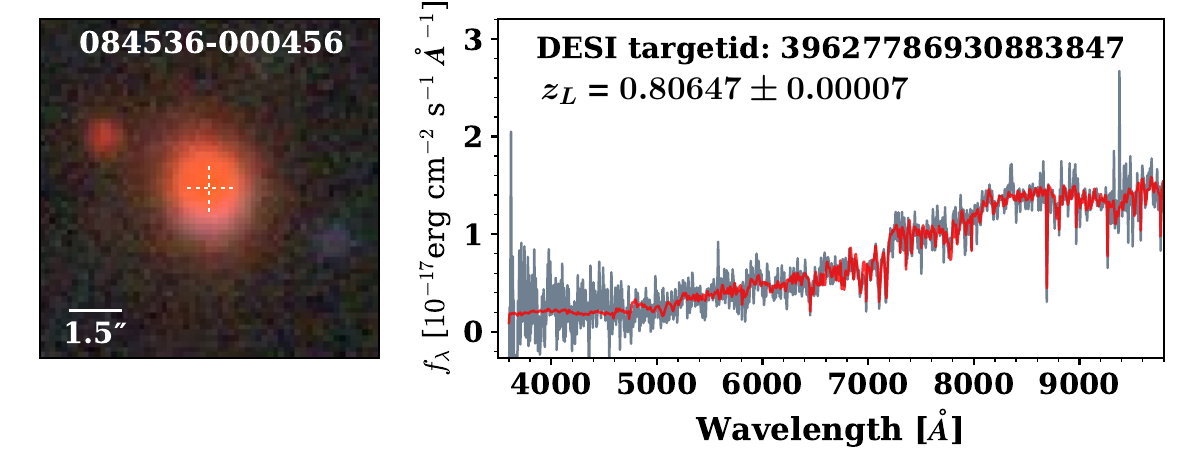}
\includegraphics[width=0.49\textwidth]
{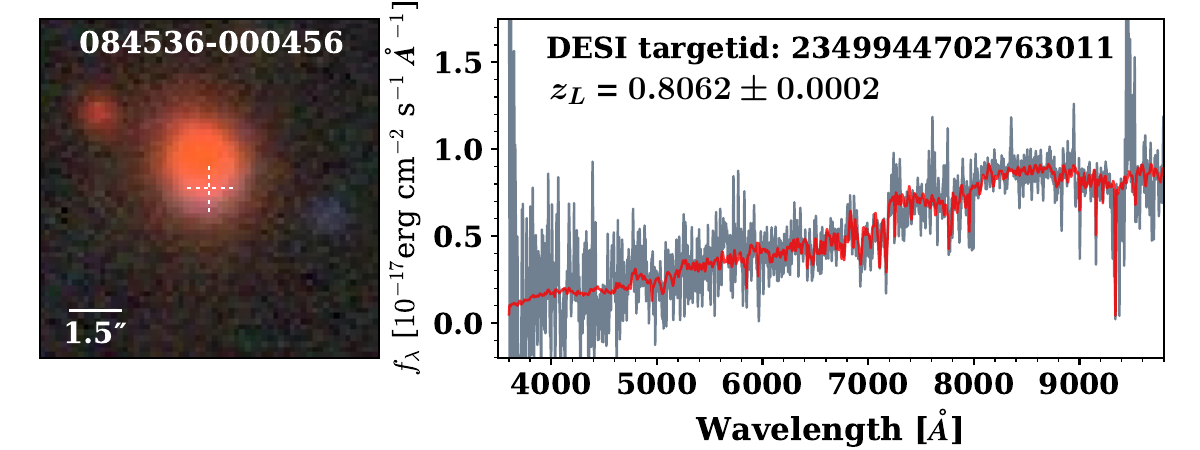}
\includegraphics[width=0.49\textwidth]
{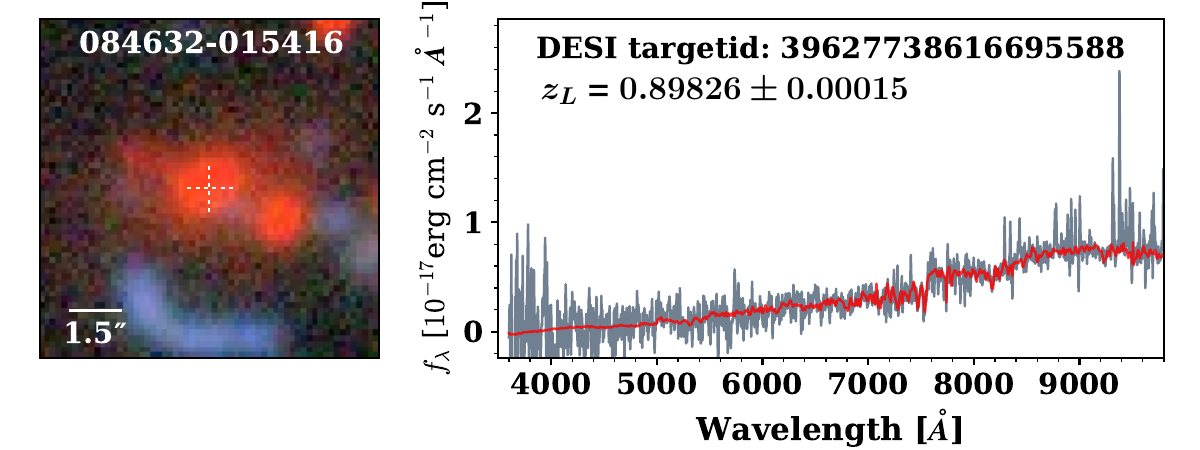}
\includegraphics[width=0.49\textwidth]
{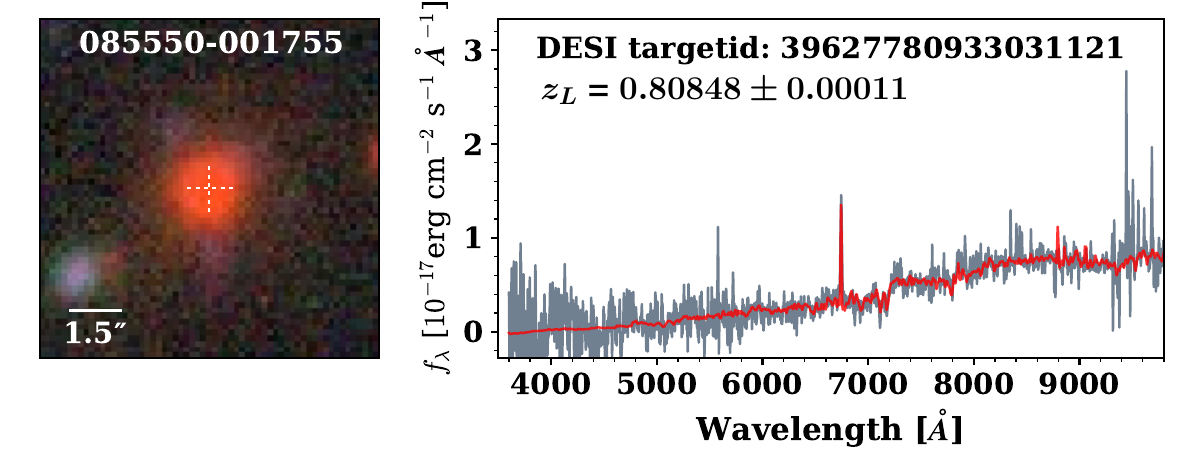}
\includegraphics[width=0.49\textwidth]
{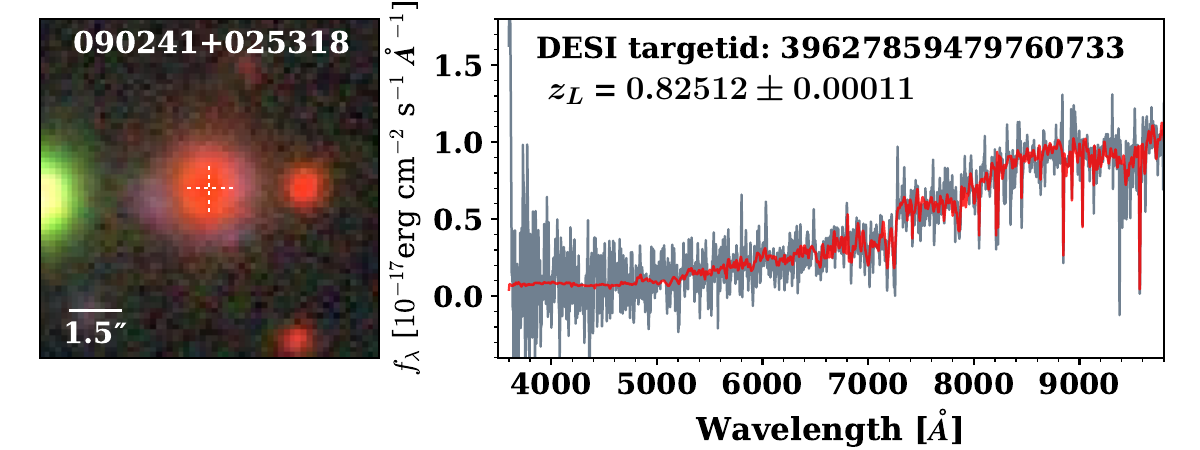}
\includegraphics[width=0.49\textwidth]
{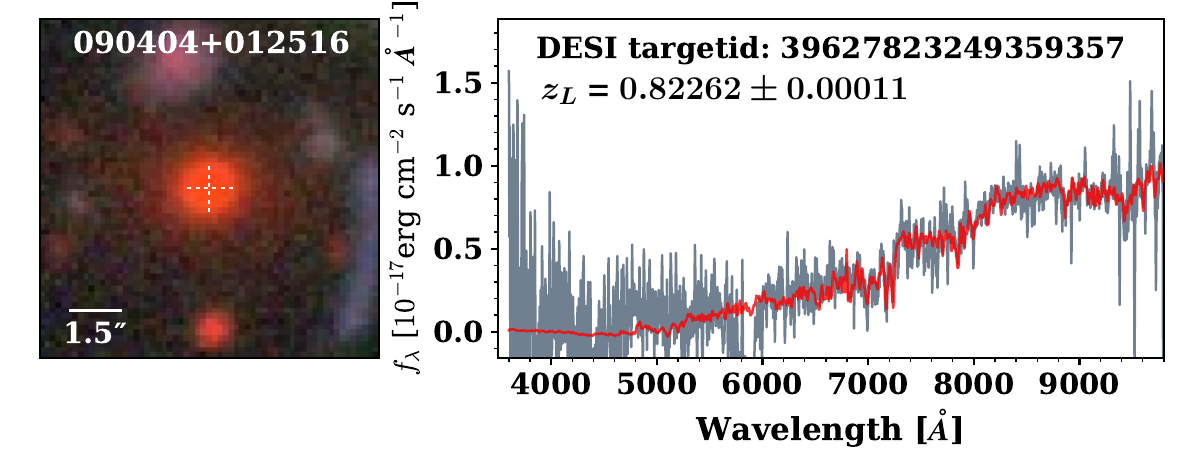}
\includegraphics[width=0.49\textwidth]
{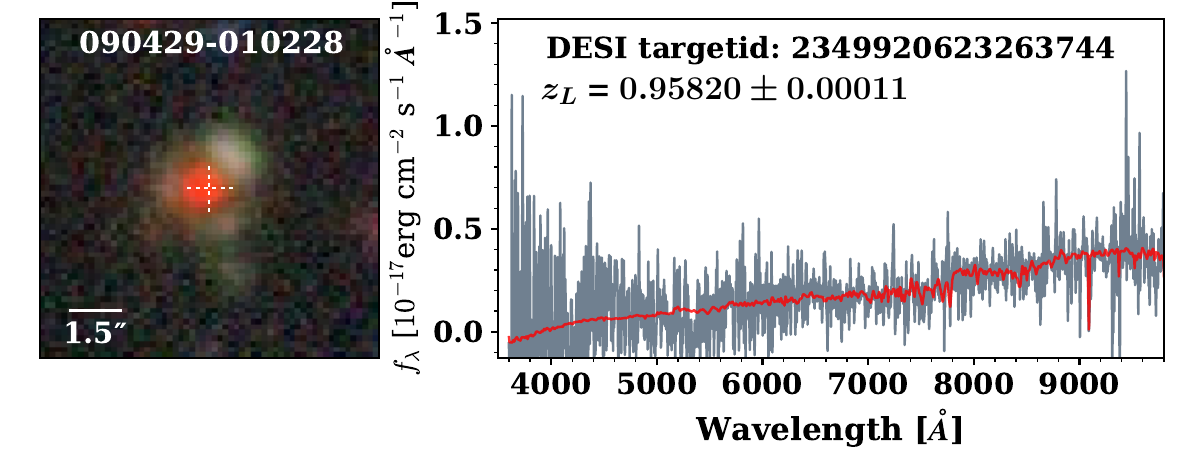}
\includegraphics[width=0.49\textwidth]
{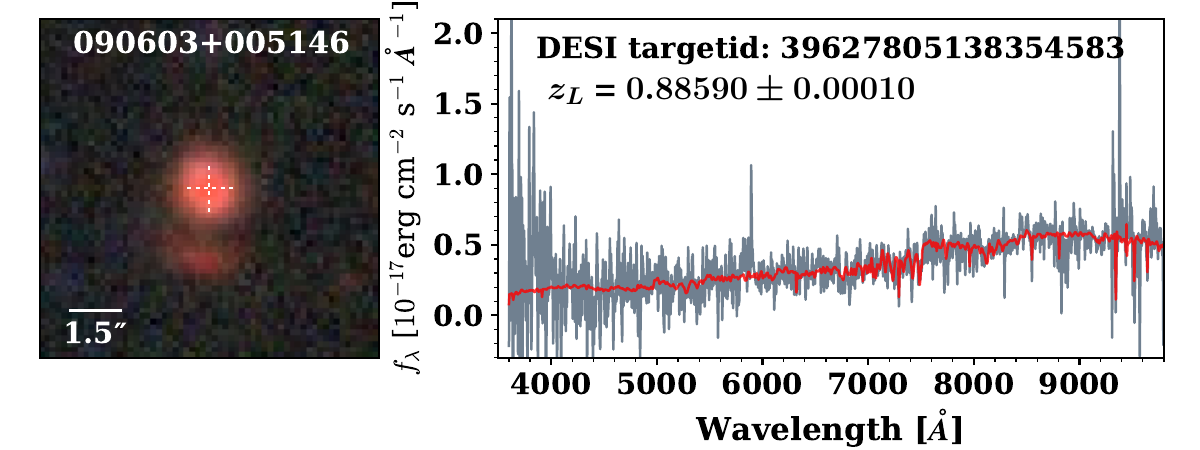}
\includegraphics[width=0.49\textwidth]
{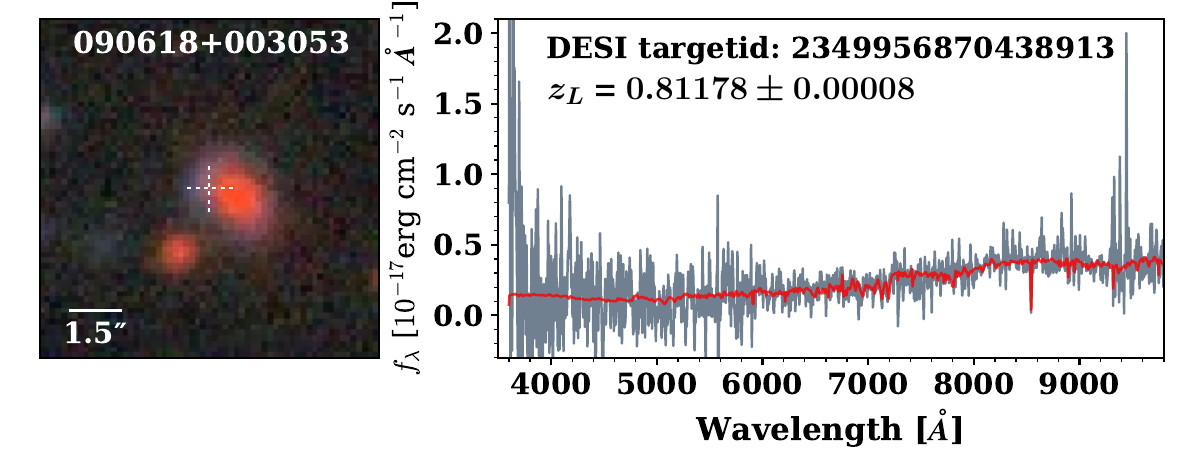}
\includegraphics[width=0.49\textwidth]
{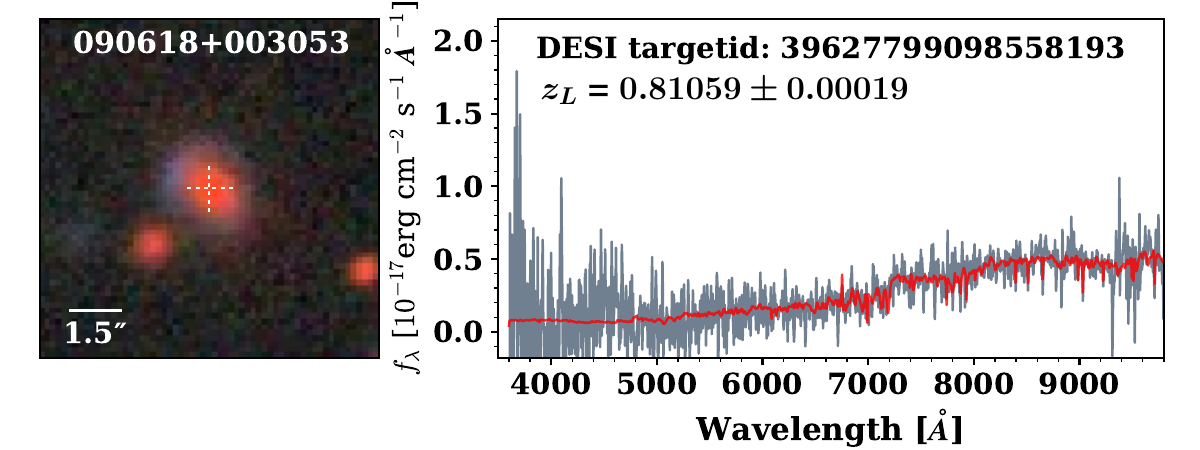}
\includegraphics[width=0.49\textwidth]
{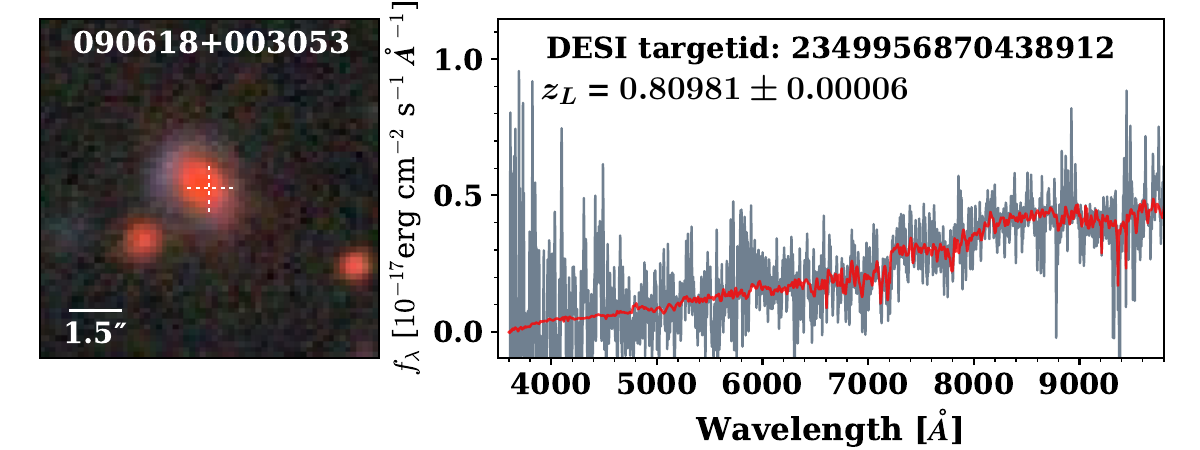}
\includegraphics[width=0.49\textwidth]
{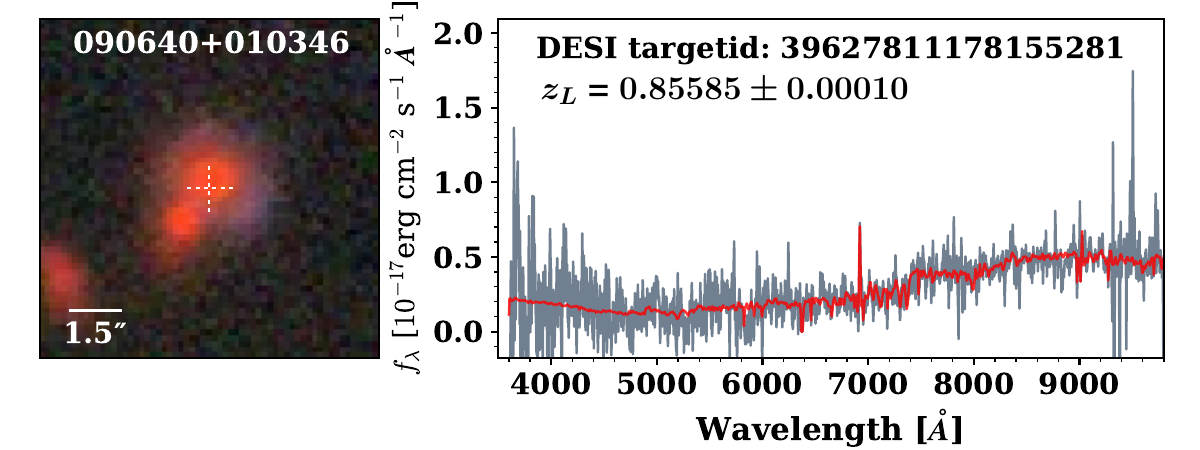}
\caption{\textit{Continued.}}
\end{figure*}

\begin{figure*}[htbp]
\ContinuedFloat
\centering
\includegraphics[width=0.49\textwidth]
{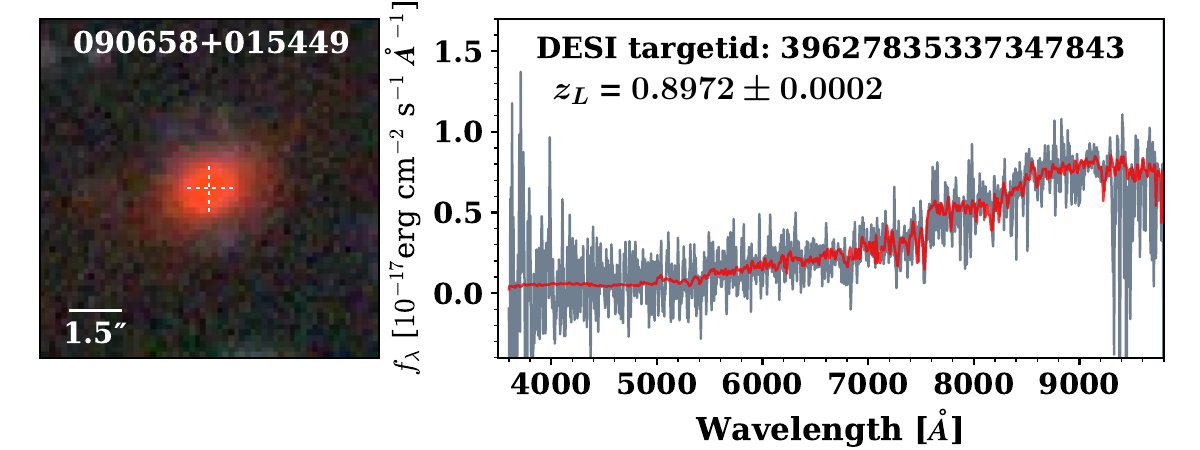}
\includegraphics[width=0.49\textwidth]
{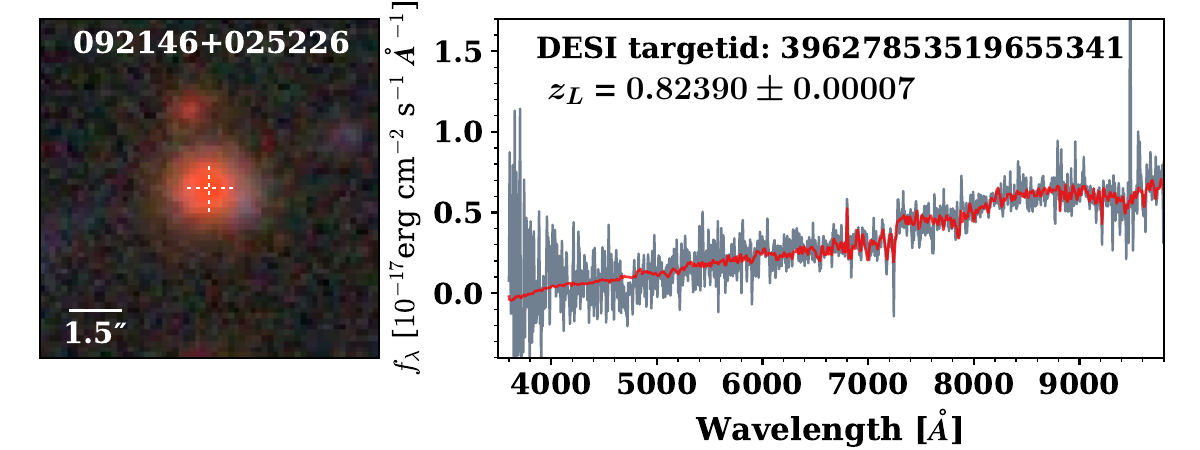}
\includegraphics[width=0.49\textwidth]
{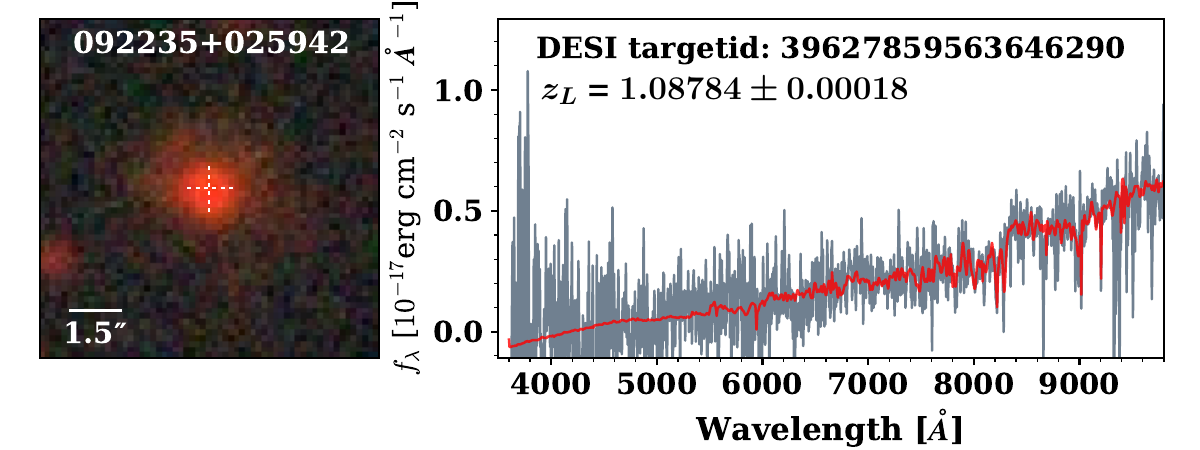}
\includegraphics[width=0.49\textwidth]
{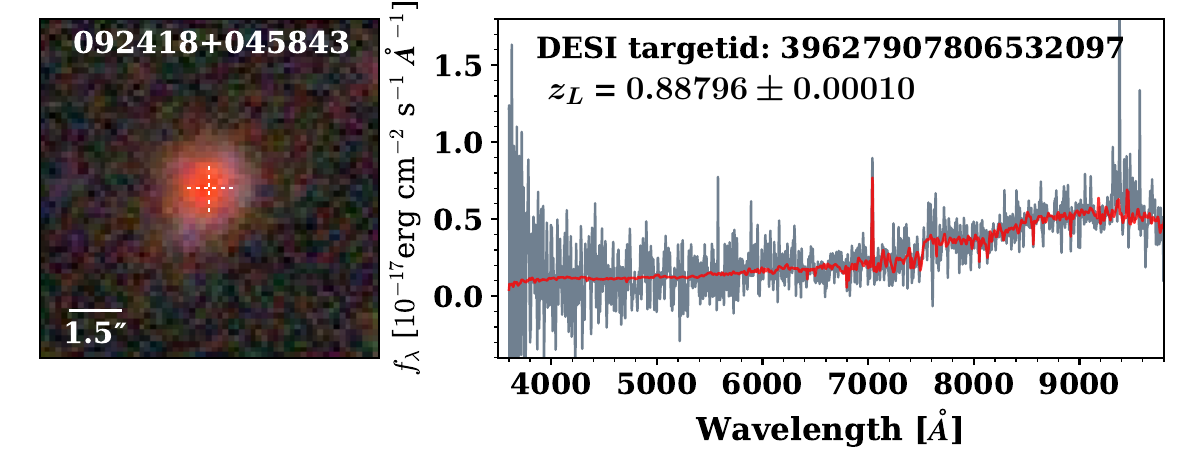}
\includegraphics[width=0.49\textwidth]
{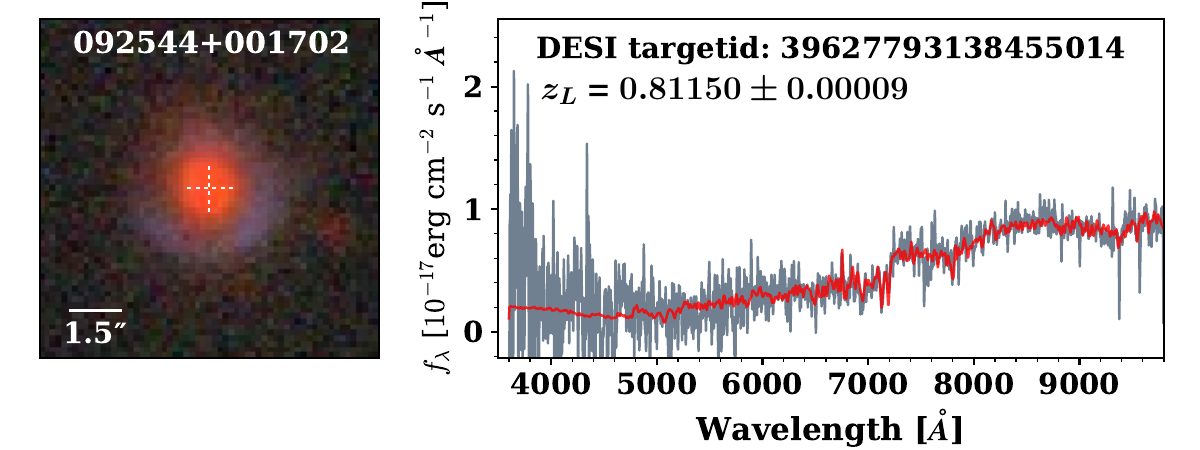}
\includegraphics[width=0.49\textwidth]
{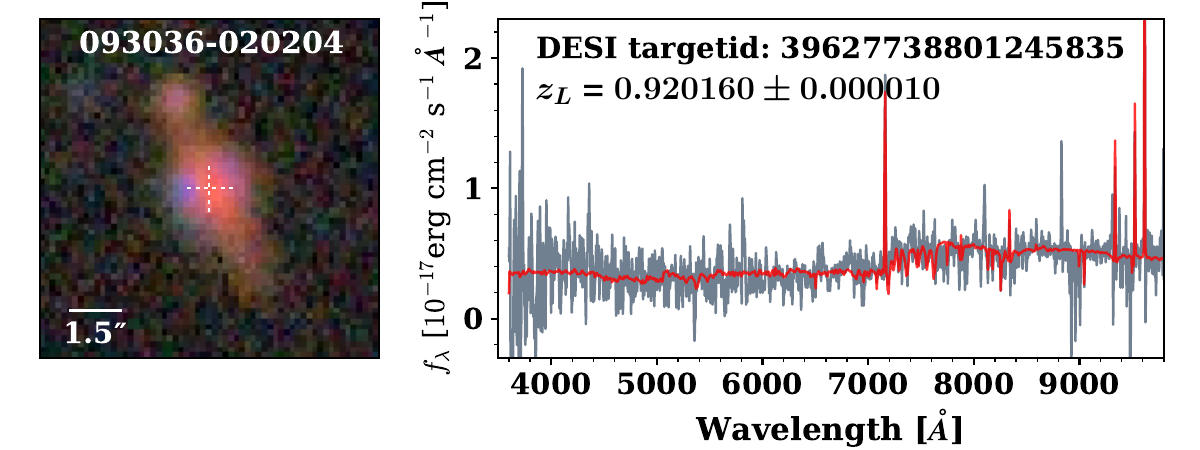}
\includegraphics[width=0.49\textwidth]
{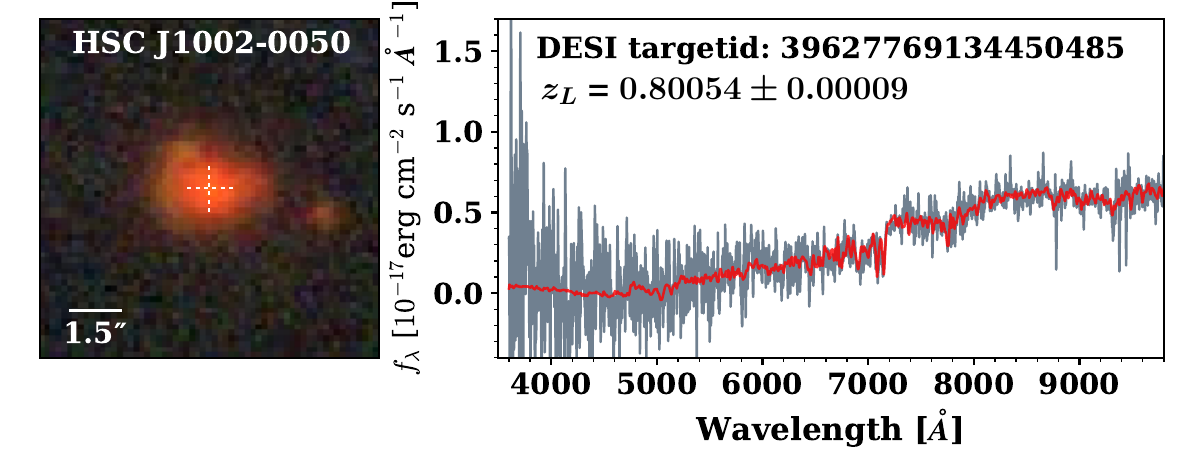}
\includegraphics[width=0.49\textwidth]
{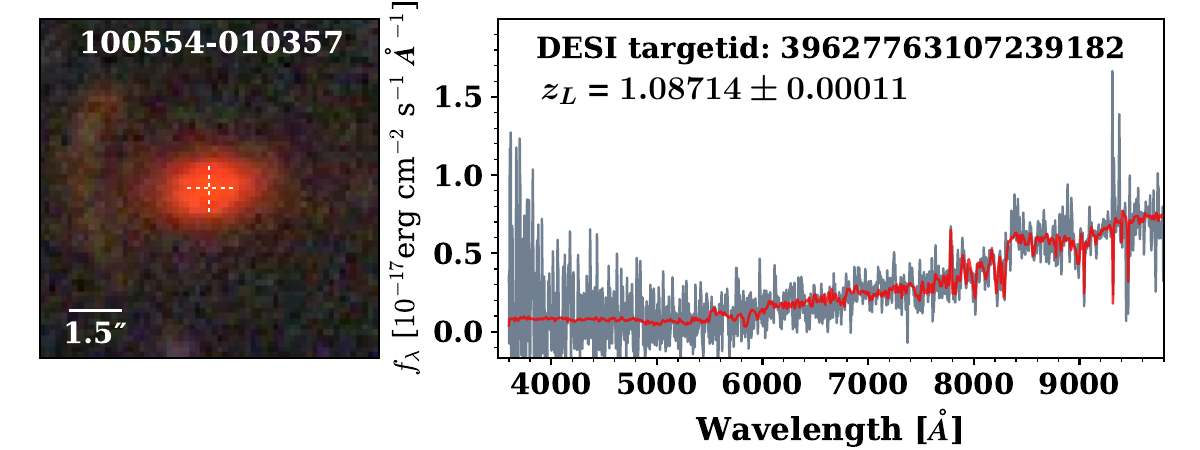}
\includegraphics[width=0.49\textwidth]
{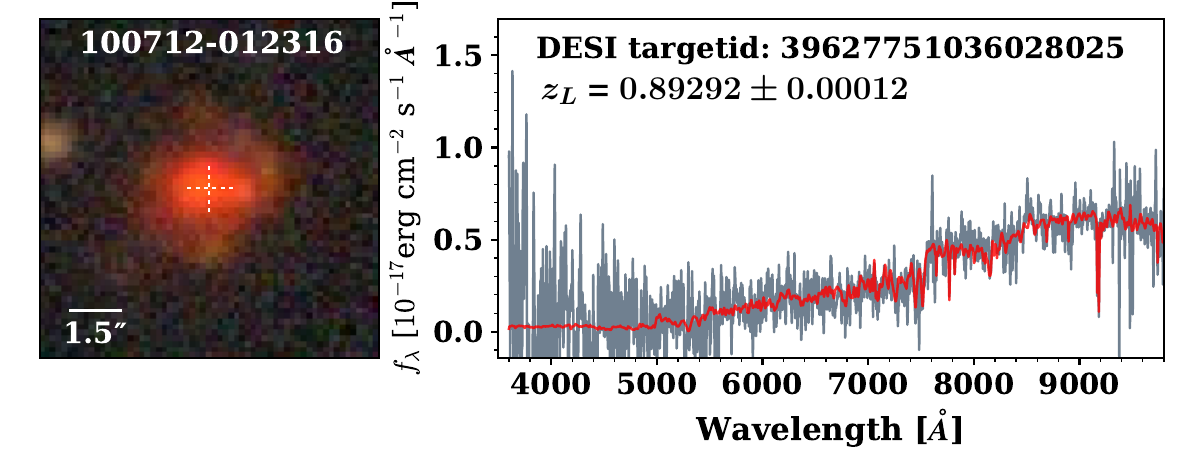}
\includegraphics[width=0.49\textwidth]
{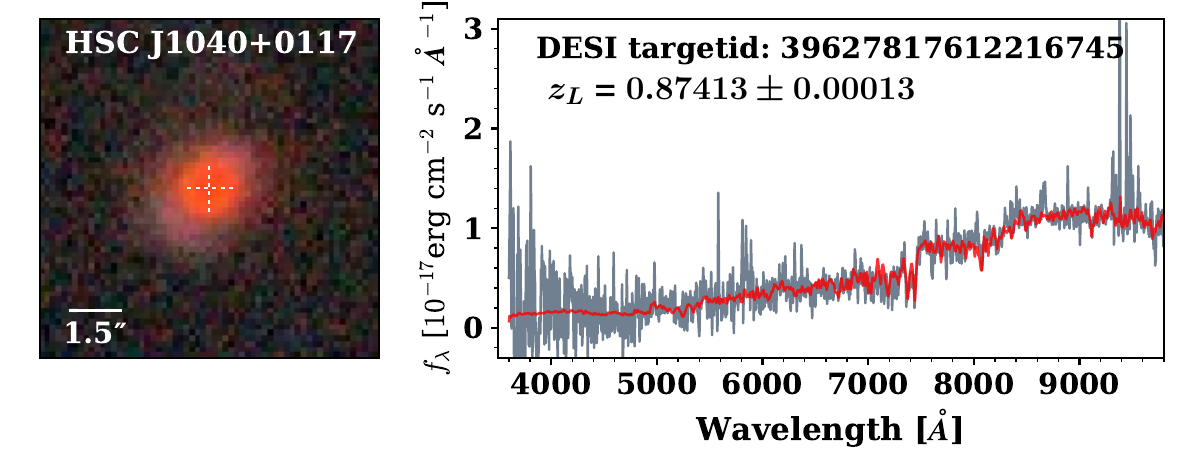}
\includegraphics[width=0.49\textwidth]
{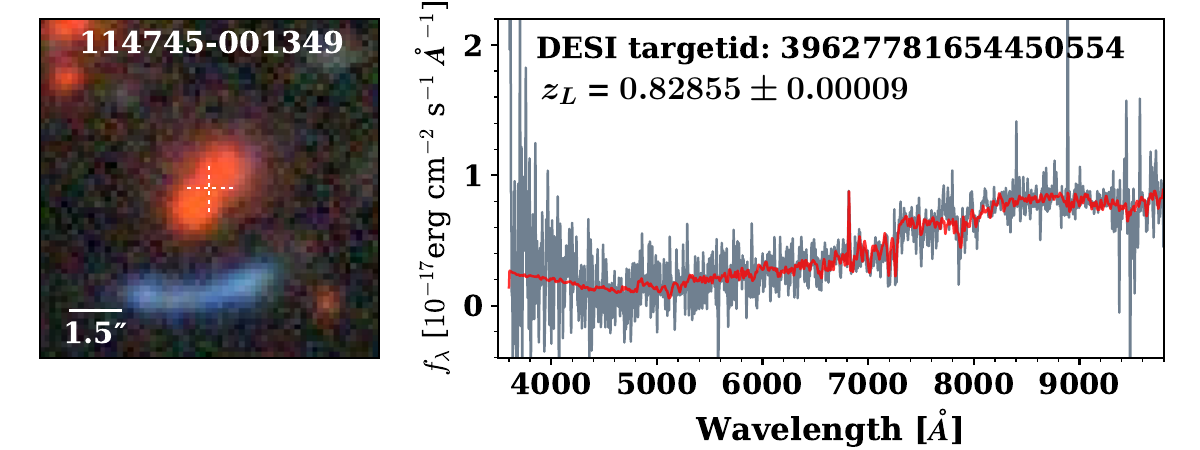}
\includegraphics[width=0.49\textwidth]
{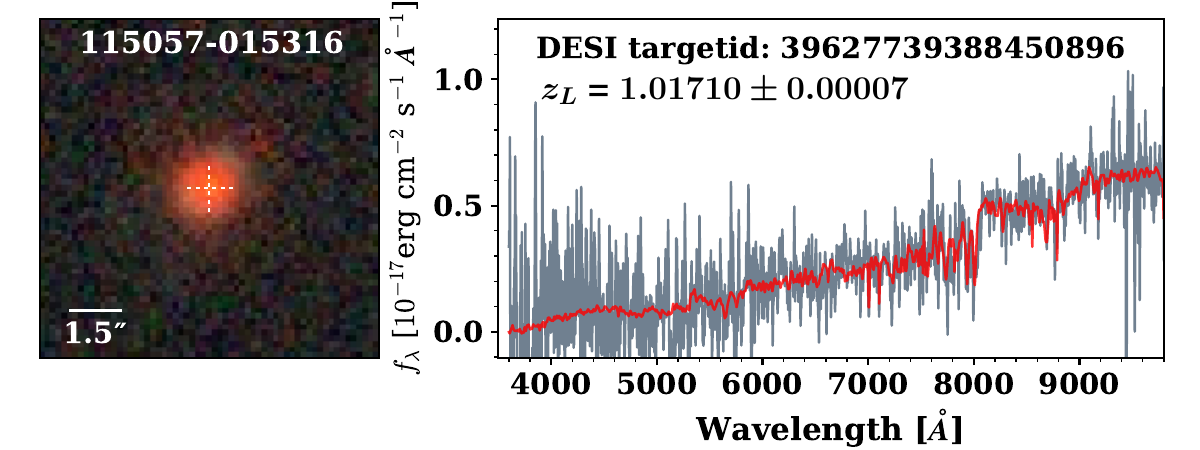}
\includegraphics[width=0.49\textwidth]
{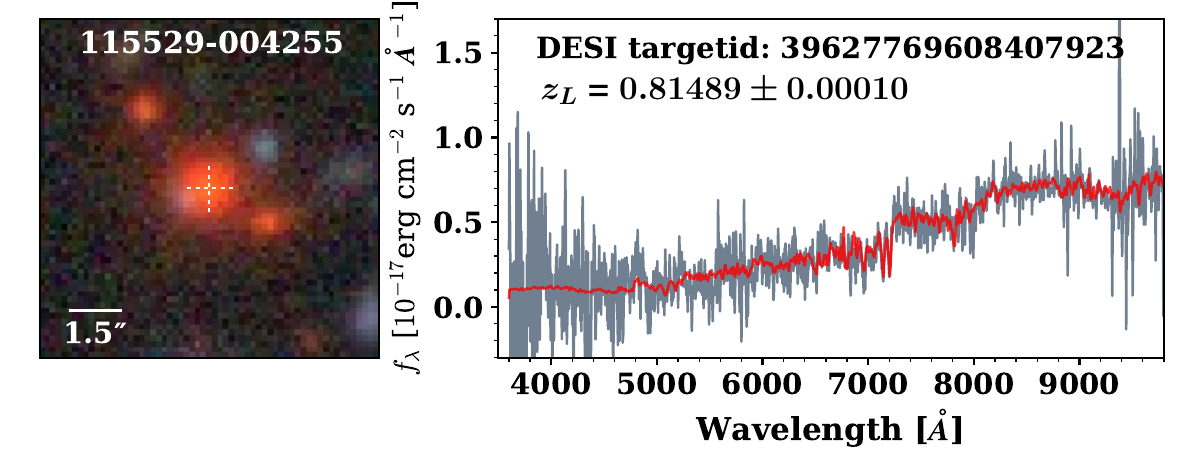}
\includegraphics[width=0.49\textwidth]
{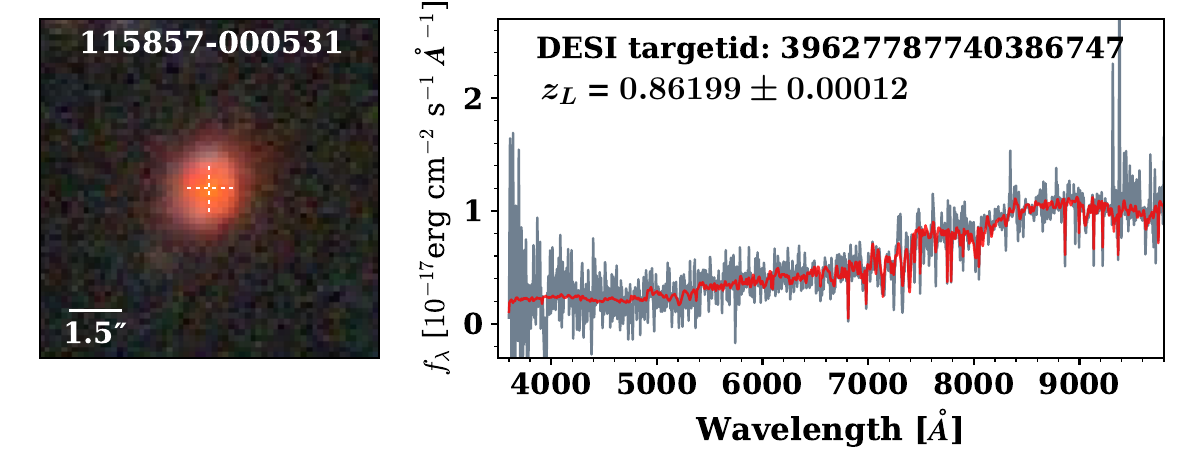}
\caption{\textit{Continued.}}
\end{figure*}

\begin{figure*}[htbp]
\ContinuedFloat
\centering
\includegraphics[width=0.49\textwidth]
{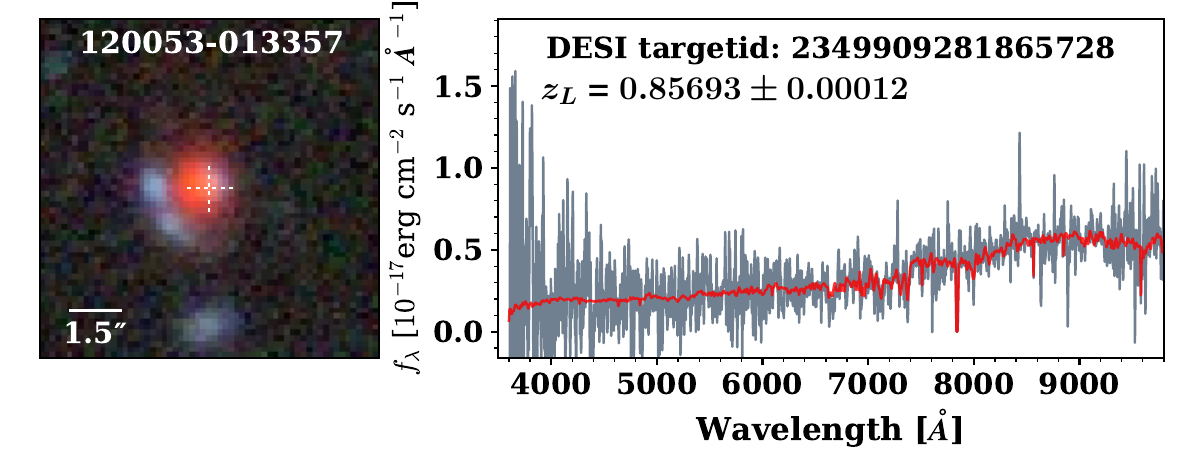}
\includegraphics[width=0.49\textwidth]
{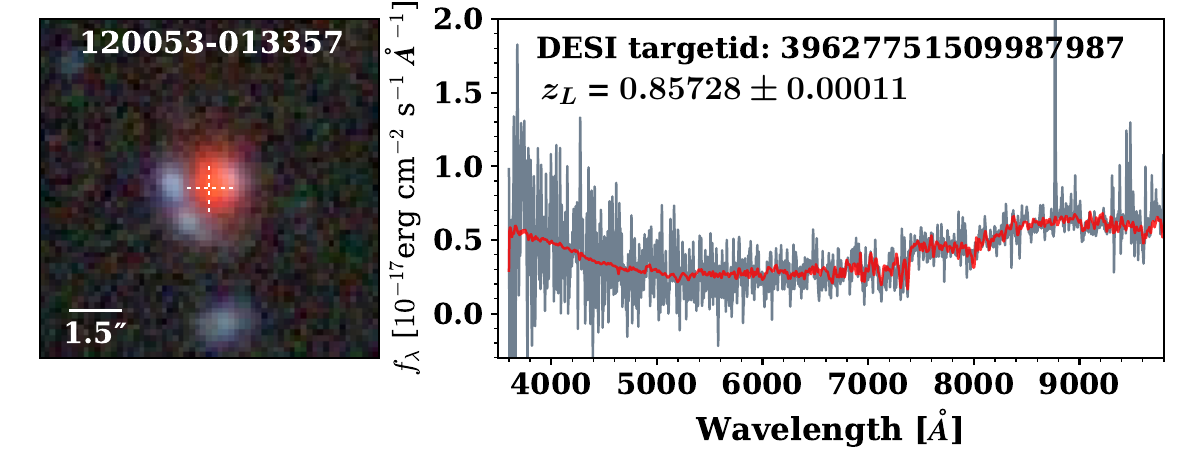}
\includegraphics[width=0.49\textwidth]
{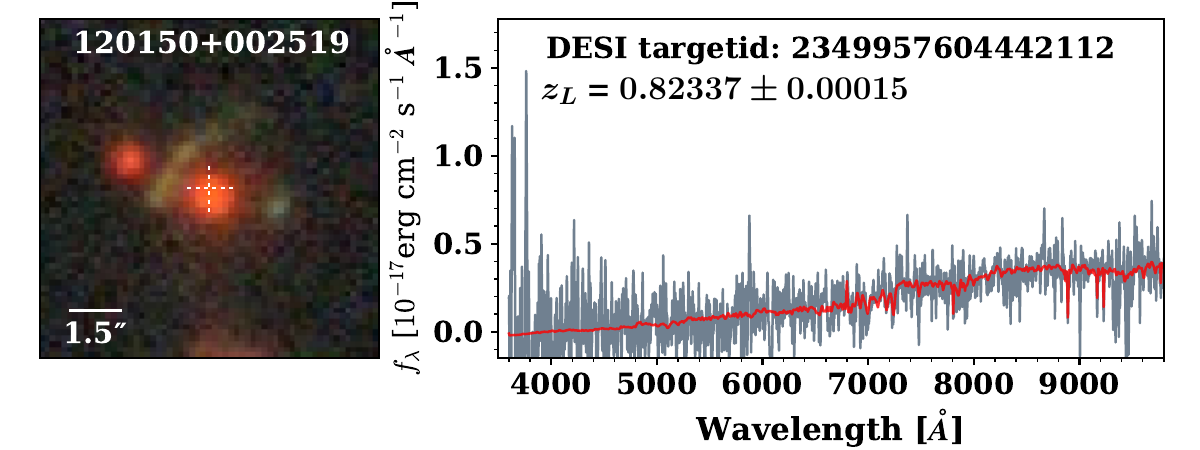}
\includegraphics[width=0.49\textwidth]
{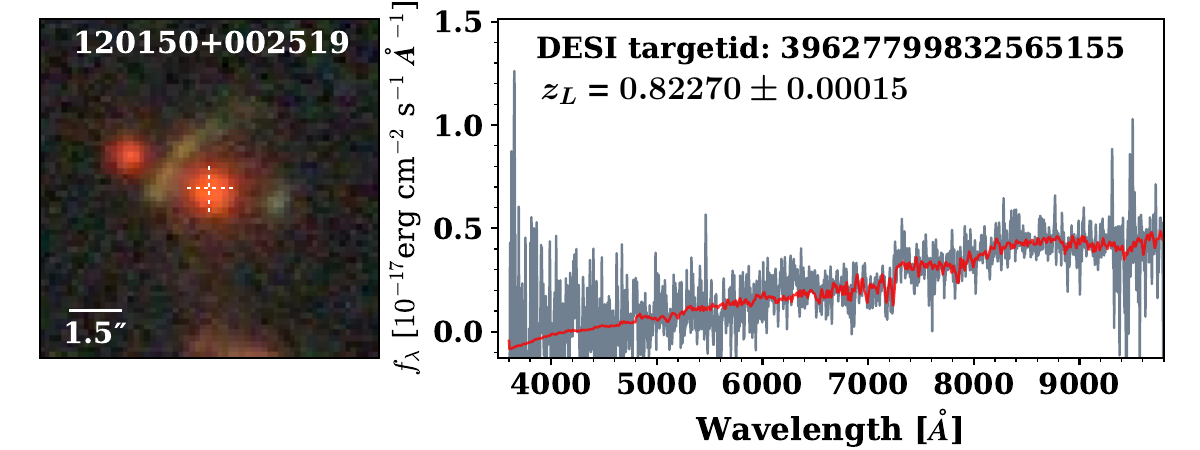}
\includegraphics[width=0.49\textwidth]
{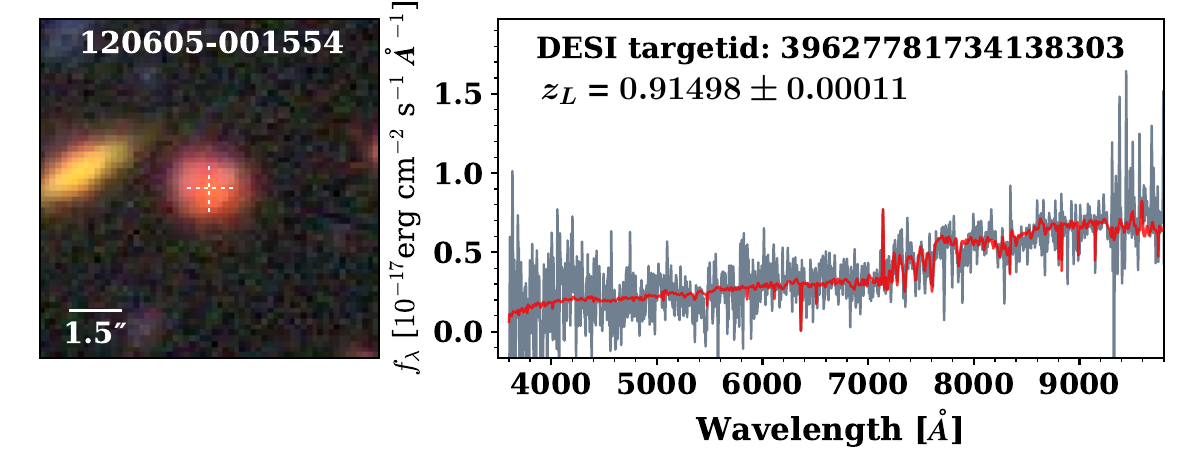}
\includegraphics[width=0.49\textwidth]
{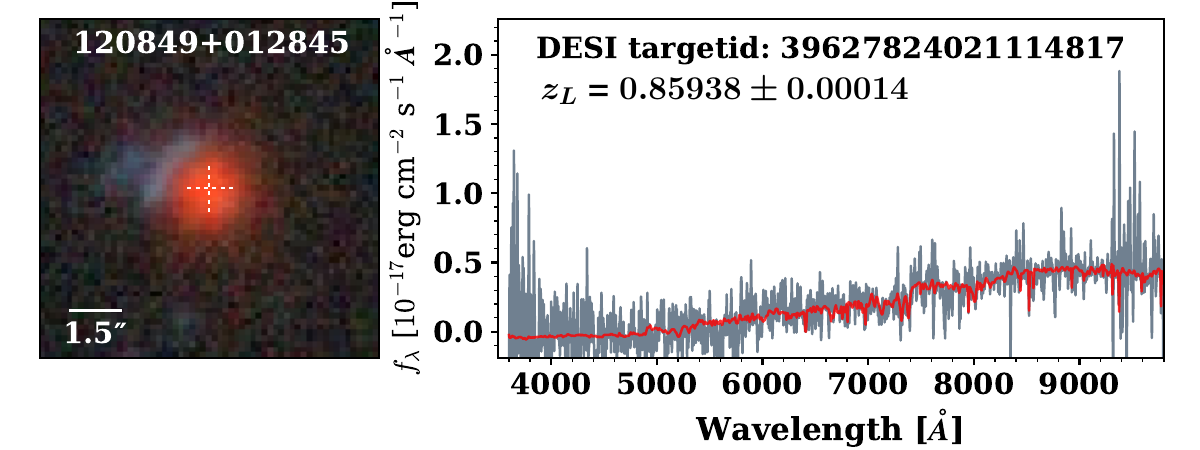}
\includegraphics[width=0.49\textwidth]
{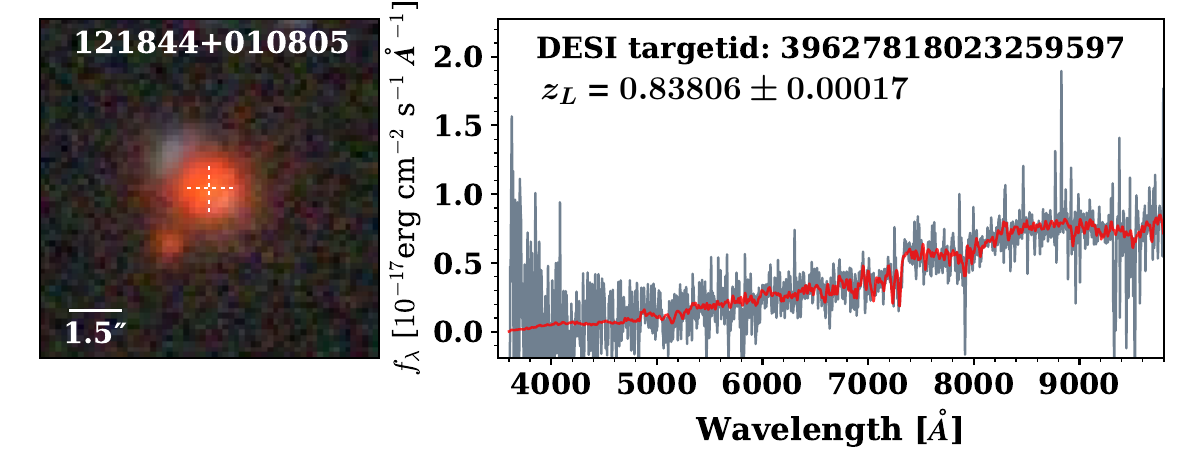}
\includegraphics[width=0.49\textwidth]
{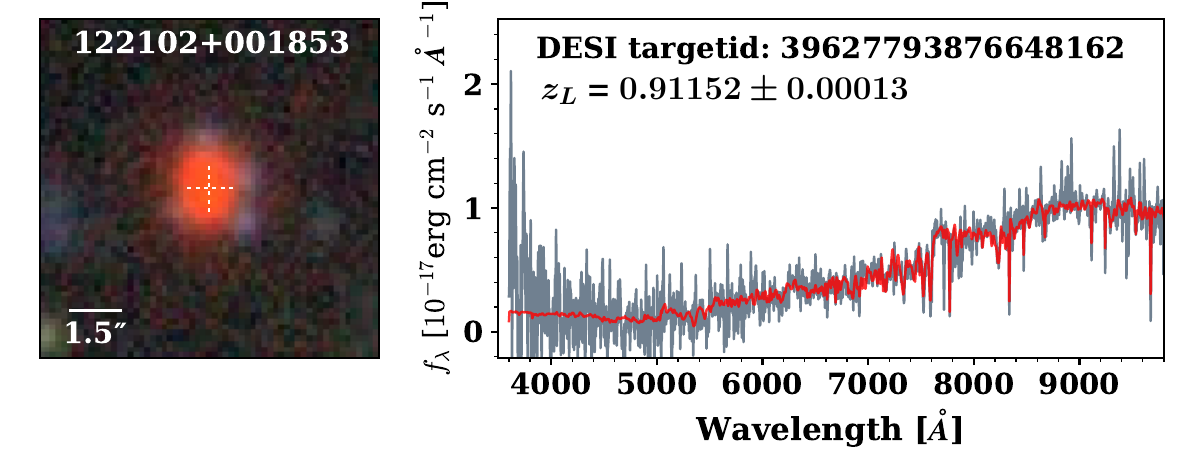}
\includegraphics[width=0.49\textwidth]
{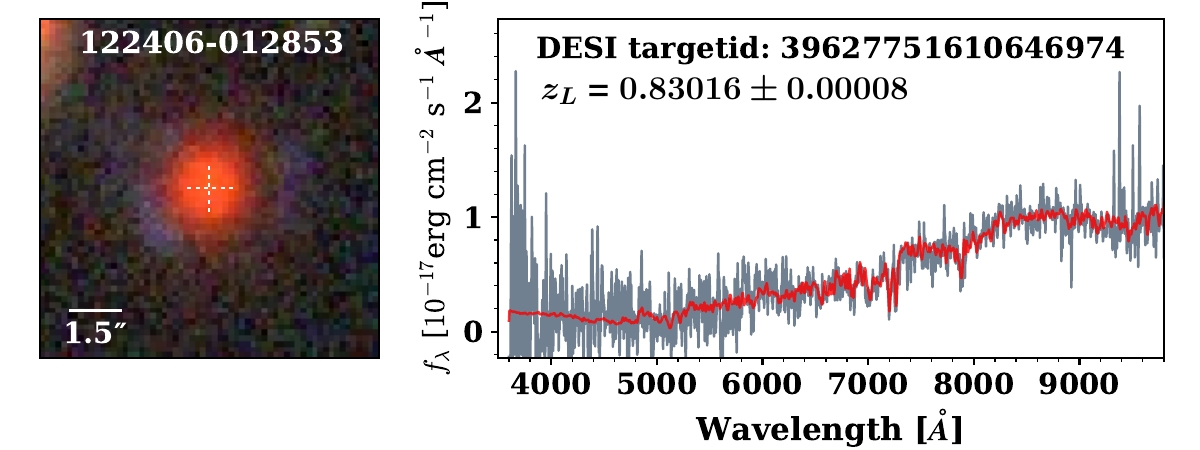}
\includegraphics[width=0.49\textwidth]
{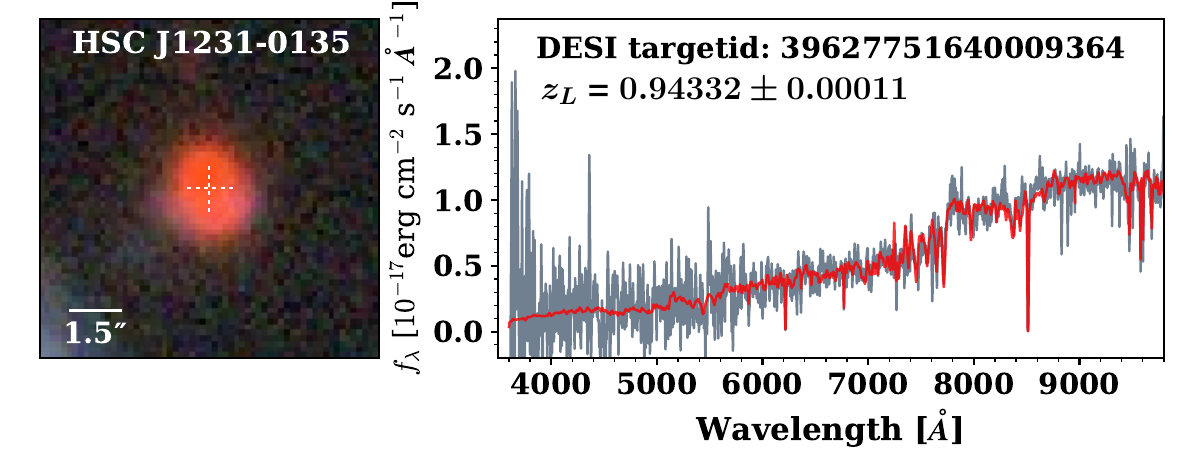}
\includegraphics[width=0.49\textwidth]
{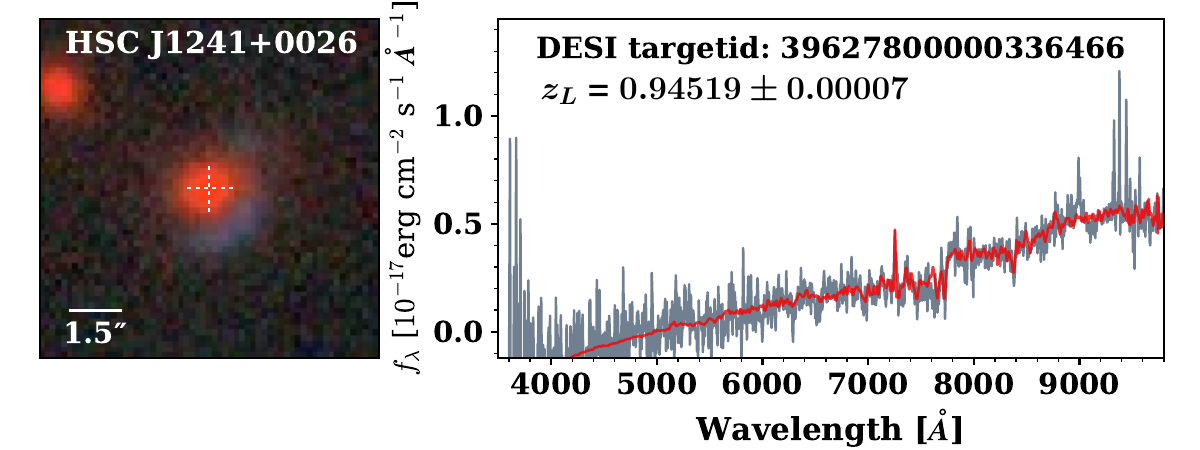}
\includegraphics[width=0.49\textwidth]
{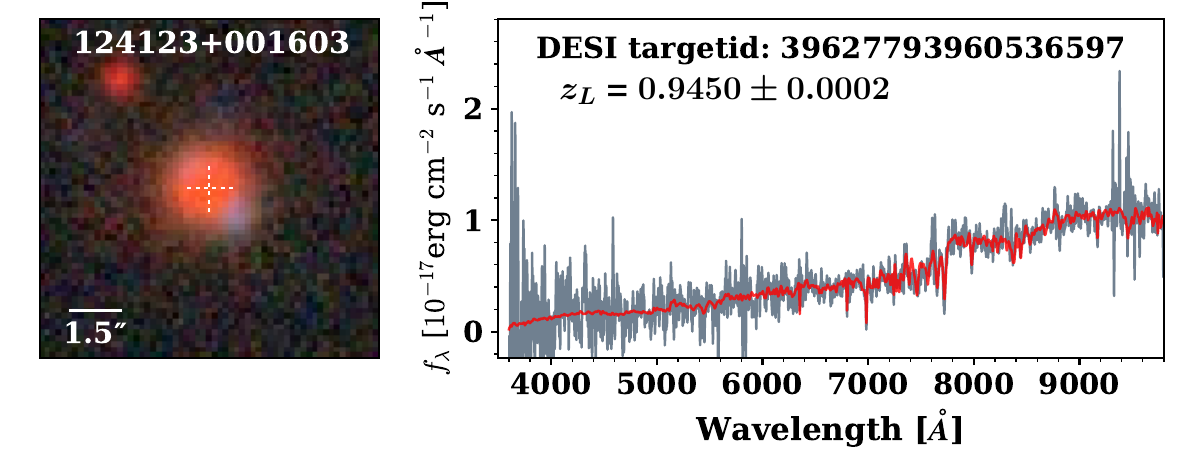}
\includegraphics[width=0.49\textwidth]
{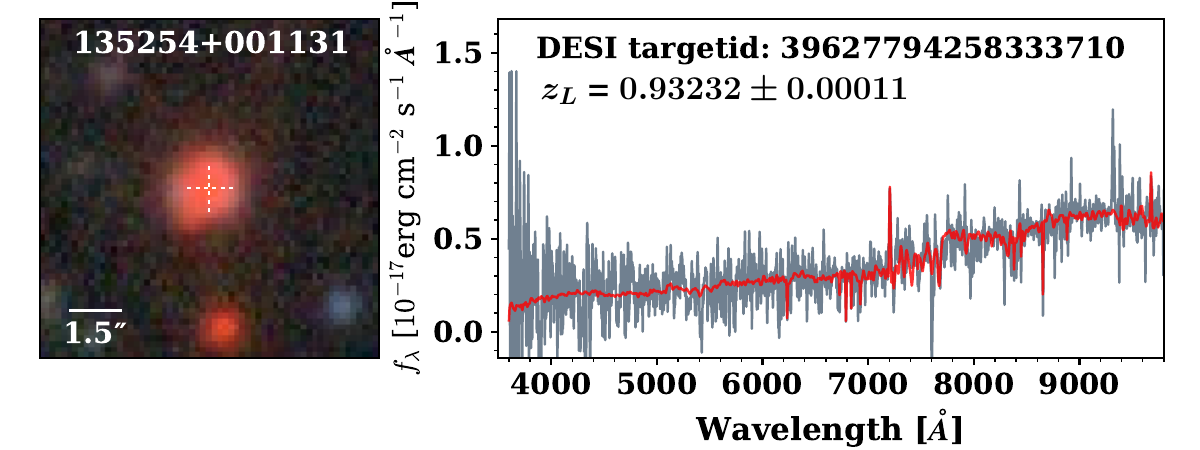}
\includegraphics[width=0.49\textwidth]
{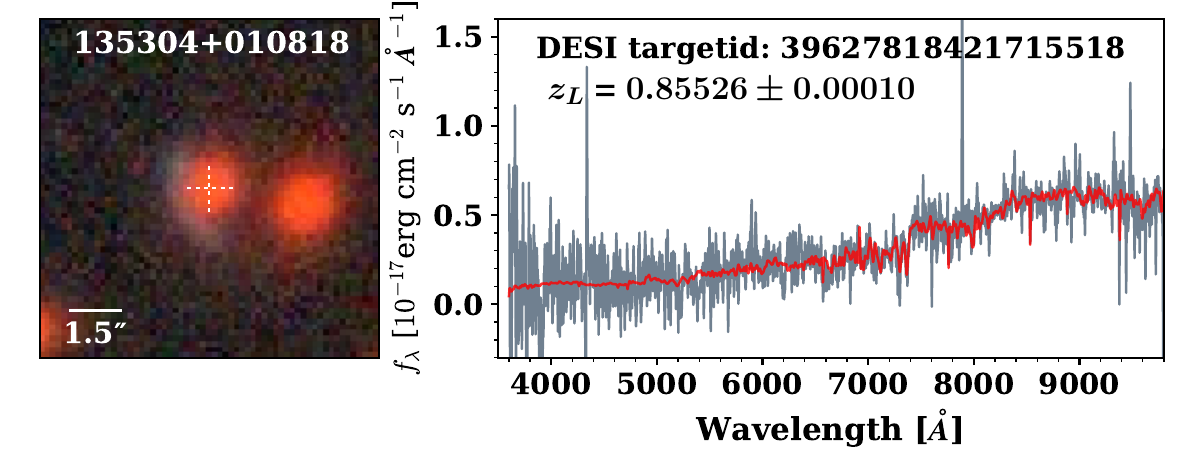}
\caption{\textit{Continued.}}
\end{figure*}

\begin{figure*}[htbp]
\ContinuedFloat
\centering
\includegraphics[width=0.49\textwidth]
{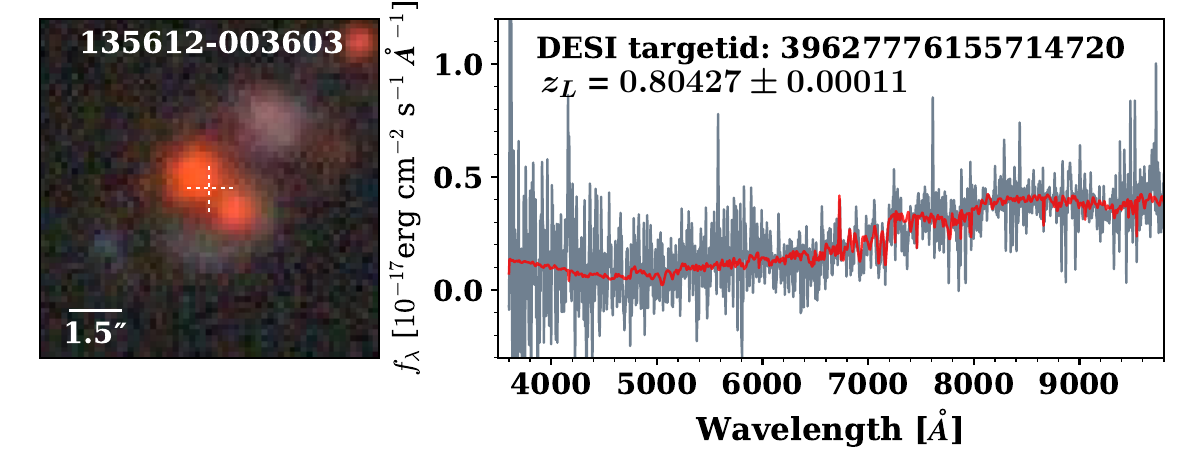}
\includegraphics[width=0.49\textwidth]
{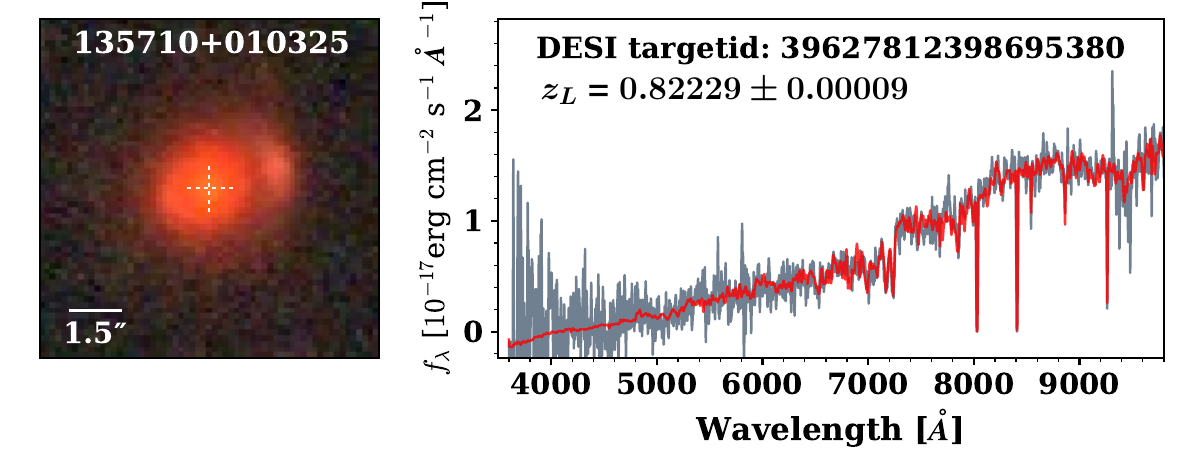}
\includegraphics[width=0.49\textwidth]
{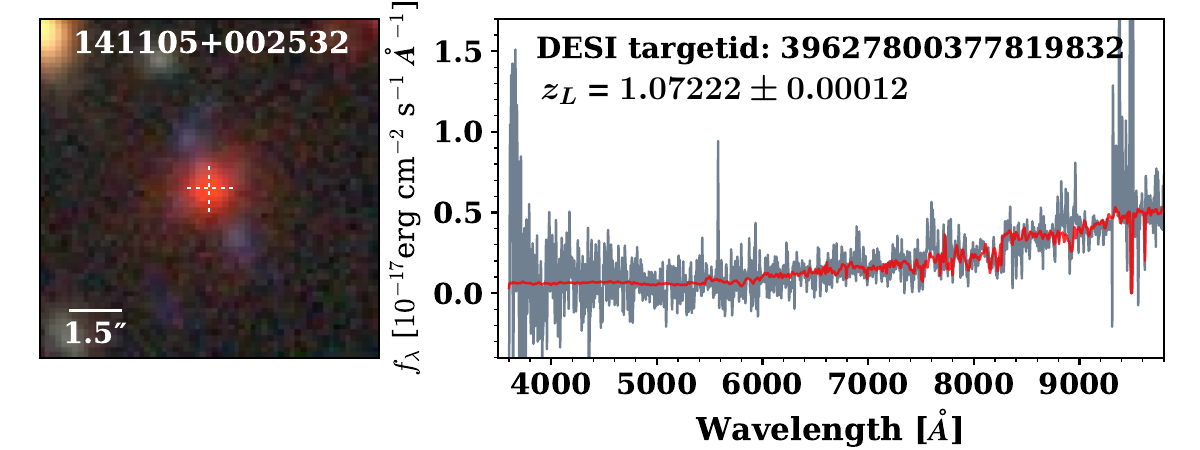}
\includegraphics[width=0.49\textwidth]
{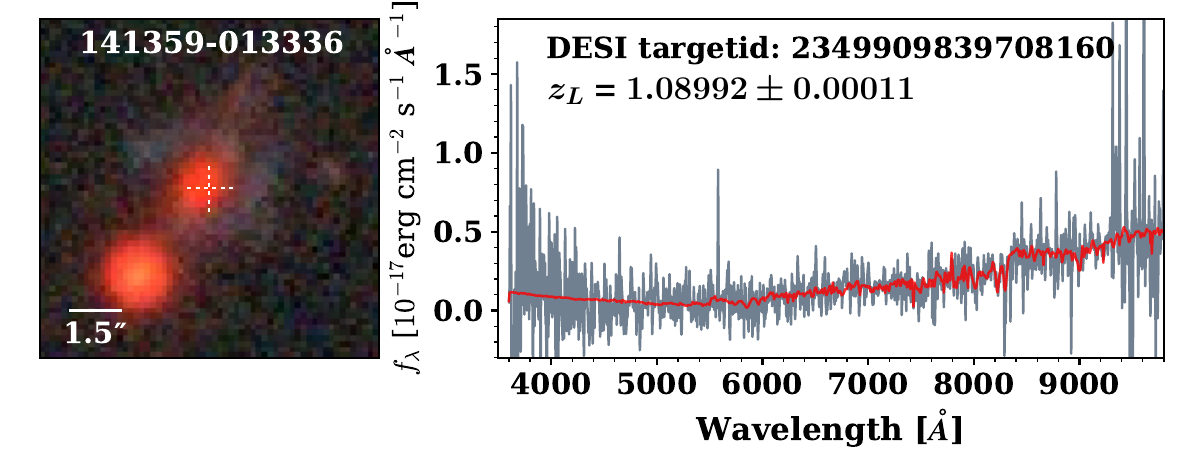}
\includegraphics[width=0.49\textwidth]
{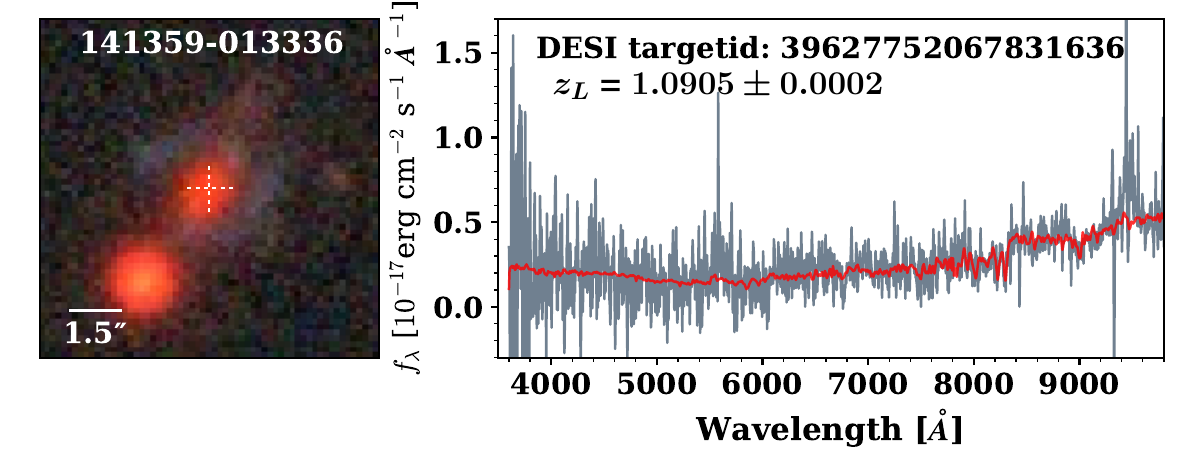}
\includegraphics[width=0.49\textwidth]
{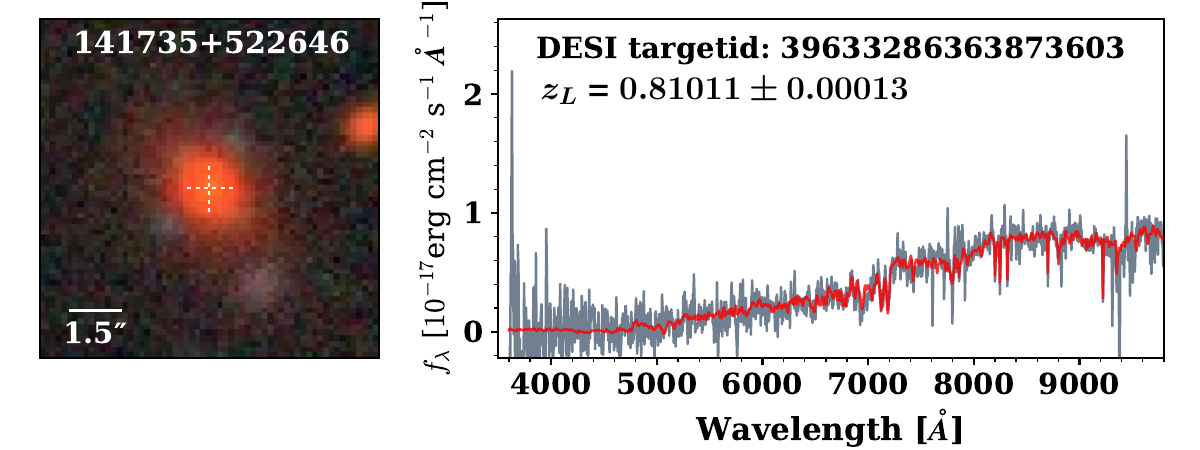}
\includegraphics[width=0.49\textwidth]
{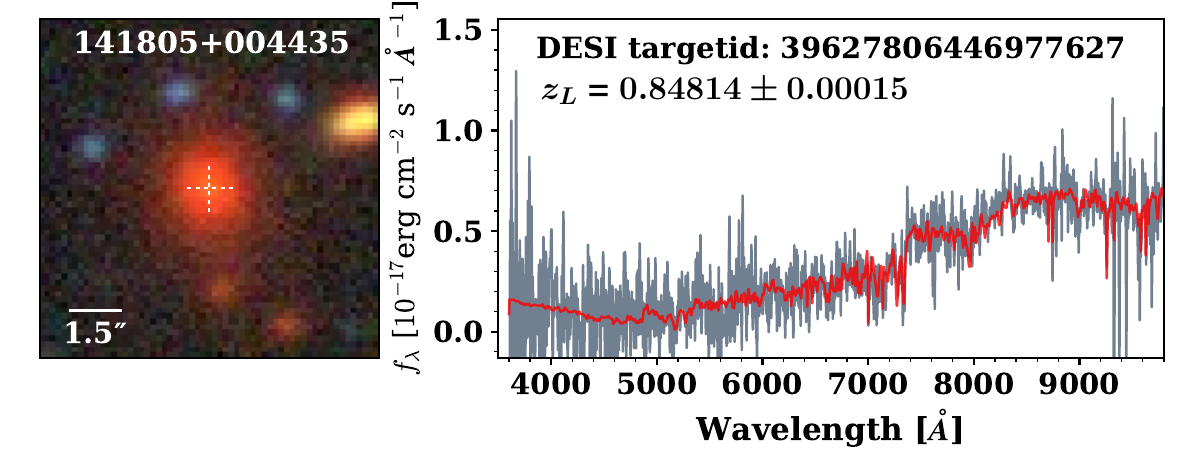}
\includegraphics[width=0.49\textwidth]
{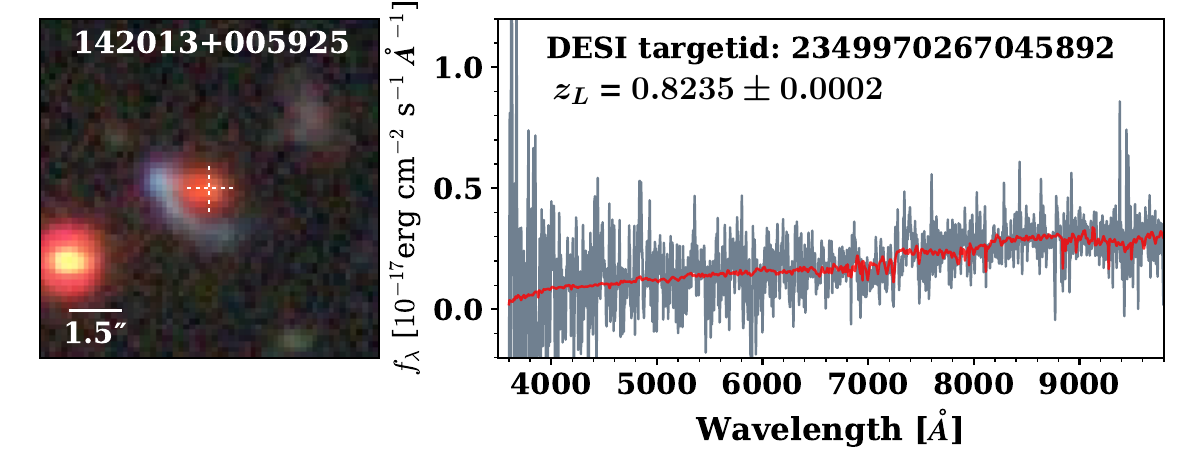}
\includegraphics[width=0.49\textwidth]
{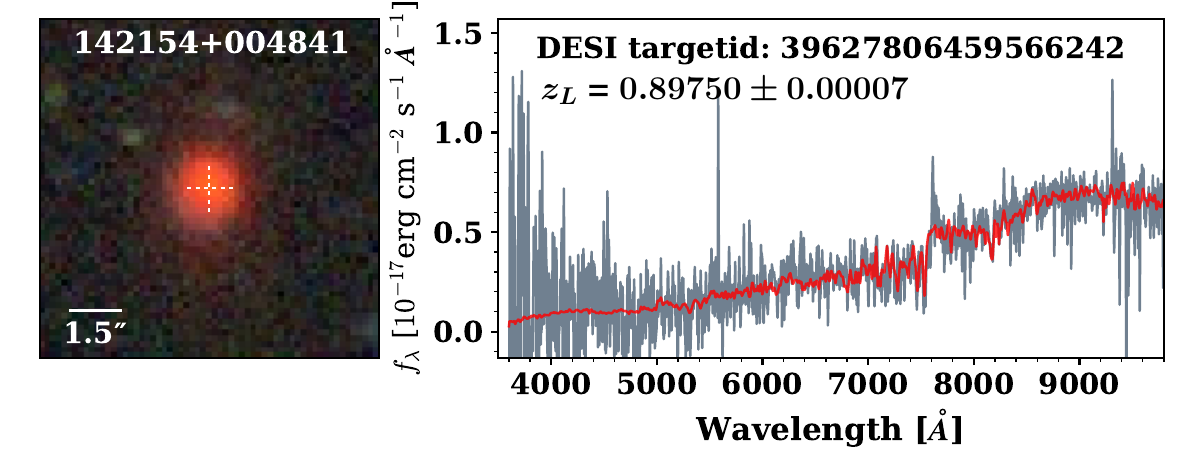}
\includegraphics[width=0.49\textwidth]
{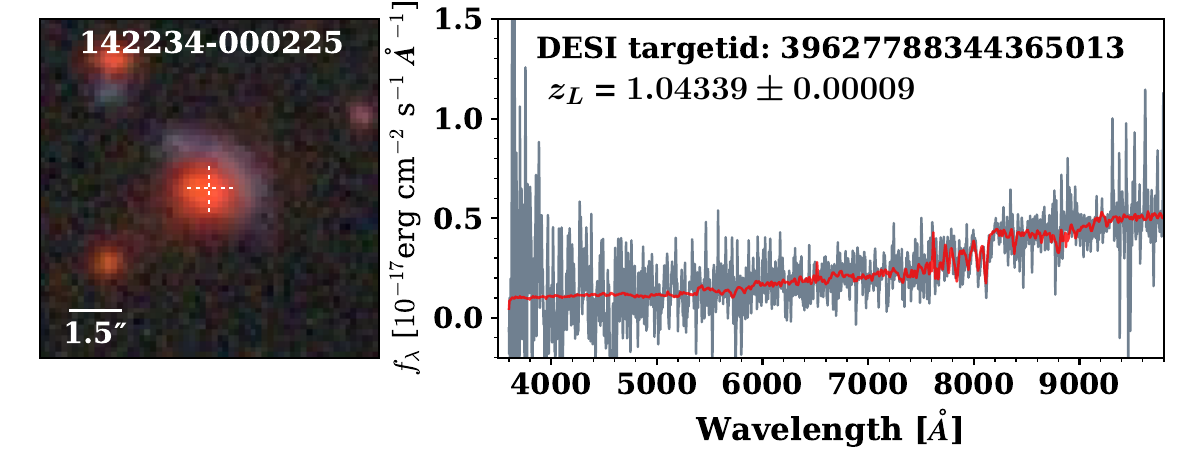}
\includegraphics[width=0.49\textwidth]
{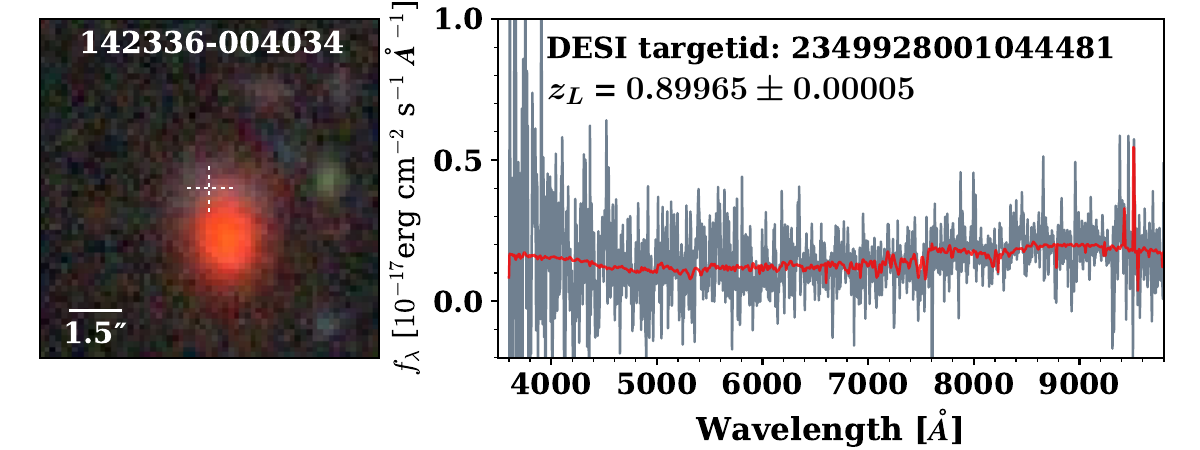}
\includegraphics[width=0.49\textwidth]
{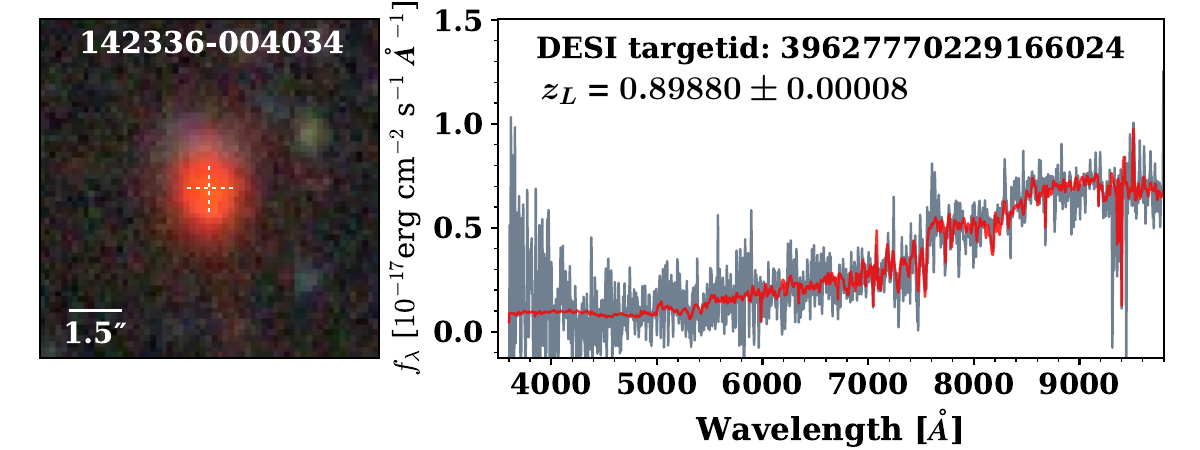}
\includegraphics[width=0.49\textwidth]
{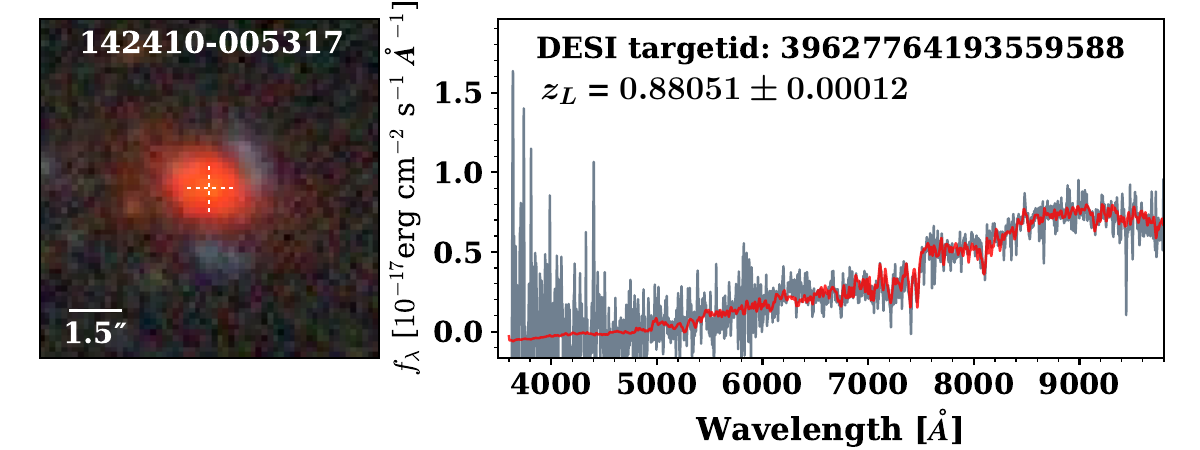}
\includegraphics[width=0.49\textwidth]
{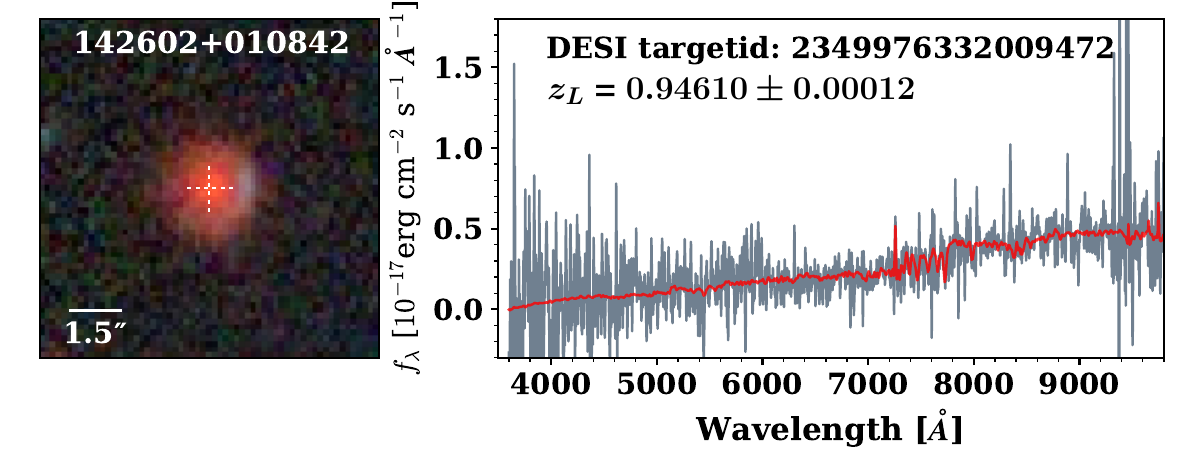}
\caption{\textit{Continued.}}
\end{figure*}

\begin{figure*}[htbp]
\ContinuedFloat
\centering
\includegraphics[width=0.49\textwidth]
{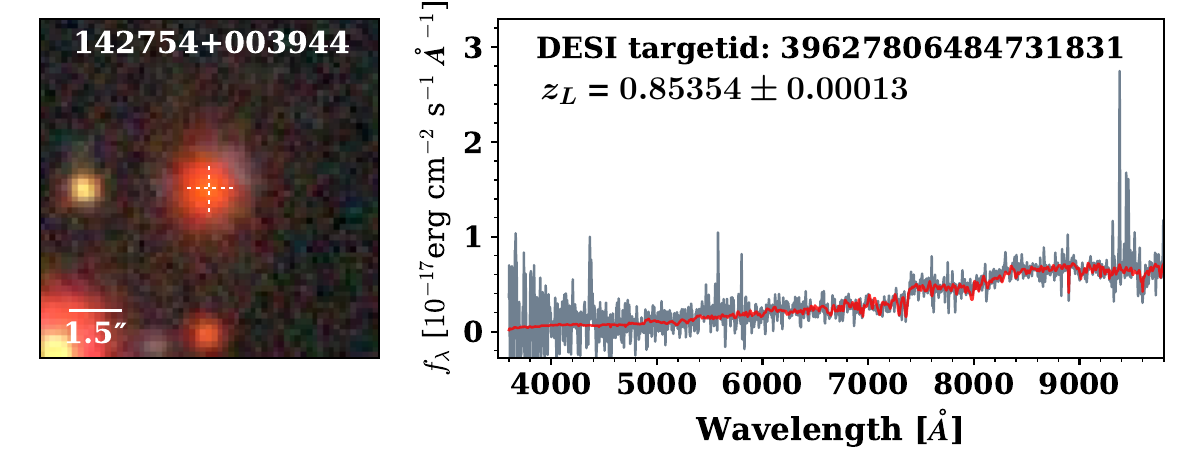}
\includegraphics[width=0.49\textwidth]
{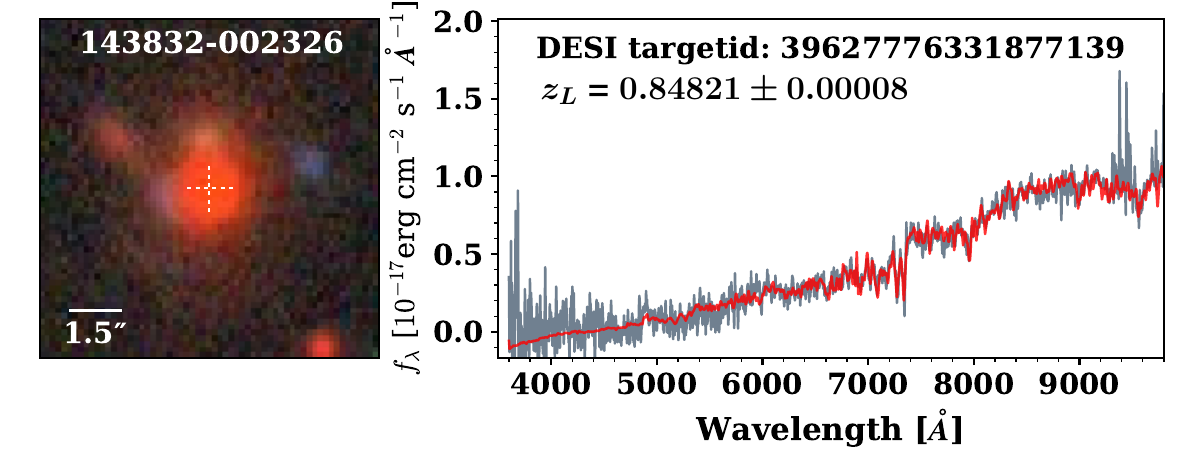}
\includegraphics[width=0.49\textwidth]
{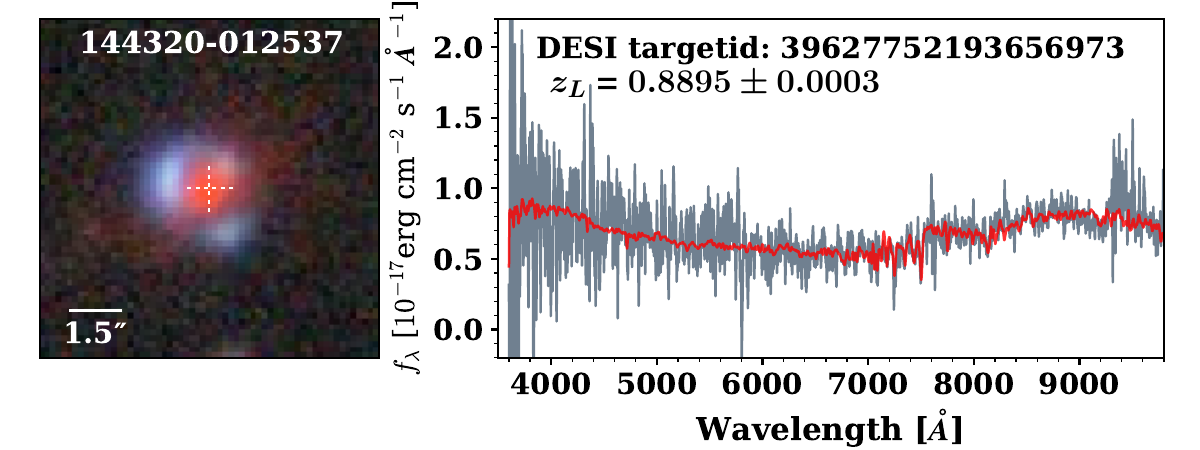}
\includegraphics[width=0.49\textwidth]
{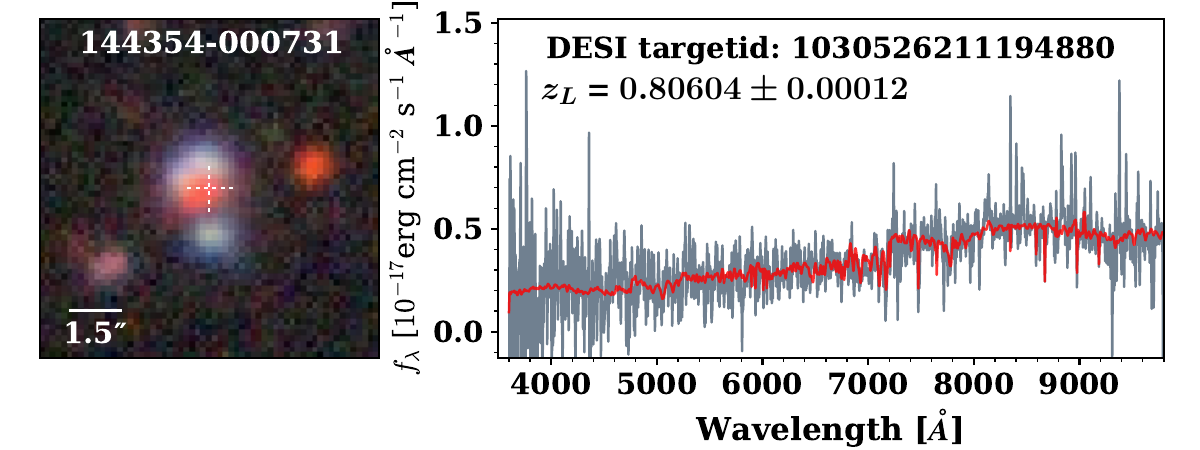}
\includegraphics[width=0.49\textwidth]
{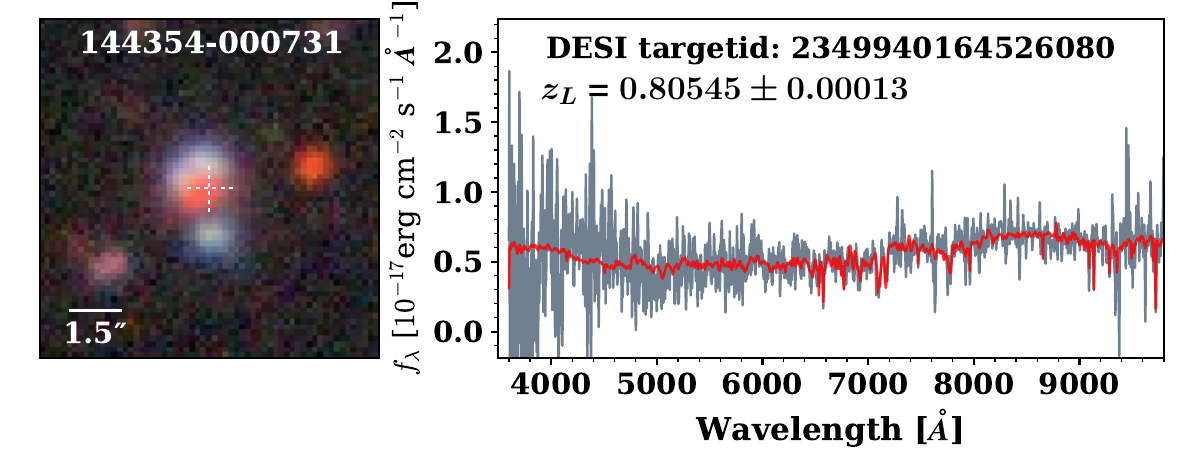}
\includegraphics[width=0.49\textwidth]
{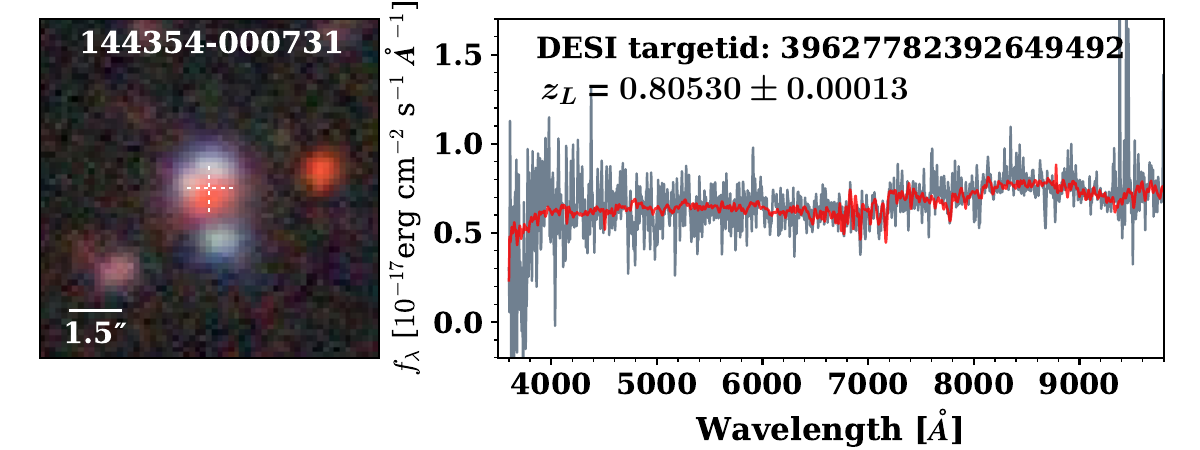}
\includegraphics[width=0.49\textwidth]
{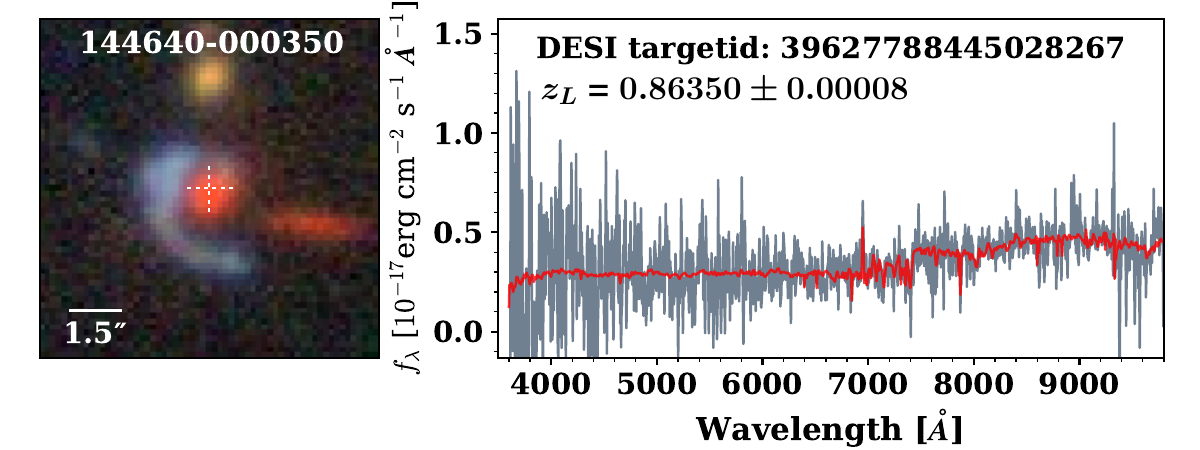}
\includegraphics[width=0.49\textwidth]
{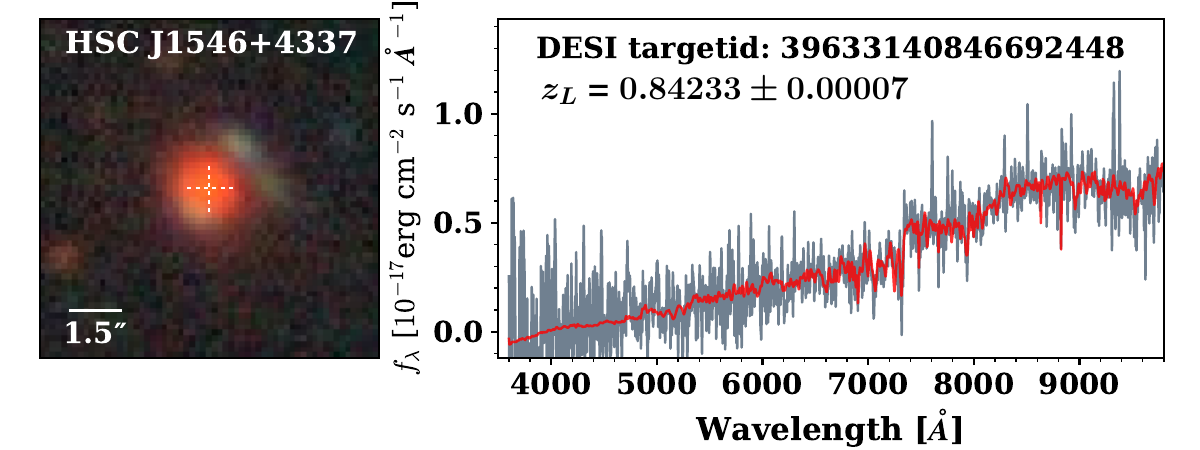}
\includegraphics[width=0.49\textwidth]
{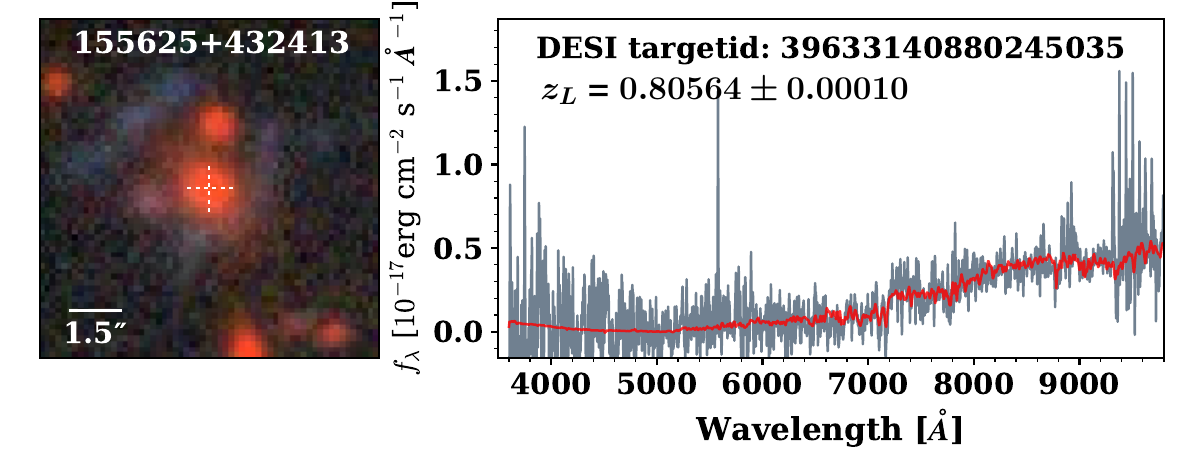}
\includegraphics[width=0.49\textwidth]
{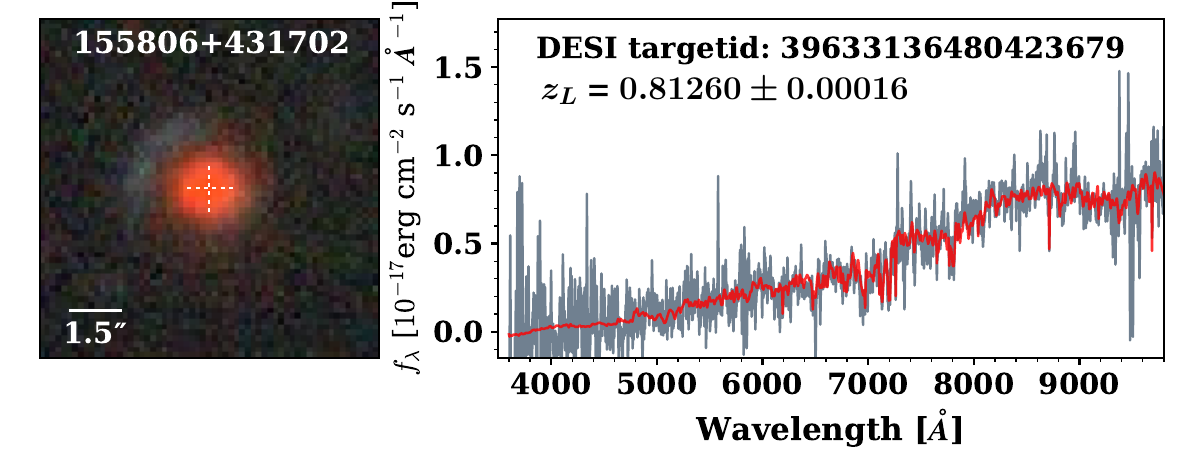}
\includegraphics[width=0.49\textwidth]
{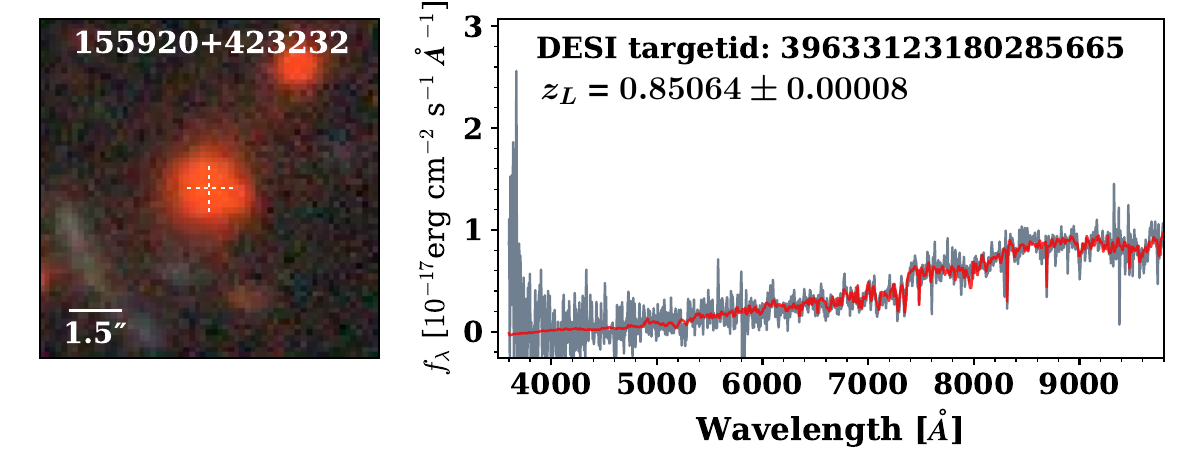}
\includegraphics[width=0.49\textwidth]
{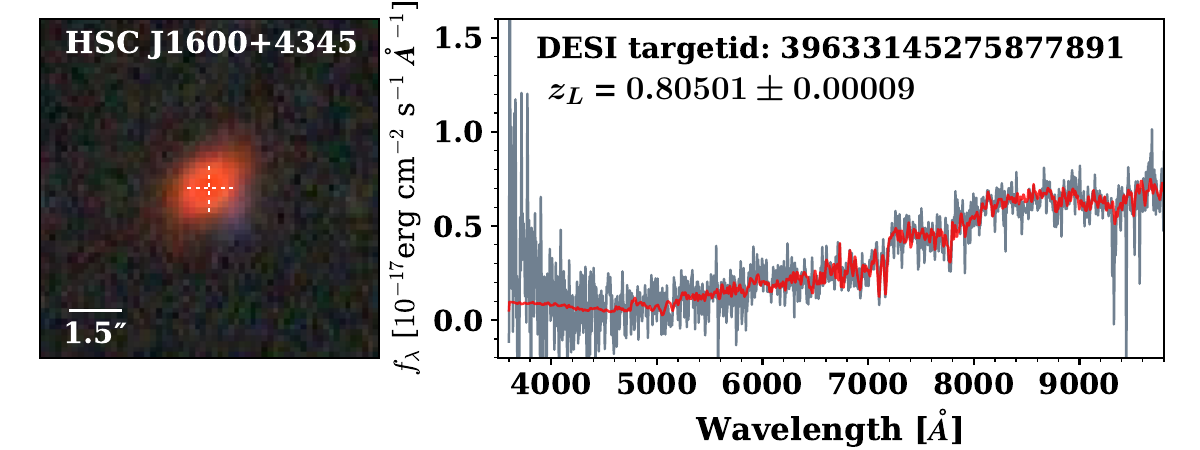}
\includegraphics[width=0.49\textwidth]
{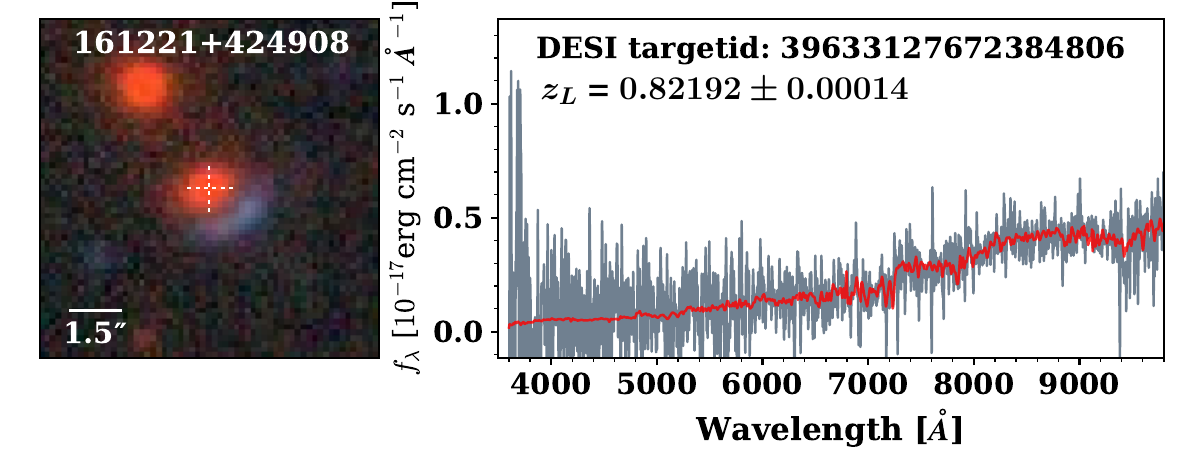}
\includegraphics[width=0.49\textwidth]
{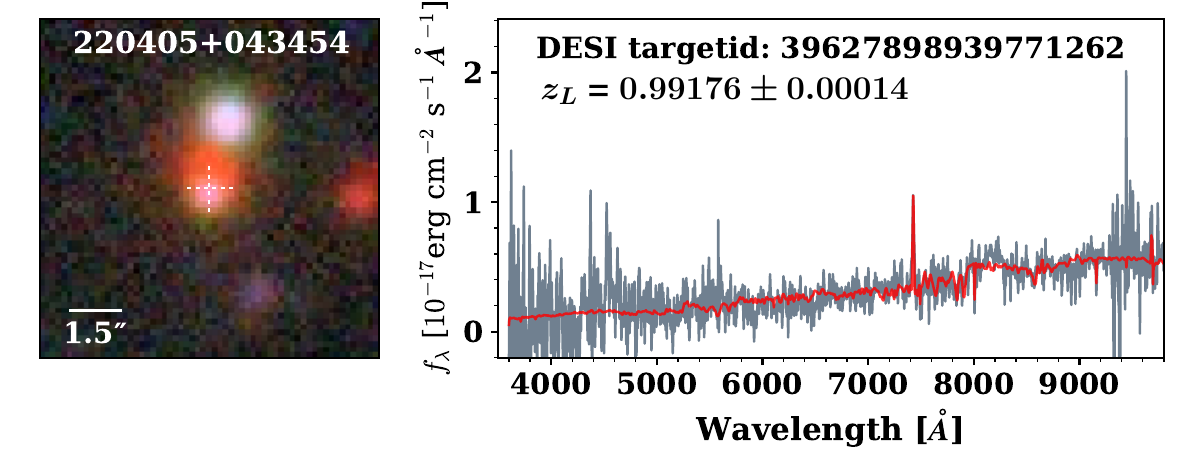}
\caption{\textit{Continued.}}
\end{figure*}

\begin{figure*}[htbp]
\ContinuedFloat
\centering
\includegraphics[width=0.49\textwidth]
{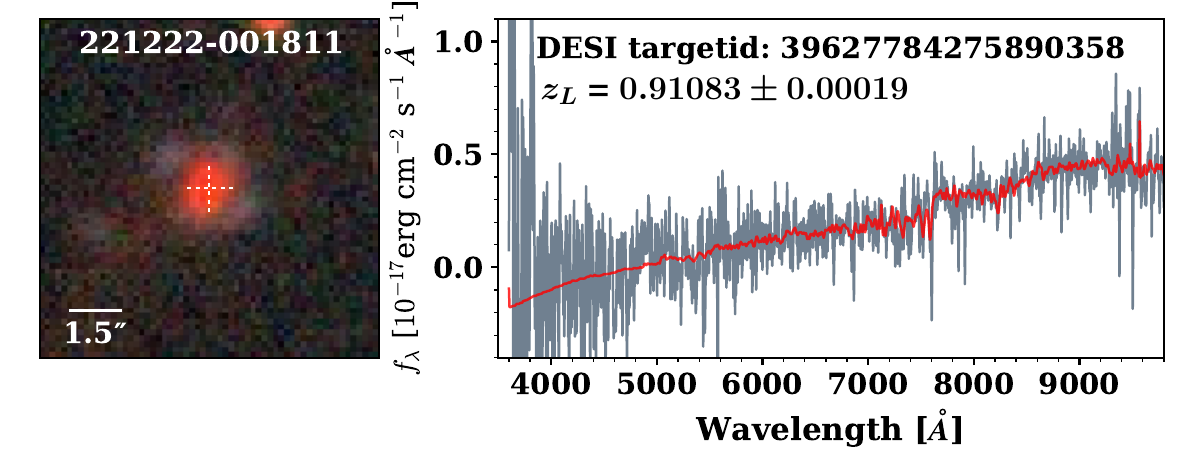}
\includegraphics[width=0.49\textwidth]
{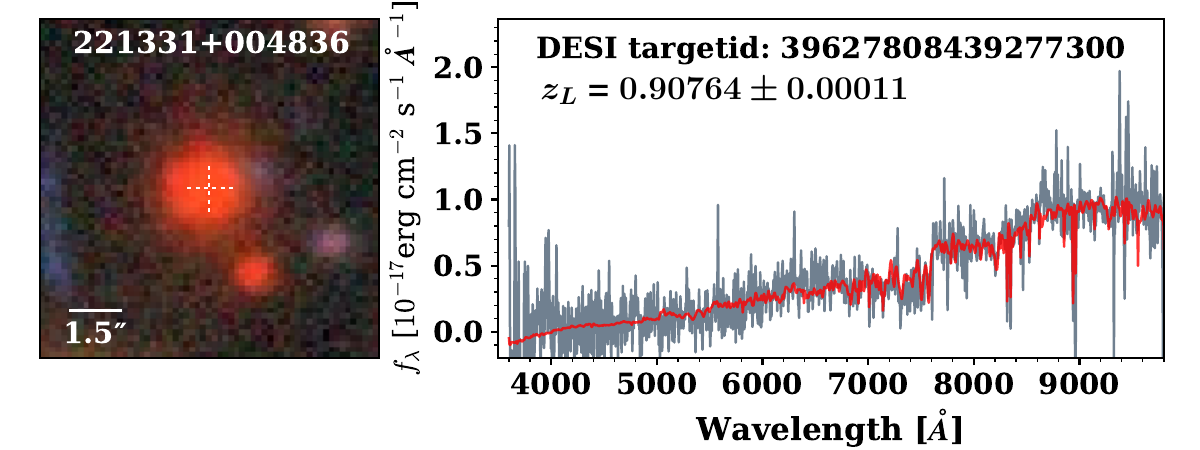}
\includegraphics[width=0.49\textwidth]
{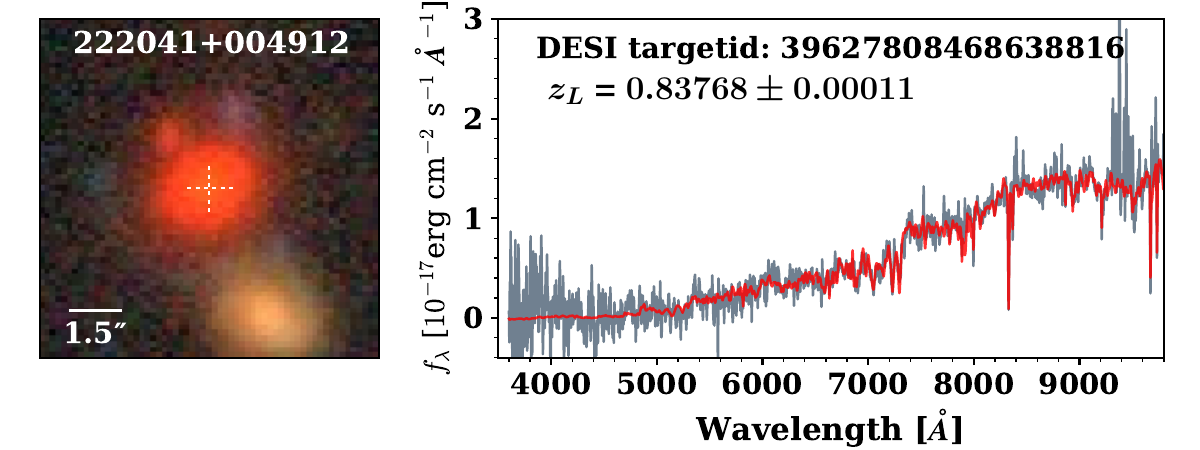}
\includegraphics[width=0.49\textwidth]
{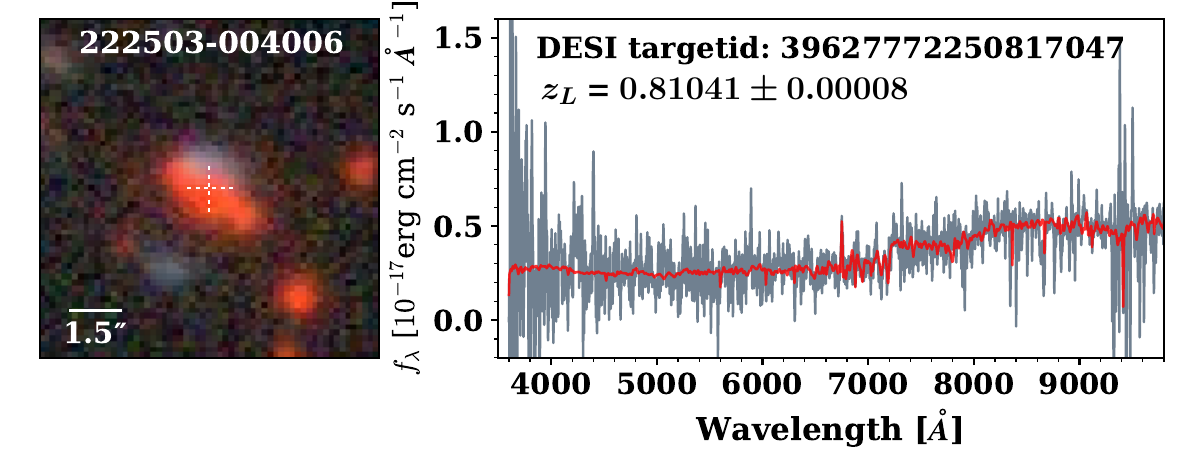}
\includegraphics[width=0.49\textwidth]
{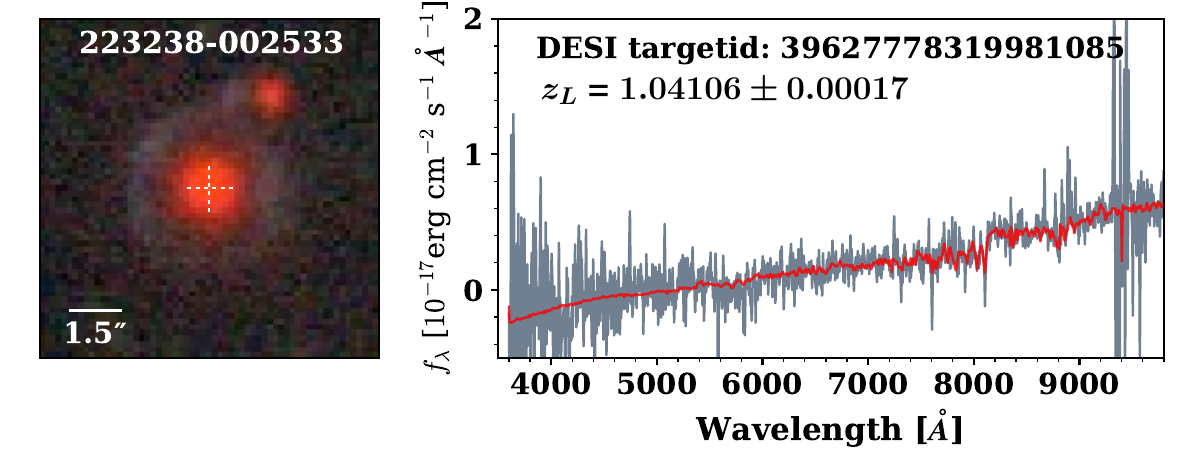}
\includegraphics[width=0.49\textwidth]
{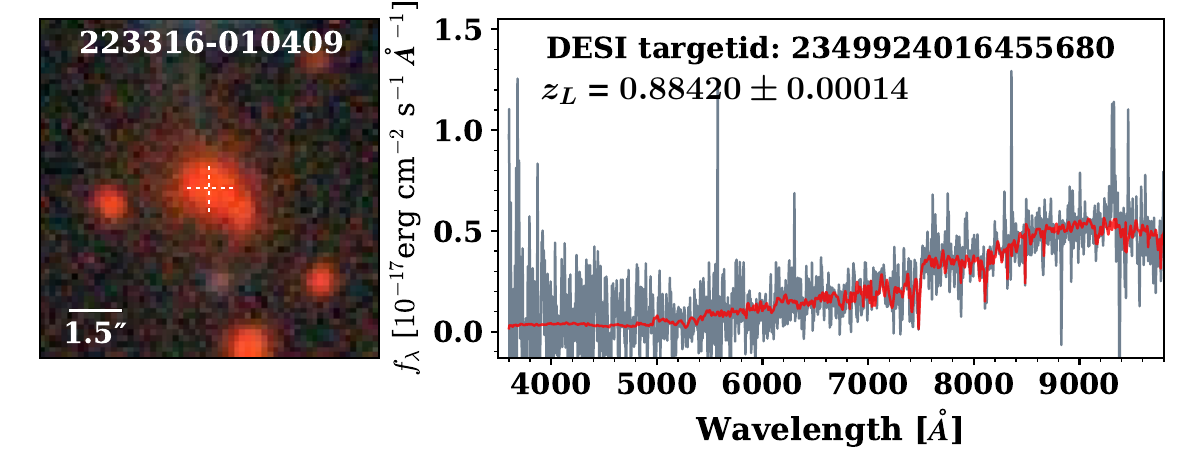}
\includegraphics[width=0.49\textwidth]
{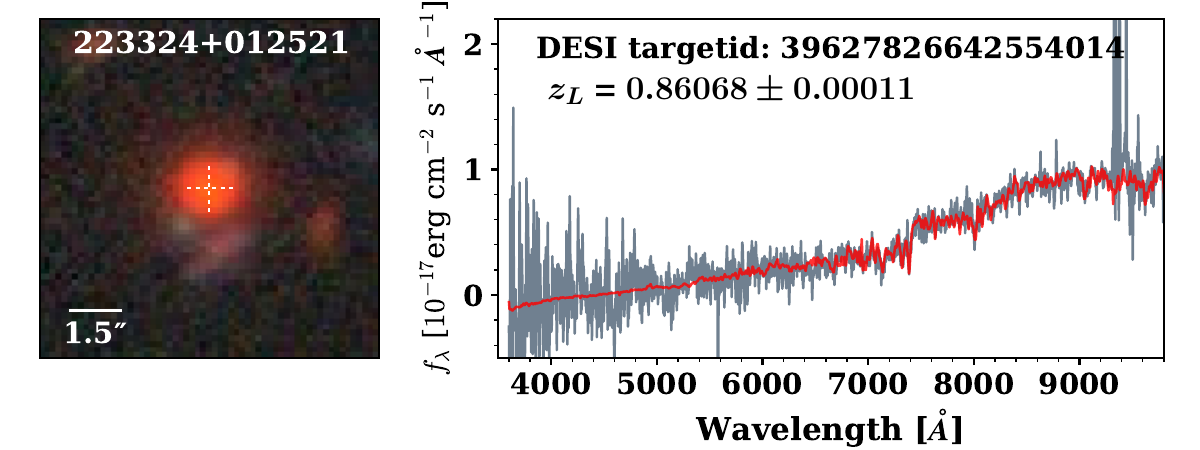}
\includegraphics[width=0.49\textwidth]
{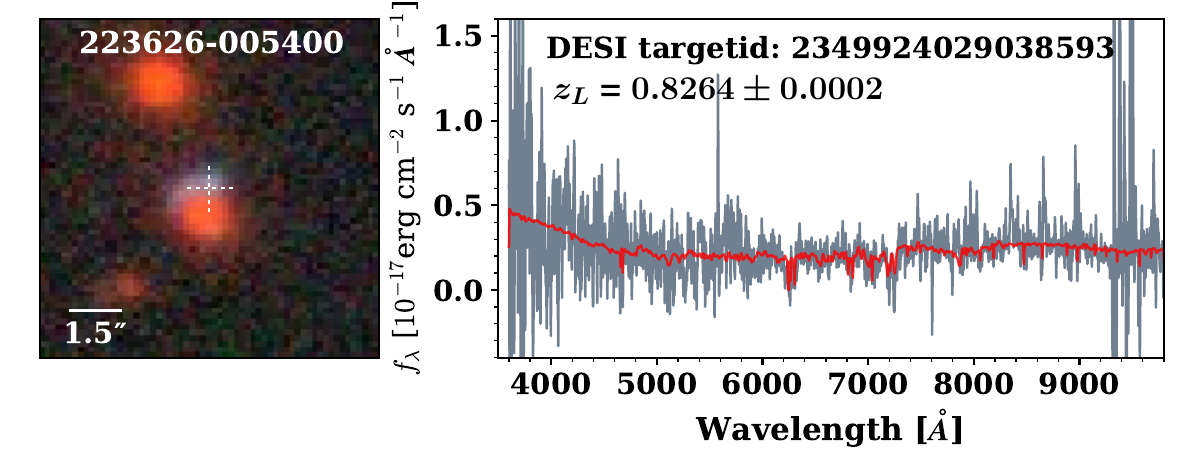}
\includegraphics[width=0.49\textwidth]
{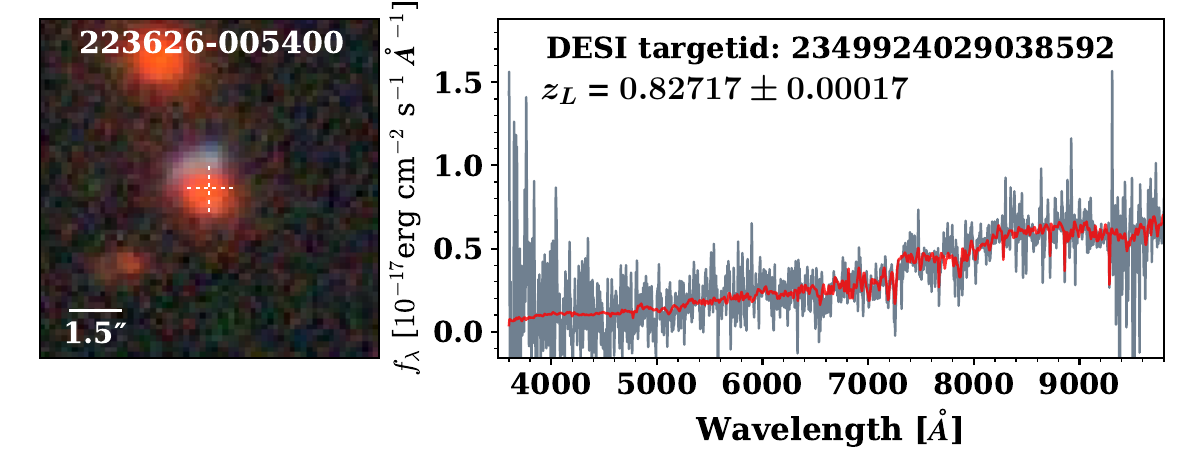}
\includegraphics[width=0.49\textwidth]
{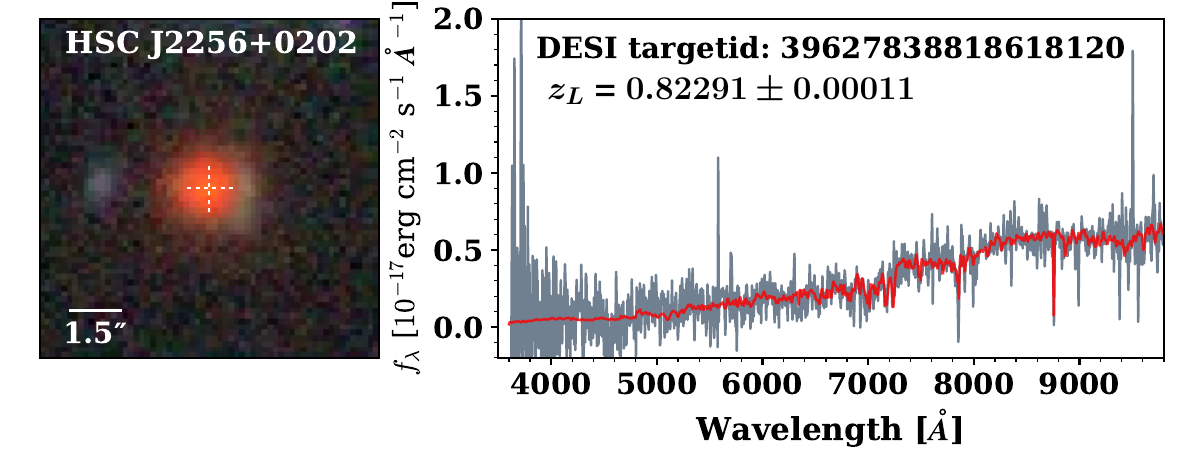}
\includegraphics[width=0.49\textwidth]
{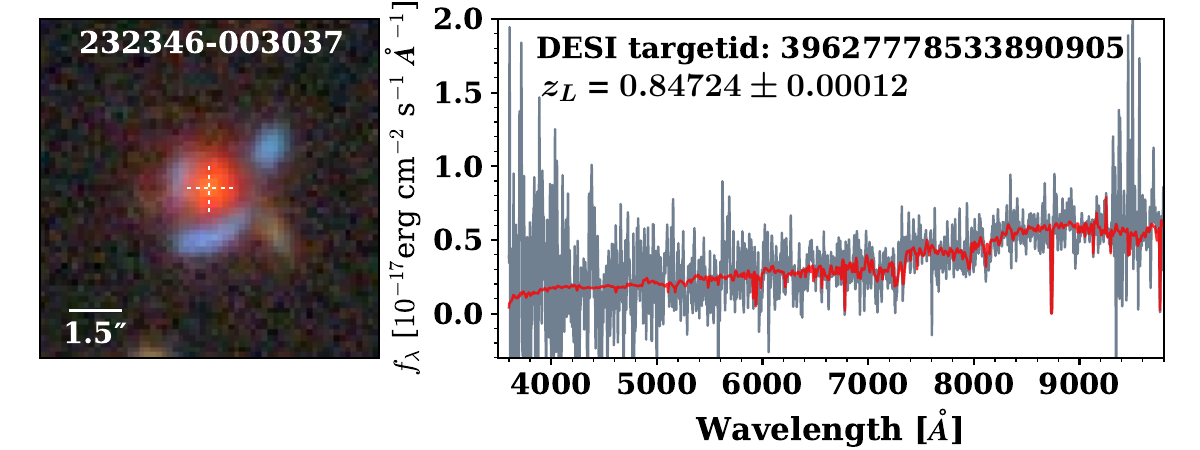}
\includegraphics[width=0.49\textwidth]
{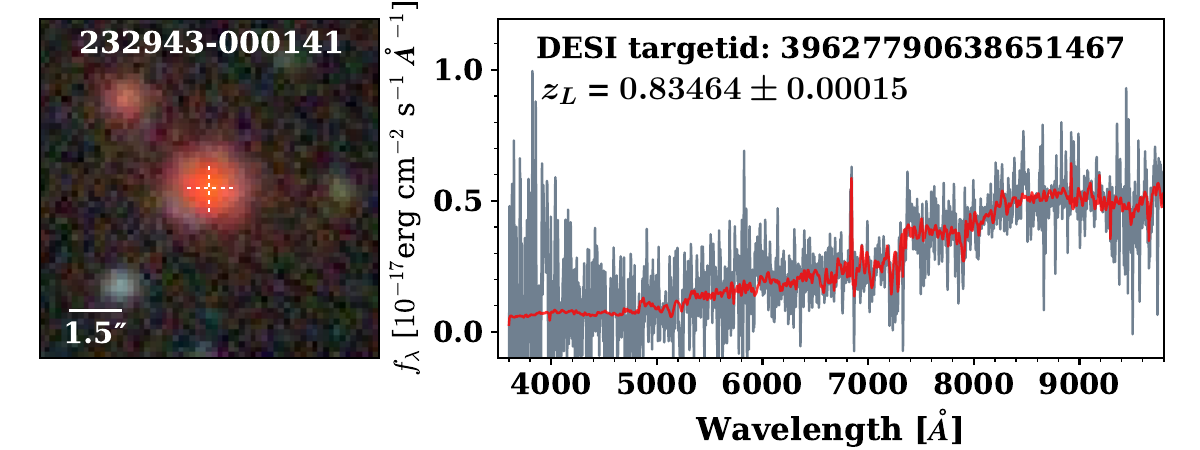}
\hspace*{0.49\textwidth}
\caption{\textit{Continued.}}
\end{figure*}

\begin{figure*}[htbp]
\centering
\includegraphics[width=0.49\textwidth]{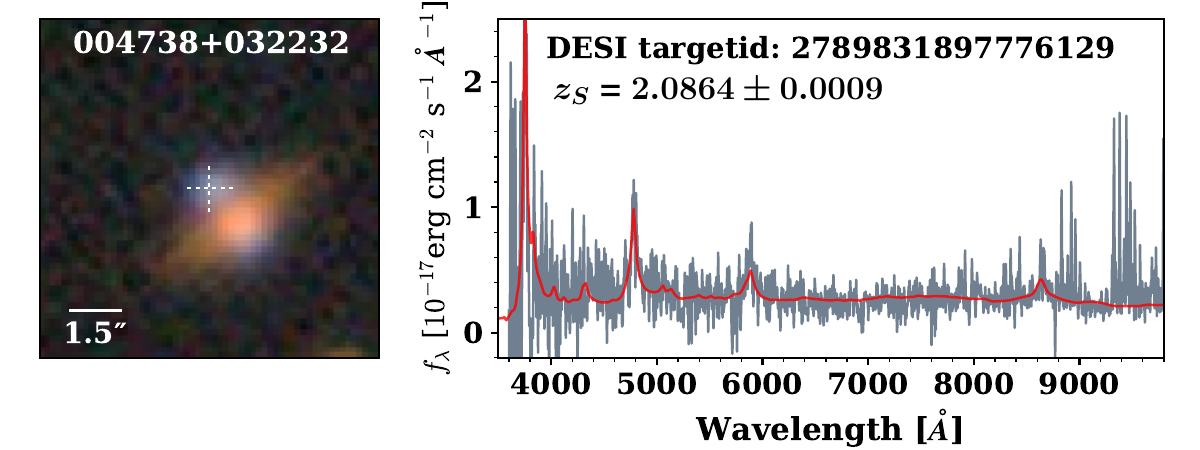}
\includegraphics[width=0.49\textwidth]{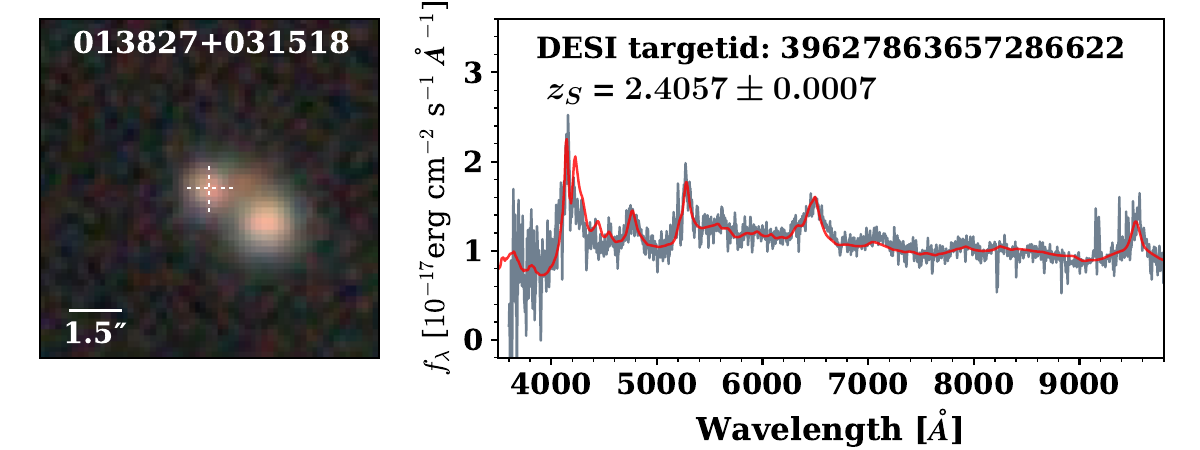}
\includegraphics[width=0.49\textwidth]{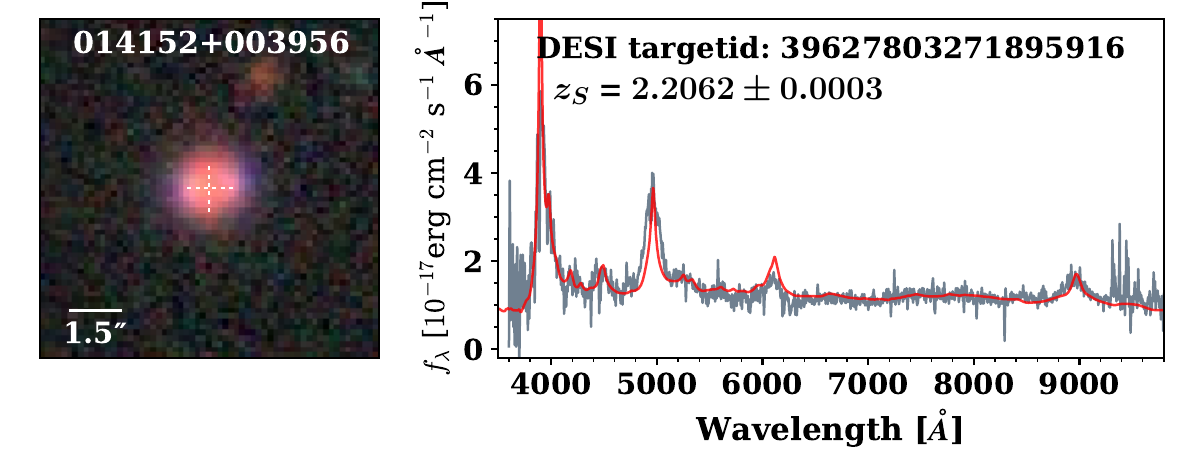}
\includegraphics[width=0.49\textwidth]{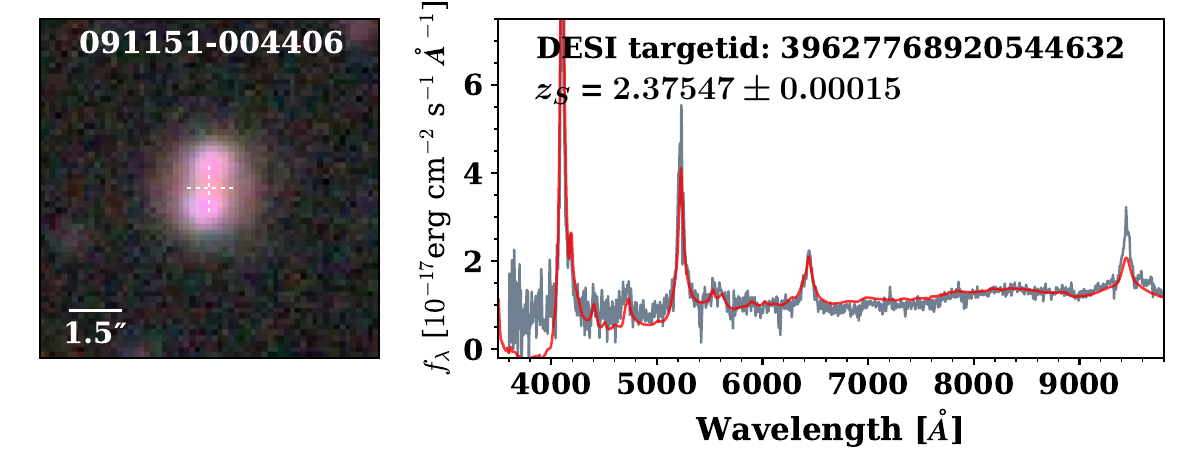}
\includegraphics[width=0.49\textwidth]{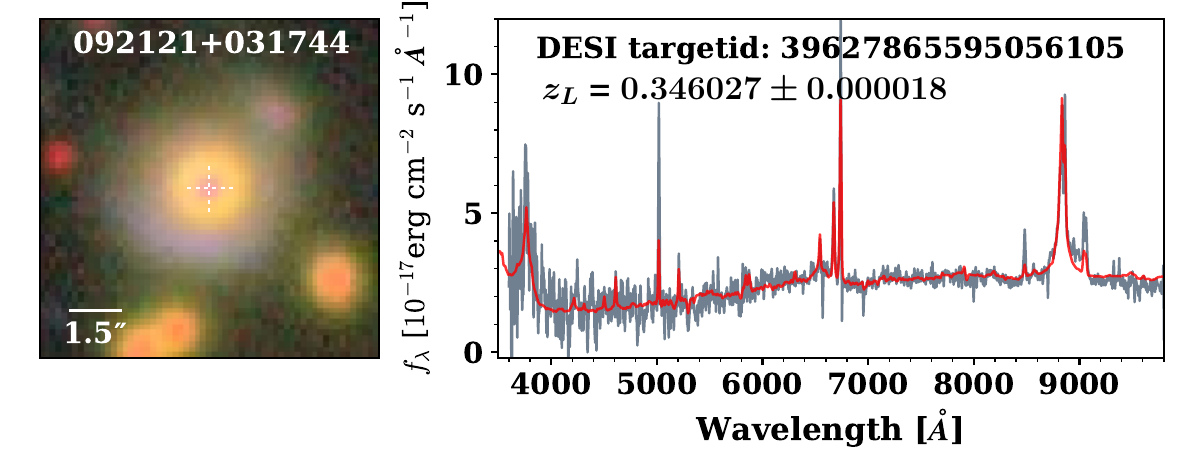}
\includegraphics[width=0.49\textwidth]{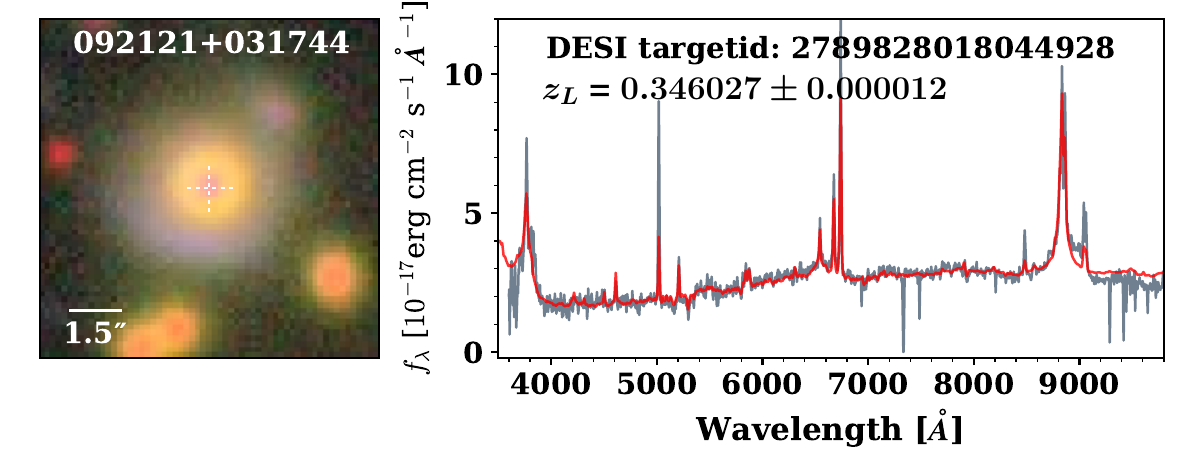}
\includegraphics[width=0.49\textwidth]{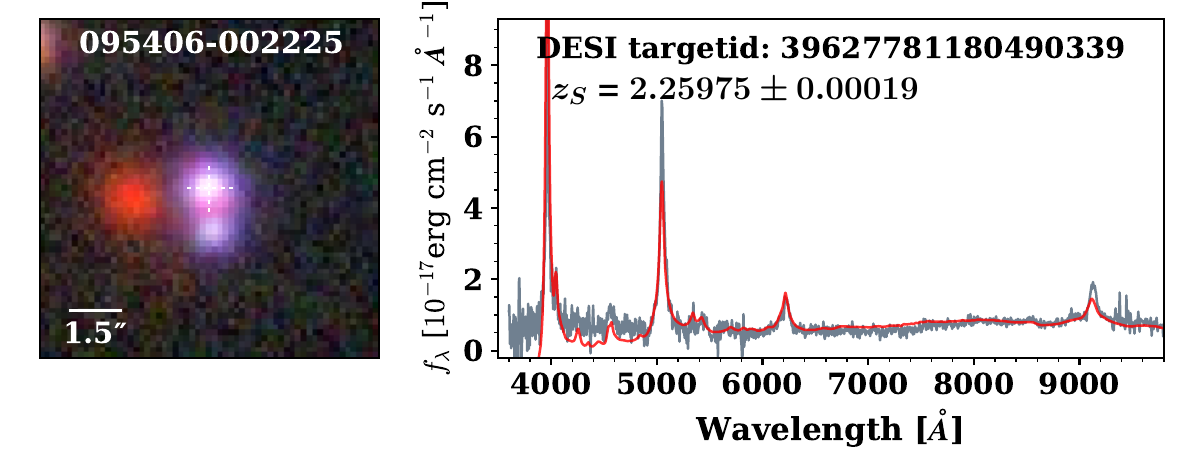}
\includegraphics[width=0.49\textwidth]{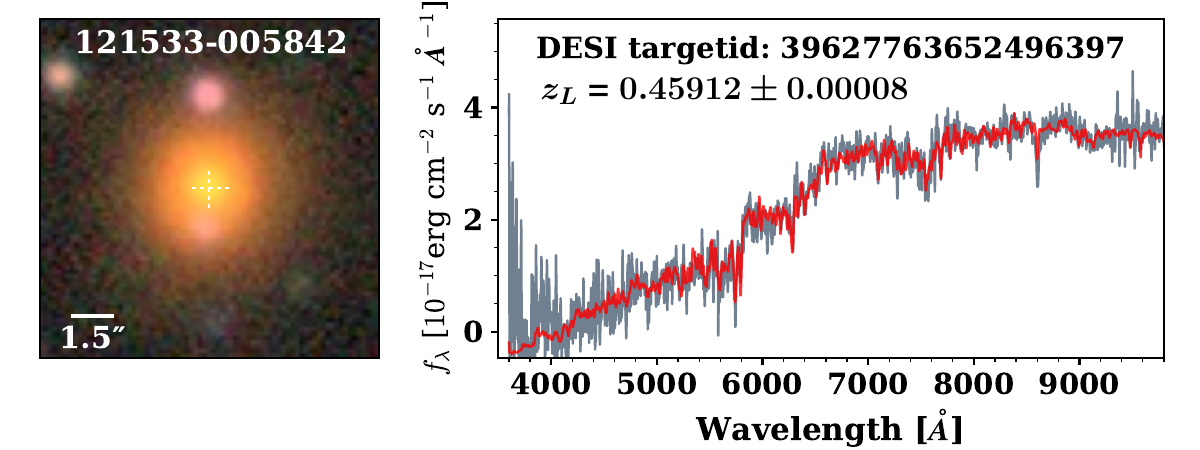}
\includegraphics[width=0.49\textwidth]{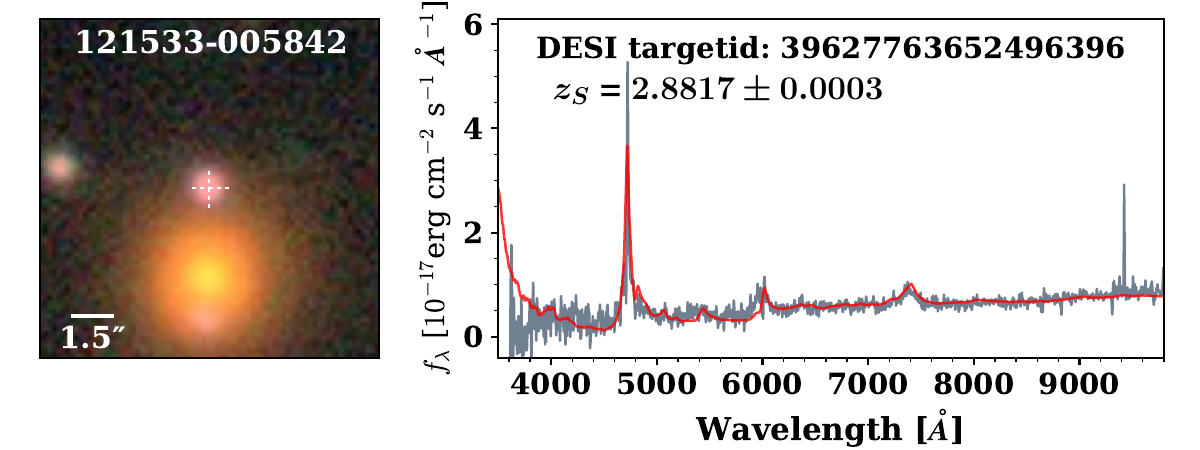}
\includegraphics[width=0.49\textwidth]{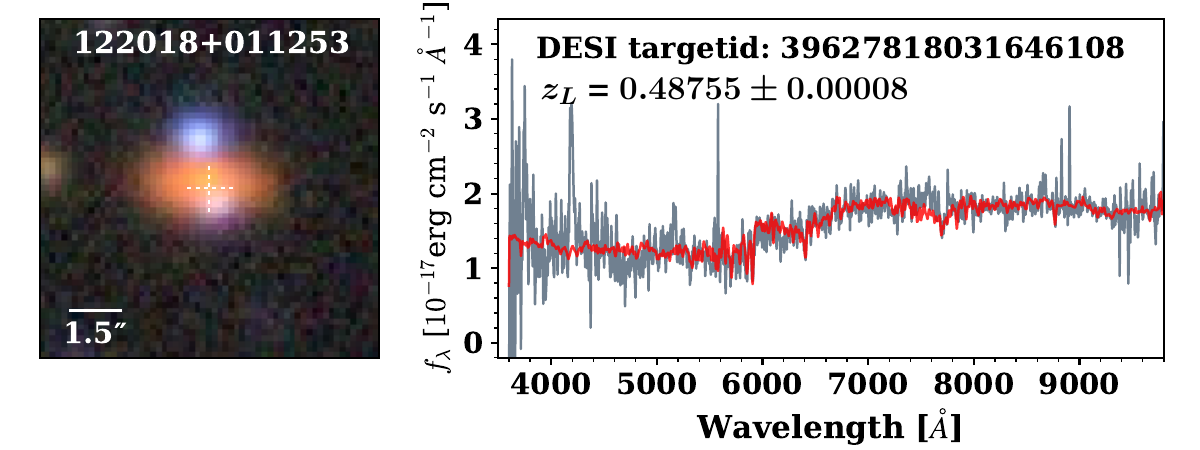}
\includegraphics[width=0.49\textwidth]{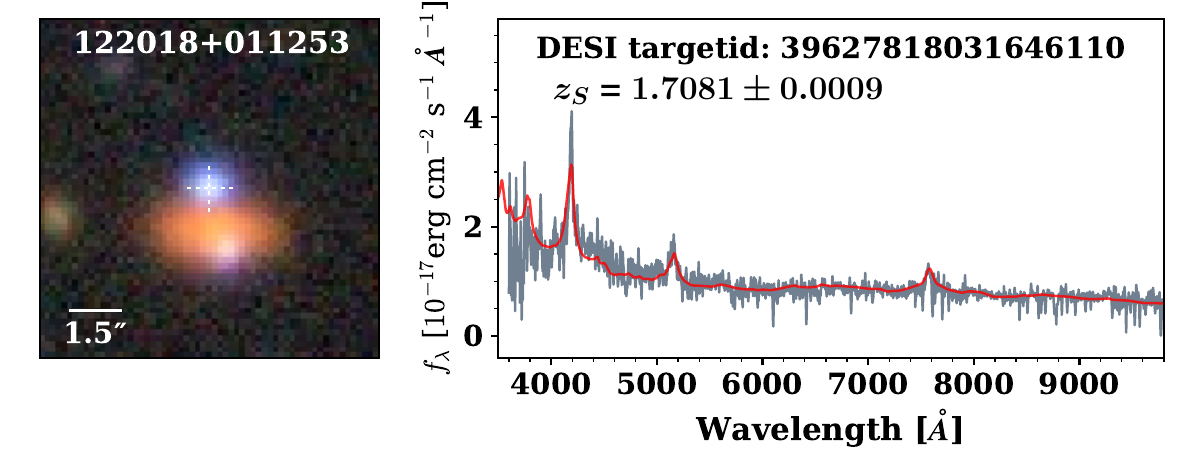}
\includegraphics[width=0.49\textwidth]{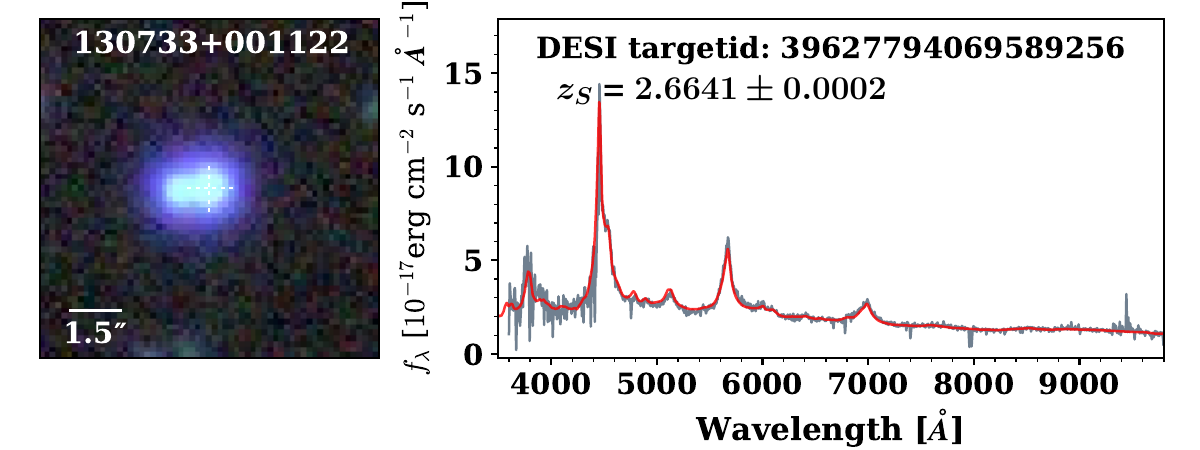}
\includegraphics[width=0.49\textwidth]{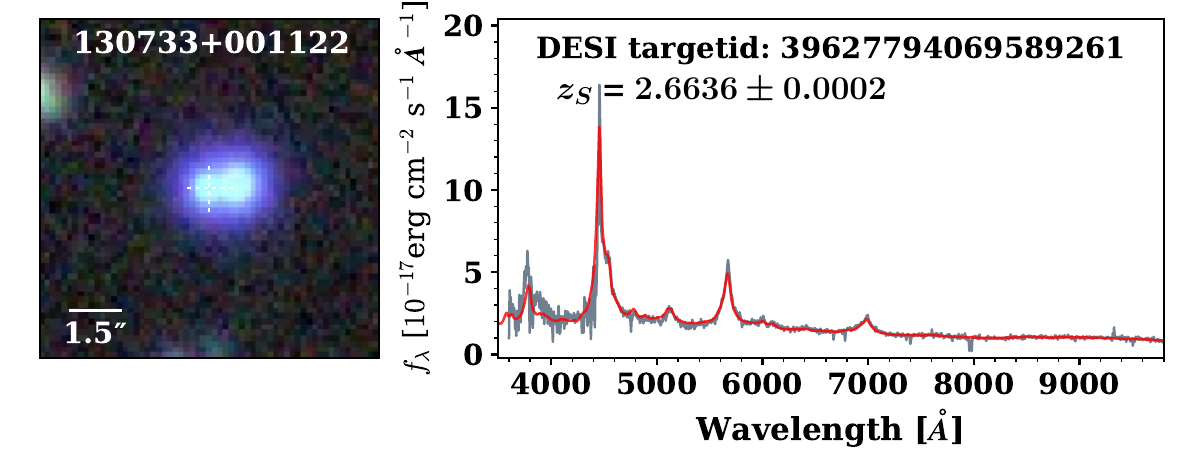}
\includegraphics[width=0.49\textwidth]{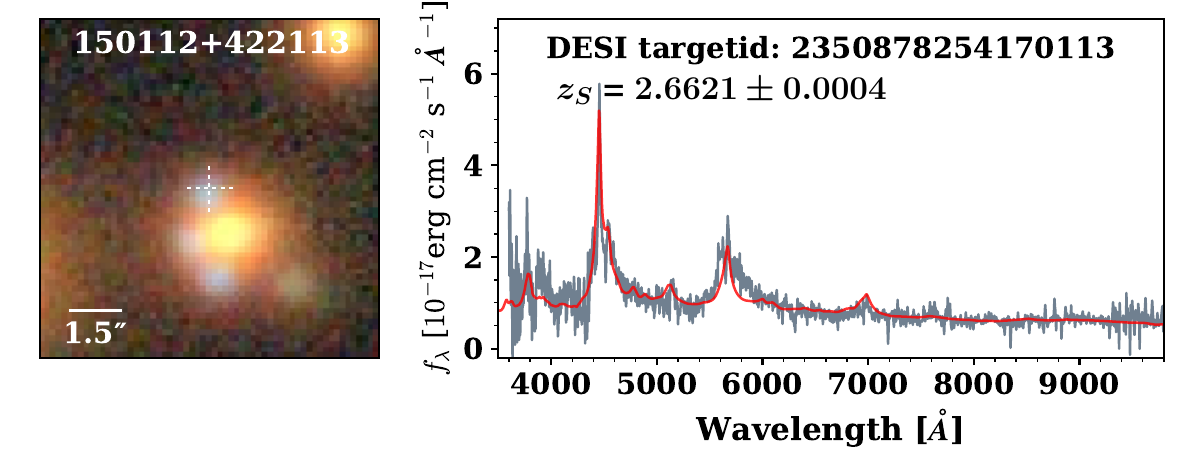}
\hspace*{0.49\textwidth}
\caption{Color-composite cutouts and DESI spectra for the 10 systems in Category 3. Symbols and lines follow the same conventions as in Figure~\ref{fig:C1}. 
}
\label{fig:C3}
\end{figure*}

\subsection{\label{sec:tables} Summary tables for the three categories}
\end{appendix}

\end{multicols}

\newpage
\renewcommand{\thetable}{A\arabic{table}}
\footnotesize
\begin{center}
\begin{longtable}{c c c c c c c}
\caption{\label{tab1} Grade-A candidates with secure lens and source redshifts. Redshifts marked with an asterisk ($^{*}$) indicate values estimated in this work rather than those reported in DESI DR1.} \\
\hline 
\multicolumn{1}{c}{Name} & \multicolumn{1}{c}{R.A.} & \multicolumn{1}{c}{Decl. } & \multicolumn{1}{c}{Grade} & \multicolumn{1}{c}{$z_L$} & \multicolumn{1}{c}{$z_S$} & \multicolumn{1}{c}{Discovery paper} \\ \hline
\endfirsthead 
\multicolumn{7}{c} {{\bfseries \tablename\ \thetable{} -- continued from previous page}} \\ 
\hline \multicolumn{1}{c}{Name} & \multicolumn{1}{c}{R.A.} & \multicolumn{1}{c}{Decl. } & \multicolumn{1}{c}{Grade} & \multicolumn{1}{c}{$z_L$} & \multicolumn{1}{c}{$z_S$} & \multicolumn{1}{c}{Discovery paper} \\ \hline
\endhead 
\hline \multicolumn{7}{r}{{Continued on next page}} \\
\hline
\endfoot 
\hline 
\endlastfoot 
HSC\,J0015$+$0137      & 3.74066 & \phantom{$-$}1.61701 & A & $0.37884 \pm 0.00005$ & $1.165247 \pm 0.000013$ & \cite{holismokesXVI} \\
012018$+$001125      & 20.07557 & \phantom{$-$}0.19049 & A & $0.59852 \pm 0.00011$ & $2.9994  \pm 0.0009 \, ^{*}$ & \cite{holismokesVIII} \\
015310$-$042315 & 28.29412 & -4.38766 & A & $0.28725 \pm 0.00007$ & $0.64816 \pm 0.00003 \, ^{*}$  & \cite{holismokesVIII}           \\
083933$-$014044      & 129.88900 & -1.67900 & A & $0.26991 \pm 0.00006$ & $0.70916 \pm 0.00005$ & \cite{sugohiV}           \\
085046$+$003905      & 132.69421 & \phantom{$-$}0.65149 & A & $0.8844 \pm 0.0003$ & $2.83777 \pm 0.00011 \, ^{*}$ & \cite{sugohiVI} \\
090430$+$042648      & 136.12750 & \phantom{$-$}4.44680 & A & $0.45646 \pm 0.00004$ & $1.01730 \pm 0.00003$ & \cite{sugohiV} \\
090938$+$002842      & 137.41082 & \phantom{$-$}0.47849 & A & $0.77951 \pm 0.00009$ & $2.5459 \pm 0.0003 \, ^{*}$  & \cite{sugohiVI} \\
094348$+$005926      & 145.95070 & \phantom{$-$}0.99060 & A & $0.43817 \pm 0.000007$ & $1.098909 \pm 0.000004$ & \cite{sugohiV} \\
HSC\,J1004$-$0031      & 151.21577 & -0.52915 & A & $0.56877 \pm 0.00012$ & $1.44286 \pm 0.00004$ & \cite{holismokesVI} \\
HSC\,J1014$+$0332      & 153.63908 & \phantom{$-$}3.54713 & A & $0.67770 \pm 0.00008$ & $2.165 \pm 0.002 \, ^{*} $ & \cite{holismokesXVI}  \\
114311$-$013935      & 175.79700 & -1.65983 & A & $0.6436 \pm 0.0002 \, ^{*} $ & $1.38763 \pm 0.00002$ & \cite{sugohiII} \\
122450$-$004215      & 186.20950 & -0.70420 & A & $0.40355 \pm 0.00005$ & $1.17846 \pm 0.00004\, ^{*}$ & \cite{Petrillo19} \\
123636$-$003539      & 189.15056 & -0.59418 & A & $0.51016 \pm 0.00009$ & $0.763224 \pm 0.000013\, ^{*}$ & \cite{holismokesVIII}      \\
141649$+$013822      & 214.20757 & \phantom{$-$}1.63953 & A & $0.43371 \pm 0.00006$ & $1.31300 \pm 0.00002\, ^{*}$ & \cite{sugohiVI} \\
155957$+$441543      & 239.98978 & \phantom{$-$}44.26217 & A & $0.59761 \pm 0.00009$ & $1.52903 \pm 0.00007$ & \cite{sugohiI} \\
222638$-$003449      & 336.65950 & -0.58030 & A & $0.40425 \pm 0.00010$ & $1.34701 \pm 0.00007\, ^{*}$ & \cite{sugohiV} \\
224154$+$000331      & 340.47761 & \phantom{$-$}0.05878 & A & $0.49660 \pm 0.00005$ & $1.61508 \pm 0.00005$ & \cite{sugohiVI} \\
224221$+$001144      & 340.58993 & \phantom{$-$}0.19575 & A & $0.38531 \pm 0.00007$ & $1.23193 \pm 0.00003$ & \cite{sugohiI} \\
\end{longtable}
\end{center}

\begin{center}
\begin{longtable}{c c c c c c c}
\caption{\label{tab2} Grade-A/B candidates with $z > 0.8$ lensing galaxies.  Redshifts marked with an asterisk ($^{*}$) indicate values estimated in this work rather than those reported in DESI DR1.} \\
\hline 
\multicolumn{1}{c}{Name} & \multicolumn{1}{c}{R.A.} & \multicolumn{1}{c}{Decl. } & \multicolumn{1}{c}{Grade} & \multicolumn{1}{c}{$z_L$} & \multicolumn{1}{c}{$z_S$} & \multicolumn{1}{c}{Discovery paper} \\ \hline
\endfirsthead 
\multicolumn{7}{c} {{\bfseries \tablename\ \thetable{} -- continued from previous page}} \\ 
\hline \multicolumn{1}{c}{Name} & \multicolumn{1}{c}{R.A.} & \multicolumn{1}{c}{Decl. } & \multicolumn{1}{c}{Grade} & \multicolumn{1}{c}{$z_L$} & \multicolumn{1}{c}{$z_S$} & \multicolumn{1}{c}{Discovery paper} \\ \hline
\endhead 
\hline \multicolumn{7}{r}{{Continued on next page}} \\
\hline
\endfoot 
\hline 
\endlastfoot 
083651$+$003038      & 129.21623 & \phantom{$-$}0.51056 & B & $0.83603 \pm 0.00010$ & $3.0629 \pm 0.0003\, ^{*}$ & \cite{holismokesVIII} \\
090402$+$031403      & 136.00876 & \phantom{$-$}3.23422 & B & $0.84681 \pm 0.00010$ & $3.3528 \pm 0.0002\, ^{*}$ & \cite{sugohiX} \\
090548$+$004743      & 136.45329 & \phantom{$-$}0.79538 & B & $0.85352 \pm 0.00008$ & $2.91165 \pm 0.00010\, ^{*}$ & \cite{sugohiVI}  \\
HSC\,J1104$-$0052      & 166.23010 & -0.87796 & B & $0.87299 \pm 0.00017$ & $2.45593 \pm 0.00019\, ^{*}$ & \cite{holismokesXVI}   \\
HSC\,J1107$-$0037      & 166.88870 & -0.63251 & B & $0.81443 \pm 0.00010$ & $1.4106 \pm 0.0003\, ^{*}$ & \cite{holismokesXVI}   \\
115630$-$020027      & 179.12894 & -2.00767 & B & $0.9983 \pm 0.0002$ & $2.2298 \pm 0.0005\, ^{*}$ & \cite{sugohiVI} \\
120657$-$010241      & 181.74033 & -1.04476 & B & $0.83004 \pm 0.00015$ & $2.7732 \pm 0.0003\, ^{*}$ & \cite{sugohiX}  \\
HSC\,J0102$-$0028    & 15.59200 & -0.48117 & B & $0.94932 \pm 0.00016$ & --- &\cite{holismokesXVI}    \\
020810$-$040220      & 32.04497 & -4.03891 & A & $0.9963 \pm 0.0002$ & --- & \cite{More2016} \\
022400$-$034625      & 36.00383 & -3.77384 & A & $0.83816 \pm 0.00009$ & --- & \cite{More12} \\
023103$-$021050      & 37.76462 & -2.18061 & B & $0.81166 \pm 0.00008$ & --- & \cite{sugohiVI} \\
023331$-$032801      & 38.38310 & -3.46700 & B & $1.04868 \pm 0.00013$ & --- & \cite{sugohiV}     \\
084423$-$001738      & 131.09584 & -0.29401 & B & $0.81546 \pm 0.00012$ & --- &  \cite{holismokesVI} \\
084536$-$000456      & 131.40025 & -0.08246 & B & $0.80634 \pm 0.00011$ & --- & \cite{sugohiVI} \\
084632$-$015416      & 131.63580 & -1.90470 & B & $0.89826 \pm 0.00015$ & --- & \cite{sugohiVI} \\
085550$-$001755      & 133.96162 & -0.29863 & B & $0.80848 \pm 0.00011$ & --- & \cite{sugohiVI} \\
090241$+$025318      & 135.67125 & \phantom{$-$}2.88838 & A & $0.82512 \pm 0.00011$ & --- & \cite{sugohiVI}  \\
090404$+$012516      & 136.01790 & \phantom{$-$}1.4211 & B & $0.82262 \pm 0.00011$ & --- & \cite{sugohiV}  \\
090429$-$010228      & 136.12390 & -1.04110 & A & $0.95820 \pm 0.00011$ & --- & \cite{sugohiV}      \\
090603$+$005146      & 136.51603 & \phantom{$-$}0.86283 & B & $0.88590 \pm 0.00010$ & --- & \cite{sugohiX}    \\
090618$+$003053      & 136.57887 & \phantom{$-$}0.51496 & B & $0.81073 \pm 0.00007$ & --- & \cite{sugohiVI} \\
090640$+$010346      & 136.66669 & \phantom{$-$}1.06304 & B & $0.85585 \pm 0.00010$ & --- & \cite{sugohiVI}  \\
090658$+$015449      & 136.74363 & \phantom{$-$}1.91368 & B & $0.8972 \pm 0.0002$ & --- & \cite{holismokesVIII}  \\
092146$+$025226      & 140.44574 & \phantom{$-$}2.87408 & B & $0.82390 \pm 0.00007$ & --- & \cite{sugohiVI}  \\
092235$+$025942      & 140.64640 & \phantom{$-$}2.99500 & B & $1.08784 \pm 0.00018$ & --- & \cite{sugohiV}    \\
092418$+$045843      & 141.07788 & \phantom{$-$}4.97871 & B & $0.88796 \pm 0.00010$ & --- & \cite{sugohiVI}  \\
092544$+$001702      & 141.43740 & \phantom{$-$}0.28410 & A & $0.81150 \pm 0.00009$ & --- & \cite{sugohiV} \\
093036$-$020204      & 142.65360 & -2.03445 & B & $0.920160 \pm 0.000010$ & --- & \cite{sugohiX}   \\
HSC\,J1002$-$0050      & 150.57397 & -0.83380 & B & $0.80054 \pm 0.00009$ & --- & \cite{holismokesXVI}  \\
100554$-$010357      & 151.47550 & -1.06590 & B & $1.08714 \pm 0.00011$ & --- & \cite{sugohiV}        \\
100712$-$012316  & 151.80080 & -1.38780 & B & $0.89292 \pm 0.00012$ & --- & \cite{sugohiV} \\
HSC\,J1040$+$0117      & 160.11316 & \phantom{$-$}1.29061 & B & $0.87413 \pm 0.00013$ & --- & \cite{holismokesXVI} \\
114745$-$001349      & 176.93810 & -0.23040 & B & $0.82855 \pm 0.00009$ & --- & \cite{sugohiV} \\
115057$-$015316      & 177.73919 & -1.88799 & B & $1.01710 \pm 0.00007$ & --- & \cite{sugohiVI}    \\
115529$-$004255      & 178.87266 & -0.71553 & B & $0.81489 \pm 0.00010$ & --- & \cite{sugohiVI}  \\
115857$-$000531      & 179.74062 & -0.09216 & B & $0.86199 \pm 0.00012$ & --- & \cite{sugohiVI} \\
120053$-$013357      & 180.22198 & -1.56605 & B & $0.85711 \pm 0.00008$ & --- & \cite{sugohiVI}      \\
120150$+$002519      & 180.46020 & \phantom{$-$}0.42220 & B & $0.82303 \pm 0.00011$ & --- & \cite{sugohiV}  \\
120605$-$001554      & 181.52107 & -0.26500 & B & $0.91498 \pm 0.00011$ & --- & \cite{sugohiX}    \\
120849$+$012845      & 182.20520 & \phantom{$-$}1.47920 & B & $0.85938 \pm 0.00014$ & --- & \cite{sugohiV}  \\
121844$+$010805      & 184.68560 & \phantom{$-$}1.13473 & B & $0.83806 \pm 0.00017$ & --- & \cite{holismokesVIII}   \\
122102$+$001853      & 185.25948 & \phantom{$-$}0.31493 & A & $0.91152 \pm 0.00013$ & --- & \cite{holismokesVIII} \\
122406$-$012853      & 186.02915 & -1.48158 & B & $0.83016 \pm 0.00008$ & --- & \cite{holismokesVIII} \\
HSC\,J1231$-$0135      & 187.89751 & -1.59427 & B & $0.94332 \pm 0.00011$ & --- & \cite{holismokesVI}  \\
HSC\,J1241$+$0026      & 190.41521 & \phantom{$-$}0.43597 & B & $0.94519 \pm 0.00007$ & --- & \cite{holismokesVI}  \\
124123$+$001603      & 190.34835 & \phantom{$-$}0.26773 & B & $0.9450 \pm 0.0002$ & --- & \cite{sugohiX} \\
135254$+$001131      & 208.22833 & \phantom{$-$}0.19220 & B & $0.93232 \pm 0.00011$ & --- & \cite{sugohiX}   \\
135304$+$010818      & 208.26781 & \phantom{$-$}1.13840 & B & $0.85526 \pm 0.00010$ & --- & \cite{holismokesVIII}   \\
135612$-$003603      & 209.05057 & -0.60098 & B & $0.80427 \pm 0.00011$ & --- & \cite{holismokesVI} \\
135710$+$010325      & 209.29493 & \phantom{$-$}1.05706 & A & $0.82229 \pm 0.00009$ & --- & \cite{holismokesVIII}  \\
141105$+$002532      & 212.77400 & \phantom{$-$}0.42560 & B & $1.07222 \pm 0.00012$ & --- & \cite{sugohiV}   \\
141359$-$013336      & 213.49800 & -1.56000 & B & $1.09021 \pm 0.00011$ & --- & \cite{sugohiV}    \\
141735$+$522646      & 214.39885 & \phantom{$-$}52.44613 & A & $0.81011 \pm 0.00013$ & --- & \cite{holismokesVIII} \\
141805$+$004435      & 214.52310 & \phantom{$-$}0.74310 & B & $0.84814 \pm 0.00015$ & --- & \cite{sugohiV}  \\
142013$+$005925      & 215.05590 & \phantom{$-$}0.99040 & B & $0.8235 \pm 0.0002$ & --- & \cite{sugohiV} \\
142154$+$004841      & 215.47703 & \phantom{$-$}0.81156 & B & $0.89750 \pm 0.00007$ & --- & \cite{sugohiVI} \\
142234$-$000225      & 215.64285 & -0.04038 & B & $1.04339 \pm 0.00009$ & --- & \cite{sugohiVI}    \\
142336$-$004034      & 215.90399 & -0.67631 & B & $0.89922 \pm 0.00005$ & --- & \cite{sugohiVI}    \\
142410$-$005317      & 216.04515 & -0.88809 & B & $0.88051 \pm 0.00012$ & --- & \cite{sugohiVI} \\
142602$+$010842      & 216.51222 & \phantom{$-$}1.14500 & B & $0.94610 \pm 0.00012$ & --- & \cite{sugohiVI}   \\
142754$+$003944      & 216.97867 & \phantom{$-$}0.66240 & B & $0.85354 \pm 0.00013$ & --- & \cite{sugohiVI}   \\
143832$-$002326      & 219.63530 & -0.39070 & B & $0.84821 \pm 0.00008$ & --- & \cite{sugohiV} \\
144320$-$012537      & 220.83630 & -1.42695 & A & $0.8895 \pm 0.0003$ & --- & \cite{sugohiIV} \\
144354$-$000731      & 220.97910 & -0.12530 & B & $0.80560 \pm 0.00007$ & --- & \cite{sugohiV}  \\
144640$-$000350      & 221.66716 & -0.06395 & B & $0.86350 \pm 0.00008$ & --- & \cite{sugohiVI} \\
HSC\,J1546$+$4337      & 236.52626 & \phantom{$-$}43.61835 & B & $0.84233 \pm 0.00007$ & --- & \cite{holismokesXVI}  \\
155625$+$432413      & 239.10825 & \phantom{$-$}43.40368 & B & $0.80564 \pm 0.00010$ & --- & \cite{holismokesVIII} \\
155806+431702      & 239.52558 & \phantom{$-$}43.28396 & B & $0.81260 \pm 0.00016$ & --- & \cite{sugohiVI}  \\
155920$+$423232      & 239.83650 & \phantom{$-$}42.54230 & B & $0.85064 \pm 0.00008$ & --- &  \cite{sugohiV}   \\
HSC\,J1600$+$4345      & 240.22091 & \phantom{$-$}43.75760 & B & $0.80501 \pm 0.00009$ & --- & \cite{holismokesXIII}        \\
161221$+$424908      & 243.09028 & \phantom{$-$}42.81916 & B & $0.82192 \pm 0.00014$ & --- & \cite{holismokesVI}  \\
220405$+$043454      & 331.02137 & \phantom{$-$}4.58194 & A & $0.99176 \pm 0.00014$ & --- & \cite{sugohiIX}  \\
221222$-$001811      & 333.09500 & -0.30310 & B & $0.91083 \pm 0.00019$ & --- & \cite{sugohiV} \\
221331$+$004836      & 333.38271 & \phantom{$-$}0.81004 & A & $0.90764 \pm 0.00011$ & --- & \cite{More12}    \\
222041$+$004912      & 335.17254 & \phantom{$-$}0.82011 & B & $0.83768 \pm 0.00011$ & --- & \cite{sugohiVI} \\
222503$-$004006      & 336.26300 & -0.66853 & B & $0.81041 \pm 0.00008$ & --- & \cite{sugohiVI} \\
223238$-$002533      & 338.16108 & -0.42611 & A & $1.04106 \pm 0.00017$ & --- & \cite{sugohiV}    \\
223316$-$010409    & 338.32000 & -1.06930 & B & $0.88420 \pm 0.00014$ & --- & \cite{sugohiV} \\
223324$+$012521      & 338.35267 & \phantom{$-$}1.42261 & B & $0.86068 \pm 0.00011$ & --- & \cite{sugohiVI} \\
223626$-$005400      & 339.10960 & -0.90021 & B & $0.82678 \pm 0.00013$ & --- & \cite{sugohiVI}      \\
HSC\,J2256$+$0202      & 344.10558 & \phantom{$-$}2.03446 & B & $0.82291 \pm 0.00011$ & --- & \cite{holismokesVI}  \\
232346$-$003037      & 350.94190 & -0.51050 & B & $0.84724 \pm 0.00012$ & --- & \cite{Jacobs19b} \\
232943$-$000141      & 352.42923 & -0.02826 & B & $0.83464 \pm 0.00015$ & --- & \cite{sugohiX} \\
\end{longtable}
\end{center}

\begin{center}
\begin{longtable}{c c c c c c c}
\caption{\label{tab3} Candidates with QSOs as sources/lenses. Redshifts marked with an asterisk ($^{*}$) indicate values estimated in this work rather than those reported in DESI DR1.} \\
\hline 
\multicolumn{1}{c}{Name} & \multicolumn{1}{c}{R.A.} & \multicolumn{1}{c}{Decl. } & \multicolumn{1}{c}{Grade} & \multicolumn{1}{c}{$z_L$} & \multicolumn{1}{c}{$z_S$} & \multicolumn{1}{c}{Discovery paper} \\ \hline
\endfirsthead 
\multicolumn{7}{c} {{\bfseries \tablename\ \thetable{} -- continued from previous page}} \\ 
\hline \multicolumn{1}{c}{Name} & \multicolumn{1}{c}{R.A.} & \multicolumn{1}{c}{Decl. } & \multicolumn{1}{c}{Grade} & \multicolumn{1}{c}{$z_L$} & \multicolumn{1}{c}{$z_S$} & \multicolumn{1}{c}{Discovery paper} \\ \hline
\endhead 
\hline \multicolumn{7}{r}{{Continued on next page}} \\
\hline
\endfoot 
\hline 
\endlastfoot 
004738$+$032232      & 11.91232 & \phantom{$-$}3.37556 & A    & --- & $2.0864 \pm 0.0009 $ & \cite{sugohiIX}                        \\ 
013827$+$031518      & 24.61357 & \phantom{$-$}3.25511 & A    & --- & $2.4057 \pm 0.0007 $ & \cite{sugohiIX}                        \\ 
014152$+$003956      & 25.46863 & \phantom{$-$}0.66559 & A    & --- & $2.2062 \pm 0.0003 $ & \cite{sugohiIX}                        \\ 
091151$-$004406      & 137.96487 & -0.73508 & A    & --- & $2.37547 \pm 0.00015 $ & \cite{sugohiIX}                        \\ 
092121$+$031744      & 140.34050 & \phantom{$-$}3.29560 & A    & $0.346027 \pm 0.000011$ & --- & \cite{sugohiV}                        \\
095406$-$002225      & 148.52904 & -0.37368 & A    & --- & $2.25975 \pm 0.00019 $ & \cite{sugohiIX}                        \\ 
121533$-$005842      & 183.88950 & -0.97856 & C    & $0.45912 \pm 0.00008 $ & $2.8817 \pm 0.0003$ & \cite{sugohiX}                       \\ 
122018$+$011253      & 185.07897 & \phantom{$-$}1.21497 & A    & $0.48755 \pm 0.00008$ & $1.7081 \pm 0.0009 $ & \cite{holismokesVIII} \\ 
130733$+$001122      & 196.88880 & \phantom{$-$}0.18966 & C    & --- & $2.66385 \pm 0.00014 $ & \cite{sugohiIX}                        \\
150112$+$422113      & 225.30070 & \phantom{$-$}42.35370 & A    & --- & $2.6621 \pm 0.0004 $ & \cite{sugohiV}                        \\ 
\end{longtable}
\end{center}

\end{document}